\documentclass[12pt,a4paper]{JHEP3} % 10pt is ignored!

\JHEPspecialurl{http://jhep.sissa.it/JOURNAL/JHEP3.tar.gz}

\usepackage{epsfig,multicol,bbm}

\usepackage{amsmath}
\usepackage{graphicx}
\usepackage{latexsym}
\usepackage{amsfonts}
\usepackage{amssymb}
\usepackage{epsfig}
\usepackage{psfrag}
\usepackage{dsfont}

\newcommand\fverb{\setbox\fverbbox=\hbox\bgroup\verb}
\newcommand\fverbdo{\egroup\medskip\noindent%
            \fbox{\unhbox\fverbbox}\ }
\newcommand\fverbit{\egroup\item[\fbox{\unhbox\fverbbox}]}
\newbox\fverbbox

\newcommand{\beq}{\begin{equation}}
\newcommand{\eeq}{\end{equation}}
\newcommand{\eq}{\end{equation}}
\newcommand{\beqa}{\begin{eqnarray}}
\newcommand{\eeqa}{\end{eqnarray}}

\newcommand{\mf}{\mathcal M}
\newcommand{\mmf}{\widetilde{\mathcal M}}

\newcommand{\C}{\mathbb C}
\newcommand{\IC}{\mathbb C}
\newcommand{\R}{\mathbb R}
\newcommand{\Z}{\mathbb Z}
\newcommand{\IZ}{\mathbb Z}
\newcommand{\IT}{\mathbb T}
\newcommand{\IS}{\mathbb S}
\renewcommand{\P}{\mathbb P}
\renewcommand{\O}{\mathcal O}
\newcommand{\N}{\mathcal N}

\newcommand{\ov}{\overline}

\newcommand{\fref}{Figure \ref}
\newcommand{\minv}{\widetilde\sigma}
\newcommand{\inv}{\sigma}

\newcommand{\cc}{\overline}

\newcommand{\drawsquare}[2]{\hbox{%
\rule{#2pt}{#1pt}\hskip-#2pt%  left vertical
\rule{#1pt}{#2pt}\hskip-#1pt%  lower horizontal
\rule[#1pt]{#1pt}{#2pt}}\rule[#1pt]{#2pt}{#2pt}\hskip-#2pt%  upper horizontal
\rule{#2pt}{#1pt}}% right vertical

\newcommand{\fund}{~\raisebox{-.5pt}{\drawsquare{6.5}{0.4}}~}
\newcommand{\antifund}{~\overline{\raisebox{-.5pt}{\drawsquare{6.5}{0.4}}}~}

\newcommand{\symm}{~\raisebox{-.5pt}{\drawsquare{6.5}{0.4}}\hskip-0.4pt%
        \raisebox{-.5pt}{\drawsquare{6.5}{0.4}}~}%  symmetric second rank

\newcommand{\asymm}{~\raisebox{-3.5pt}{\drawsquare{6.5}{0.4}}\hskip-6.9pt%
        \raisebox{3pt}{\drawsquare{6.5}{0.4}}~}%  antisymmetric second rank

\newcommand{\antiasymm}{~\overline{\raisebox{-3.5pt}{\drawsquare{6.5}{0.4}}\hskip-6.9pt%
        \raisebox{3pt}{\drawsquare{6.5}{0.4}}}~}%  antisymmetric second rank

\newcommand{\antisymm}{~\overline{\raisebox{-.5pt}{\drawsquare{6.5}{0.4}}\hskip-0.4pt%
        \raisebox{-.5pt}{\drawsquare{6.5}{0.4}}}~}%  symmetric second rank

\title{Dimers and Orientifolds}

\author{Sebasti\'an Franco$^1$, Amihay Hanany$^2$, Daniel Krefl$^{3,4}$, Jaemo Park$^{5,6}$,}
\author{Angel M. Uranga$^{7,8}$ and David Vegh$^9$}

\author{\\$^1$ Joseph Henry Laboratories, Princeton University \\ Princeton NJ 08544, USA
\\$^2$ Perimeter Institute for Theoretical Physics \\ 31 Caroline Street North, Waterloo Ontario N2L 2Y5 Canada
\\$^3$ Max-Planck-Institut f\"ur Physik\\F\"ohringer Ring 6, 80805 Munich, Germany
\\$^4$ Arnold Sommerfeld Center for Theoretical Physics, \\Ludwig-Maximilians-Universit\"at, Theresienstr. 37, 
80333 Munich, Germany
\\$^5$ Department of Physics, Postech, Pohang 790-784 Korea
\\$^6$ Postech Center for Theoretical Physics (PCTP), Pohang 790-784, Korea
\\$^7$ PH-TH Division, CERN\\ CH-1211 Geneva, Switzerland
\\$^8$ On leave from Instituto de F\'{\i}sica Te\'orica, Facultad de Ciencias, Madrid, Spain
\\$^9$ Center for Theoretical Physics, Massachusetts Institute of Technology, \\
77 Massachusetts Avenue, Cambridge MA 02139, USA
\\

}

\preprint{
\\CERN-PH-TH/2007-099
\\IFT-UAM/CSIC-07-34
\\LMU-ASC 41/07
\\MIT-CTP 3846
\\MPP-2007-76
\\PUPT-2238
}

\abstract{We introduce new techniques based on brane tilings to
  investigate D3-branes probing orientifolds of toric 
Calabi-Yau singularities. With these new tools, one can write down
  many orientifold models and derive the resulting 
low-energy gauge theories living on the D-branes. Using the set of
  ideas in this paper one recovers essentially all 
orientifolded theories known so far. Furthermore, new orientifolds of
  non-orbifold toric singularities are obtained.
 The possible applications of the tools presented in this paper are
  diverse. One particular application is 
the construction of models which feature dynamical supersymmetry
  breaking as well as the computation of
 D-instanton induced superpotential terms.
}

\begin{document}

%%%%%%%%%%%%%%%%%%%%%%%%%%%%%%%%%%%%%%%%%%%%%%%%%%%%%%%%%%%%%%%%%%%%%%%%%%%%%%%%
\section{Introduction}
\label{intro}

The study of D-branes at singularities and the gauge theories on them
is interesting for a variety of reasons. On the one hand, 
branes at singularities give rise to interesting extensions of the
original AdS/CFT correspondence to theories with a reduced 
amount of (super)symmetry \cite{Klebanov:1998hh, 
Morrison:1998cs}. This front has witnessed remarkable progress in
recent years: in the conformal case, with the precision 
matching of geometric properties of new infinite families of
Sasaki-Einstein metrics and their gauge theory 
duals
\cite{Gauntlett:2004yd,Gauntlett:2004hh,Gauntlett:2004hs,Martelli:2004wu,Benvenuti:2004dy};
in the presence of fractional branes, with the dictionary between geometric properties
of the singularity and strong infrared 
dynamics in the dual gauge theories \cite{Klebanov:2000hb,Franco:2005fd,Berenstein:2005xa,Franco:2005zu,Bertolini:2005di}.

On the other hand, they provide a natural setup for a bottom-up approach to string phenomenology, allowing for local 
constructions of Standard Model-like gauge theories \cite{Aldazabal:2000sa,Berenstein:2001nk,Verlinde:2005jr}. 
Many features
of the resulting models depend only on the local structure of the singularity and can be investigated without a detailed 
knowledge of the full compactification manifold.

Orientifolds are an interesting new twist in these constructions, with possibly novel features. To name a few, they can be 
used to produce interesting spectra (with new kinds of gauge factors and representations), they naturally lead to 
non-conformal theories (with orientifold charges arising as $1/N$ corrections), easily lead to supersymmetry breaking in the 
infrared (with or without runaway), and 
lead to models with interesting non-perturbative superpotential interactions
(since orientifolding eliminates superfluous zero modes on certain D-brane instantons). Thus, the construction of 
orientifolds of D-branes at singularities is an interesting direction worth being pursued.

Unfortunately, the techniques to construct such orientifolds are very rudimentary.
So far, only a very limited number of orientifolds of non-orbifold singularities has been constructed. Orientifolds of 
orbifold singularities are in principle amenable to direct construction using worldsheet techniques, although in practice 
only a few families of models have been constructed. For orientifolds of non-orbifold singularities, partial resolution of 
orientifolds of orbifolds can be used  to derive a few new models, but the approach becomes non-practical for singularities 
beyond the simplest ones. A simple classification of orientifolds is also possible for the very restricted subset of singularities that are T-dual to simple Hanany-Witten (HW) setups \cite{Hanany:1996ie}.

\medskip

The problem of finding the gauge theory on a set of branes probing an
arbitrary toric CY singularity $\mf$ was fully 
solved with the introduction of dimer model methods
\cite{Hanany:2005ve,Franco:2005rj}.\footnote{For a recent review, see \cite{Kennaway:2007tq}.} One of the main virtues of 
dimers is its computational simplicity, in sharp contrast with pre-existent alternatives such as partial resolution.

Given the striking success of dimer models in the study of branes at
singularities, it is natural to ask how to expand 
their range of applicability. A natural problem is the classification
of orientifolds of toric singularities. This paper 
extends the success of dimer models in the study of toric
singularities to their orientifolds, providing a general method 
and explicit construction of them.

The organization of the paper is as follows. In \S\ref{background}, we review some basics of dimers,
quivers and orientifolds. The main results are presented in \S\ref{points} and \S\ref{lines}, where we explain how to 
obtain orientifolds of arbitrary toric singularities corresponding to involutions with fixed points and fixed lines, 
respectively.
The mirror perspective is discussed in
\S\ref{mirrorsec}. In \S\ref{appl} we present some applications of our framework and conclude in \S\ref{conclu}. We collect additional related material in appendices.

%%%%%%%%%%%%%%%%%%%%%%%%%%%%%%%%%%%%%%%%%%%%%%%%%%%%%%%%%%%%%%%%%%%%%%%%%%%%%%%%
\section{Some background on dimers and on orientifolds}
\label{background}
%%%%%%%%%%%%%%%%%%%%%%%%%%%%%%%%%%%%%%%%%%%%%%%%%%%%%%%%%%%%%%%%%%%%%%%%%%%%%%%%

%===============================================================================
\subsection{Quiver gauge theories and dimer diagrams}
\label{introdimer}
%===============================================================================

In this section we give several aspects of quiver gauge theories that live on D3-branes at toric
Calabi-Yau singularities and their construction in terms of dimer diagrams (a.k.a. brane
tilings). Brane tilings give the most efficient construction of such quiver gauge theories and
in the math literature one can find them under the name of periodic bipartite tilings of the two
dimensional plane. For the physical application of dimers to D-branes at singularities, see
e.g. \cite{Hanany:2005ve}-\cite{Imamura:2007dc}. A review of the mathematical aspects of dimers can be found in \cite{Kenyon2003,Kenyon:2003uj}.

The gauge theory living on D3-branes probing a toric threefold singularity $\mf$ is
determined by a set of unitary gauge factors (vector multiplets with a unitary gauge group), chiral multiplets in 
bi-fundamental representations, and a superpotential
given by a sum of traces of products of such bi-fundamental fields.
The gauge group and matter content of such gauge theories can be encoded
in a quiver diagram, with nodes corresponding to gauge factors, and
arrows to bi-fundamentals.\footnote{It is 
also possible to encode adjoint fields in quiver diagrams. They are represented by arrows starting and ending at the same node.}
The superpotential terms are gauge invariant operators and hence correspond to closed oriented loops of arrows in 
the quiver, however not all such loops in the quiver are superpotential terms.

A powerful tool that encodes all the gauge theory information,
including the gauge group, the matter content and 
the superpotential, is given by the so-called brane tiling or dimer
diagrams \cite{Hanany:2005ve,Franco:2005rj}. 
This is a tiling of $\IT^2$ defined by a bipartite graph, namely one
whose nodes can be colored black and white, 
with no edges connecting nodes of the same color. There is a dictionary 
between the tiling data and the gauge theory data that associates
faces in the dimer diagram to gauge factors 
in the field theory, edges with bi-fundamental fields (fields in the adjoint in the case that the
same face is at both sides of the edge), and nodes with superpotential
terms. The bipartite character of the diagram 
is important in that it defines an orientation for edges (e.g. from
black to white nodes), which determines 
the chirality of the bi-fundamental fields (for example, going
clockwise around white nodes and counter-clockwise 
around black ones). The color of a node determines the sign of the
corresponding superpotential term, 
by convention $+(-)$ sign for a white (black) node.

The inverse procedure, of computing the dimer from the gauge theory
data, was discussed in detail in \cite{Hanany:2005ss} 
and is given a new light with a recipe which is provided in
\S\ref{Dchains}. Different techniques allow to systematically 
construct the dimer diagram (and hence recover the gauge theory) for
any given toric 
singularity
\cite{Franco:2005sm,Hanany:2005ss,Garcia-Etxebarria:2006aq}. One can
also directly obtain 
the geometry of the singularity from the information of the dimer
diagram 
(i.e. of the gauge theory) \cite{Hanany:2005ss,Feng:2005gw}. A novel way to achieve this is introduced in Appendix \ref{Schains}.
\\\\
\noindent{\bf Mesonic operators and paths in the dimer}
\\\\
A set of objects which are of interest below are the gauge invariant
BPS mesonic operators. These objects, 
which are elements of the chiral ring are the simplest gauge invariant
operators in the gauge theory. 
They are naturally defined in the dimer diagram and encode the
interplay between the gauge theory and 
the singular geometry. There is a special subset of the mesonic
operators which generate the chiral ring and 
they are used in the algebraic description of the manifold. In this
paper they are sometimes called
 ``fundamental mesons". See \fref{Cfig1} for a simple example of such
 fundamental mesons. 
As discussed since the early days of D-branes at singularities
\cite{Douglas:1996sw,Douglas:1997de}, 
the geometry transverse to a D-brane can be regarded as the (mesonic)
moduli space of the gauge theory 
living on its world-volume. The gauge invariant mesonic operators are
then good complex coordinates on the 
Calabi-Yau singularity at which the D-branes are located. The full set
of mesonic operators, recently 
studied in
\cite{Benvenuti:2005ja,Benvenuti:2005cz,Benvenuti:2006qr,Feng:2007ur},
has a natural realization 
in the dimer diagram. They correspond to closed oriented paths in the
dimer, with orientation determined 
by the orientation of the edges.

Since each mesonic operator corresponds to an algebraic function on
the singular geometry, the complete set 
of gauge invariant mesonic operators provides the complete set of
algebraic functions, which in the spirit of 
algebraic geometry is an equivalent description of the geometry
itself. The use of mesonic operators as 
coordinates of the singular geometry are exploited in \S\ref{geomaction}.

A last important comment concerns the global symmetries of the gauge theory. In general, gauge theories living on D3-branes 
at toric singularities have a generic 
$U(1)^3$ global symmetry, associated to the toric action on the geometry.\footnote{In the absence of fractional branes this symmetry is not broken by quantum effects. In the presence of fractional branes, the symmetry is still present but may be broken by quantum effects. In addition, there are also baryonic $U(1)$ symmetries which are not used in this paper.}
This includes the $U(1)_R$ symmetry, and two global symmetries, which are sometimes called ``flavor symmetries" in the 
literature. The charges of the different fields under $U(1)_R$ can be represented geometrically in the dimer
diagram as the angles of the corresponding edges when the dimer diagram is drawn in the so-called isoradial 
embedding \cite{Hanany:2005ss}. The two remaining global symmetries are associated to the non-trivial 1-cycles in the 
two-torus of the dimer diagram. A possible convention for these two
global flavor charges can be as follows: given a 
gauge invariant mesonic operator, represent it in the universal cover of the dimer as an open path between one fundamental 
domain to another, assign lattice coordinates which specify the position of each fundamental domain, and assign the 
lattice difference between the starting fundamental domain to the ending fundamental domain as the two global flavor charges.

%===============================================================================
\subsection{Orientifolds}
\label{ofolds}

%===============================================================================

In this paper we are interested in studying systems of D3-branes that
probe a (toric) CY singularity $\mf$ 
in the presence of orientifold quotients. Namely, we consider the quotient of the system by the orientifold action
\beq
\omega\sigma (-1)^{F_L},
\eeq
where $\omega$ reverses the orientation of the worldsheet, while $\sigma$ is an involution of $\mf$ i.e. a discrete 
isometric diffeomorphism, and $F_L$ is the left-moving fermion number. The fixed-point loci of $\inv$ are orientifold planes. 
They are denoted by $O^-$-planes or $O^+$-planes, according to their RR charge (in the convention that the corresponding 
D-branes carry positive charge). Usually we work in the covering space description (which is referred to as the parent theory).

In order that the orientifold quotient preserves a common supersymmetry with the D3-brane system (and the Calabi-Yau geometry), 
the involution must act  on the globally defined holomorphic 3-form $\Omega$ as
\beq\label{geneq1}
\Omega\rightarrow -\Omega.
\eeq

In later sections we show how to construct systematically large classes of orientifold models using a simple set of rules 
in the dimer diagram, which allow to easily read off the resulting field theories on the D3-brane world-volume. 
Before doing this, it is convenient to review the current state of the art in the literature:

For D-branes at orbifolds, the construction of orientifold theories can be carried out using CFT techniques. 
Most of the models studied are compact toroidal orientifolds, 
see e.g.  \cite{Pradisi:1988xd,Pradisi:1988ue,Angelantonj:1996uy,Gimon:1996rq,Dabholkar:1996zi,Gimon:1996ay,Aldazabal:1998mr}. 
The study of non-compact singularities has not been extensive, but a few large classes of such orientifold models are 
known \cite{Ibanez:1998qp}, and related models using brane constructions 
can be found e.g. in \cite{Brunner:1998jr, Erlich:1999rb, Hanany:1999sj,Feng:1999fw, Feng:1999zv}.

For non-orbifold singularities, there is no systematic construction of orientifold quotients in the literature. 
The only available tool so far is to start with the orientifold of a
larger orbifold singularity and perform partial 
resolutions that result in Higgsing the corresponding theory. This
recipe, which generalizes \cite{Morrison:1998cs} 
for orientifold singularities, is very involved in practice, and only a few simple cases have been worked out in \cite{Park:1999ep}.

For a class of toric singularities (known as $L^{aba}$ theories in modern terminology
\cite{Benvenuti:2005ja,Franco:2005sm,Butti:2005sw}), it is possible to perform a T-duality
\cite{Uranga:1998vf,Dasgupta:1998su} to a Hanany-Witten (HW) setup \cite{Hanany:1996ie} and
introduce orientifold planes \cite{Landsteiner:1997vd,Landsteiner:1997ei,Uranga:1998uj}. Using
such tools, several classes of orientifolds of $\IC^2/\IZ_N\times \IC$ were constructed in
\cite{Park:1999eb} and recovered using CFT tools. Also several orientifolds of the SPP theory
and the conifold were constructed in \cite{Park:1999ep} and confirmed using the partial
resolution method mentioned above.

In this paper we show that dimer diagrams allow a systematic construction of orientifold quotients for arbitrary 
toric singularities.

%%%%%%%%%%%%%%%%%%%%%%%%%%%%%%%%%%%%%%%%%%%%%%%%%%%%%%%%%%%%%%%%%%%%%%%%%%%%%%%%
\subsection{Orientifolding dimers}
%%%%%%%%%%%%%%%%%%%%%%%%%%%%%%%%%%%%%%%%%%%%%%%%%%%%%%%%%%%%%%%%%%%%%%%%%%%%%%%%

\label{dimersec}

\begin{figure}
\begin{center}
\psfrag{a}[cc][][1]{a)}
\psfrag{b}[cc][][1]{b)}
\psfrag{1}[cc][][1]{1}
\psfrag{2}[cc][][1]{2}
\includegraphics[scale=1]{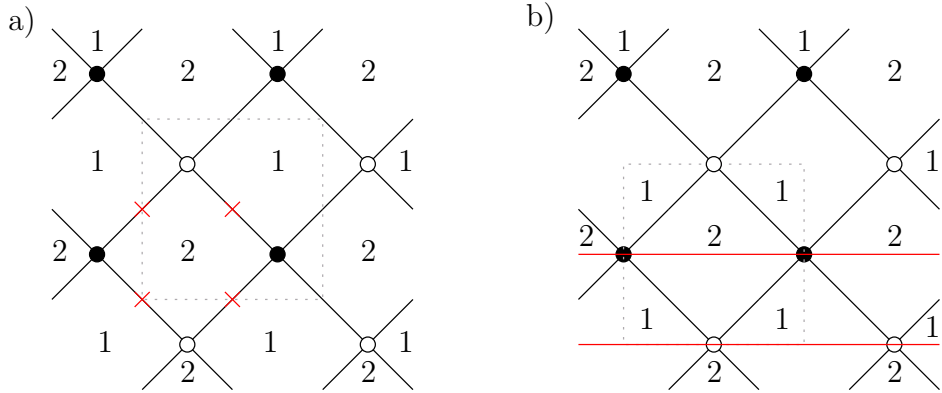}
\caption{The two basic examples of possible reflections as applied to the dimer of the conifold: a) A point reflection. 
b) A line reflection. The dotted box marks the unit cell of the periodic tiling and defines the $\IT^2$ of the dimer. 
The red crosses mark the fixed points under the point reflection in the $\IT^2$, while the red lines mark the fixed 
lines under the line reflection. Under point reflections white nodes are exchanged with black nodes, indicating arrow 
(orientation) reversal in the quiver.}
\label{odimersfig1}
\end{center}
\end{figure}

Dimer diagrams encode both the geometry of toric singularities and the field theories arising on D3-branes probing 
the singularity. Hence, we may expect the dimers to be natural objects to both  implement orientifold quotients of 
the system, and to read off the effect of the orientifold on the field theory.

Given a system of D-branes at singularities, the orientifold field theory is obtained from the parent theory by a 
certain $\IZ_2$ identification of gauge groups, chiral multiplets and superpotential couplings (for detailed descriptions 
see \S\ref{points}). This implies that the orientifold operation should be manifest as a $\IZ_2$ symmetry of the dimer diagram. 
The mapping of the faces give the gauge group identifications and the mappings of the edges give the chiral multiplet 
identifications while the mapping of the vertices give identifications of the superpotential monomials.\footnote{Although sometimes equivalent, this procedure should not be confused with the na\"ive quotient of a $\IZ_2$ symmetric quiver. As explained in \S\ref{points} and \S\ref{lines}, various constraints follow from the dimer construction. Direct orientifolding of the quiver usually leads to theories without interpretation as quotients by a geometric action and a Chan-Paton action.}

We may consider several kinds of such involutions, with different sets of fixed points. A given theory may admit several (or none) of such symmetries, corresponding to different possible orientifolds of the system. We consider two main classes:

\begin{itemize}

\item A possible involution corresponds to an inversion of the two
  $\IT^2$ coordinates of the dimer diagram, 
as shown in \fref{odimersfig1}a for the conifold dimer diagram. This
  is a reflection with respect to a point. 
It has several features:
\begin{enumerate}
\item This involution commutes with the $U(1)^3$ Cartan subgroup of
  the mesonic flavor symmetry of the theory: 
It preserves the angles in the dimer 
 and it preserves distances between different fundamental domains.
\item This involution exists for $\IT^2$ of arbitrary complex structure, hence we may expect it to exist for large classes 
of dimer diagrams.
\end{enumerate}
This kind of orientifold projection is discussed in detail in \S\ref{points}.

\item A second possible involution corresponds to an inversion of one of the $\IT^2$ directions. This is a reflection with 
respect to a fixed line. An example is shown in Figure \ref{odimersfig1}b. The two features listed above for point 
reflection are now different:
\begin{enumerate}
\item This involution does not commute with global symmetries and differs from the action in the case of point reflections. 
The parent theory has two mesonic flavor $U(1)$ symmetries. This involution breaks the two mesonic $U(1)$'s into a $U(1)$ 
subgroup which is some linear combination of the two.
\item Such a $\IZ_2$ symmetry exists only for specific $\IT^2$ complex structures, hence we may expect it to exist for a 
more restricted class of dimers (a necessary requirement is that the dimer, described in the isoradial embedding 
in \cite{Hanany:2005ss}, defines a $\IT^2$ with a complex structure that is compatible with the $\IZ_2$ action).
\end{enumerate}

This kind of orientifolds are a special case of involutions which keep a linear combination of global $U(1)$ symmetries 
intact and is presented in \S\ref{lines}.

\end{itemize}

%%%%%%%%%%%%%%%%%%%%%%%%%%%%%%%%%%%%%%%%%%%%%%%%%%%%%%%%%%%%%%%
\section{Orientifolds from dimers with fixed points}
\label{points}
%%%%%%%%%%%%%%%%%%%%%%%%%%%%%%%%%%%%%%%%%%%%%%%%%%%%%%%%%%%%%%%

%==============================================================
\subsection{Orientifold rules}
\label{section_generalities}
%==============================================================

In this section we consider systems of D3-branes at singularities in
the presence of orientifold actions which 
commute with the $U(1)^3$ subgroup of the mesonic flavor symmetry. As
described above, they should be manifest 
as $\IZ_2$ symmetries of the dimer diagram 
corresponding to reflections of the two $\IT^2$ coordinates.\footnote{
An additional requirement is that these reflections of the dimer
diagram must map black to white nodes and viceversa. 
Since the color of nodes determines the orientation of arrows
(bi-fundamentals), such reflections produce the 
orientation revearsal expected from an orientifold. Further, this
implies that about half of the superpotential 
terms would appear in the orientifolded theory.} A description in
terms of the mirror geometry is provided 
in \S\ref{mirrorsec}.

There are four points in the dimer diagram which are fixed under such $\IZ_2$ actions, and which correspond to objects 
(gauge factors or chiral multiplets) of the field theory which are mapped to themselves under the orientifold action. 
Using the interpretation of the dimer diagram as a brane tiling \cite{Franco:2005rj,Franco:2005sm}, namely as a physical 
configuration of D5-branes suspended on open discs with boundaries given by the faces of a web of NS-branes, such fixed 
points correspond to orientifold planes in the configuration. This interpretation introduces a possible sign, which can 
be assigned to each orientifold plane, which determines the specific orientifold projection on the vector multiplets or 
chiral multiplets as they are mapped to themselves. We refer to them as positive or negative orientifold planes, denoted 
as $O^+$, $O^-$. As is argued later, there is a global constraint
(related to supersymmetry) which restricts the number 
of orientifold planes of same sign to be either even or odd, depending
on the dimer diagram under consideration. 
This constraint is similar to a global constraint on signs of orientifold planes as in \cite{Keurentjes:2001sr, HJK}

Let us set notation by denoting a face with an index $a$ and its image
face under the orientifold action 
by $a'$ ($a=a'$ for faces on top of an orientifold plane). The
orientifold field theory is read out from 
the parent field theory and the orientifold action according to the following rules:

\begin{itemize}

\item Each face $a\not=a'$ gives a gauge factor $U(N_a)$ (with the face $a'$ being its image)

\item Each face $a=a'$ on top of an $O^+$ or $O^-$ plane gives a gauge factor $SO(N_a)$ or $Sp(N_a/2)$, 
respectively (clearly one needs even $N_a$ in the latter case).

\item A bi-fundamental chiral multiplet $(\fund_a,\antifund_b)$ of the parent theory, for $b\neq a'$, 
gives a bi-fundamental $(\fund_a,\antifund_b)$ of the orientifold theory (with its image $(\fund_{b'},\antifund_{a'})$ ).

\item A bi-fundamental chiral multiplet $(\fund_a,\antifund_{b'})$ of the parent theory gives 
a bi-fundamental $(\fund_a,\fund_b)$ of the orientifold theory. Similarly a bi-fundamental 
$(\fund_{a'},\antifund_b)$ gives a $(\antifund_a,\antifund_b)$.

\item An edge with $a=a'$ leading to bi-fundamentals $(\fund_a,\antifund_{a'})$ on top of an $O^+$ or 
$O^-$ plane, is projected down to a representation $\symm_a$ or $\asymm_a$, respectively. Similarly bi-fundamentals 
$(\fund_{a'},\antifund_a)$ lead to the conjugate representations, $\antisymm_a$ or $\antiasymm$, respectively.

\item In some cases additional (anti)fundamental fields are added in order to cancel chiral anomalies. An example of such a case is given in \S\ref{conifoldO}.

\item The superpotential of the orientifold theory is obtained from the parent superpotential by projecting 
out half of the terms and replacing in the surviving terms the parent fields by their images under the orientifold projection.

\end{itemize}

The two rules providing the orientifold projection on elements mapped to themselves under the orientifold 
action are not independent. They are related to each other by removing/adding an edge on top of an orientifold plane, 
see \S\ref{Higgsing}. This operation is well known and corresponds to Higgsing/unHiggsing in the parent 
theory \cite{Franco:2005rj}.

%================================================================
\subsubsection{Example: Orientifolds of $\IC^3$}
\label{C3}
%================================================================

Explicit examples of orientifold field theories obtained using this procedure are described in coming sections, 
but it may be convenient to illustrate the concepts with a simple example.
Let us consider the set of all possible point reflection orientifolds for the system of D3-branes on $\IC^3$. 
The dimer diagram is shown in Figure \ref{Cfig1}. There is one orientifold plane on top of the face 
(denoted $a$ in the figure), while the other three (b, c, d in the figure) are on top of edges. 
The different cases are as follows:

\begin{enumerate}

\item Consider the configuration where the signs of the orientifold planes are chosen to be $(a,b,c,d)=(+---)$. 
Using our rules above, the gauge group projects down to $SO(N)$ while each of the three chiral multiplets projects down to 
the antisymmetric (adjoint) representation. There are two superpotential terms which are inherited from the parent theory 
and give rise to a commutator term. The resulting theory corresponds to the ${\cal N}=4$ supersymmetric orientifold theory 
with gauge group $SO(N)$ obtained from D3-branes in flat space sitting on top of an O3$^-$ plane. Denote the three 
antisymmetric fields by $A_{1,2,3}$, then the projected superpotential is given by
\begin{equation}
W = A_1 A_2 A_3.
\label{Nis4}
\end{equation}
\item Similarly, the configuration with $(a,b,c,d)=(-+++)$ reproduces the orientifold with ${\cal N}=4$ supersymmetry and 
a gauge group $Sp(N/2)$, obtained by placing $N/2$ D3-branes on top of an O3$^+$ plane. 
The superpotential is as in Equation \eqref{Nis4} with the antisymmetrics replaced by symmetric representations.
\item The choices $(a,b,c,d)=(+-++)$, $(++-+)$, $(+++-)$ lead to an $SO(N)$ gauge theory, one chiral multiplet in the antisymmetric 
(adjoint) representation, and two chiral multiplets in the symmetric representation. Denote the adjoint representation by $\Phi$ and 
the two fields in the symmetric representation by $S_{1,2}$. Then the superpotential is
\beq
W = S_1 \Phi S_2.
\eeq
This reproduces the ${\cal N}=2$ supersymmetric orientifold field
theory living on $N$ D3-branes in $\IC^3$ that 
sit on an $O7^+$ plane.
\item Similarly the choices $(a,b,c,d) = (-+--)$, $(--+-)$, $(---+)$
  lead to the ${\cal N}=2$ supersymmetric 
$Sp(N/2)$ field theory that 
lives on a stack of $N$ D3-branes sitting on top of an $O7^-$ plane.
\end{enumerate}
In cases 3 and 4 the direction transverse to the O7-plane is related to the location of the adjoint chiral 
multiplet ($-$ sign for case 3 and $+$ sign for case 4). The relation between the geometric action of the orientifold and 
the sign choices in the dimer is discussed in \S\ref{geomaction}.

The 4 cases above refer to 8 out of 16 possible orientifold sign
assignments. Other sign choices, with an even number 
of orientifold planes of the same sign, do not lead to consistent supersymmetric orientifolds of D3-branes on $\IC^3$.
Hence we encounter for the first time a global constraint that seems
to constrain the set of possible consistent orientifolds. 
In further examples it turns out that this is a general feature of
orientifold dimers. For a given dimer diagram, 
consistent supersymmetric orientifolds are obtained for sign choices
that have a fixed parity for the number of 
orientifold planes with the same sign. Henceforth, we refer to the
even/odd character of the number of orientifold planes 
of the same sign as the `sign parity' of the configuration. We are now ready to state the Sign Rule. 
The sign parity is determined by the number of the superpotential terms $N_W$ in the parent theory 
(i.e. the number of nodes in the unit cell of the parent dimer diagram) according to:

\medskip
\bigskip
\noindent{\bf \underline{Sign rule} :}
{\it
The sign parity of a given orientifold is equal to $N_W/2 \mod 2$. That is, the number of orientifold planes with the same 
sign is even(odd) for $N_W/2$ even(odd).}
\bigskip

This statement is supported in \S\ref{section_constraint_signs}, after studying the orientifold action on mesonic operators and using observations made there.
$N_W$ is even for every toric singularity. From the dimer point of view, this is a result of the bipartiteness of the dimer 
graph. An alternative way to determine the correct sign parity that reproduces all the consistent orientifolds of a given 
dimer diagram uses the Higgsing/unHiggsing procedure, see \S\ref{Higgsing}.

% zzz

\subsubsection{A comment on tadpoles/anomalies and extra flavors}
\label{extraflavors}

Using the above rules one can construct large classes of orientifold field theories, as we describe in examples below. An important point is that the resulting field theories are in general chiral (even in cases where the parent theory is non-chiral, see the following 
examples) and have potential gauge anomalies. As usual (see \cite{Leigh:1998hj} for the orbifold case), they are canceled 
as a consequence of cancellation of RR tadpoles (with mixed $U(1)$ anomalies involving a Green-Schwarz 
mechanism \cite{Ibanez:1998qp}). The tadpole conditions can be analyzed by following the procedures in 
\cite{Imamura:2006ub, Imamura:2006ie}. This is discussed in some detail in the mirror geometry perspective 
in \S\ref{mirrorsec}. For the rest of this section and the next we ignore this issue and assume that a suitable 
choice of ranks for the gauge groups, and a possible addition of extra ingredients like non-compact D7-branes leading to fundamental flavors, are sufficient to render the systems consistent. It is straightforward to 
determine the choice that leads to a consistent theory with a simple gauge theory analysis.
In some cases, more than one of these alternatives are possible.

% zzz 

Let us give a more explicit description of the introduction of extra D7-branes and fundamental flavors. D7-branes wrapped on holomorphic 4-cycles passing through the singularity can be added in the parent theory as described in the appendix of \cite{Franco:2006es}. The D7-branes are naturally associated to the bi-fundamental edges in the dimer. Namely, for each bi-fundamental $X_{ij}$ it is possible to introduce a D7-brane, whose  D3-D7 open string sectors lead to flavors ${\tilde Q}_i$, $Q_j$ in the representations $\antifund_i$, $\fund_j$, respectively, with a cubic superpotential coupling ${\tilde Q}_i X_{ij} Q_j$. In the orientifold theory, cancellation of tadpoles/anomalies may require the introduction of D7-branes associated to bi-fundamentals mapped to themselves under the orientifold action. In such situation, the gauge factors $i,j$ are identified, the flavors ${\tilde Q}$ are identified with the flavors $Q$, and the bi-fundamental $X_{ij}$ projects down to a two-index tensor representation $X$. The superpotential is $XQQ$. It is important to point out that the addition of D7-branes may break some of the global symmetries of the parent theory preserved by the orientifold quotient.

%===============================================================================
\subsection{Geometric action}
\label{geomaction}
%===============================================================================

In the previous subsection we provided tools to construct different
orientifold quotients of a given system of D3-branes 
at singularities. These correspond to quotients $\omega\sigma
(-1)^{F_L}$ with different geometric actions $\sigma$ 
on the CY manifold. In this subsection we describe how to relate
$\sigma$ to the corresponding orientifold quotient 
which is given by the choice of signs in the dimer diagram.

Our strategy is to propose a set of rules for computing $\sigma$ by
looking at a series of examples of increasing 
complexity. As a corollary, we derive the sign rule in
\S\ref{section_constraint_signs}. The discussion shows 
a clear correlation between the sign rule and the condition on having supersymmetry.

In order to compute the action of the orientifold on the parent
geometry, we describe the geometry as the moduli space 
of vacua for the parent theory. It is enough to consider the
orientifold action on the set of mesonic generators 
of the moduli space. The action on other mesonic operators is then
inferred by the way they are represented 
in a product form. For a systematic discussion of mesonic operators
and dimer diagrams, 
see
\cite{Benvenuti:2005ja,Benvenuti:2005cz,Benvenuti:2006qr,Feng:2007ur}.
Mesonic operators are transparent to 
the action on Chan-Paton (gauge) indices, since all fundamental
indices are contracted with anti-fundamental indices. 
This implies that sign choices which differ by an overall flip of the
Chan-Paton projection should correspond 
to the same orientifold action. As mentioned in \S\ref{dimersec} the
orientifold quotient preserves 
the mesonic flavor symmetries. As a result this orientifold action
usually corresponds at most 
to picking up a $\pm 1$ sign. This is interpreted as the
transformation of the corresponding coordinate 
in the geometry under the orientifold action.

We begin our discussion with the simplest example: The orientifolds of $\mathbb C^3$.

%===============================================================================
\subsubsection{On the orientifolds of $\mathbb C^3$}
%===============================================================================

The gauge theory for $\mathbb C^3$ has $\N=4$ supersymmetry. The
mesonic operators, which correspond 
to the coordinates of the geometry are the adjoint fields\footnote{We
 consider only single-trace operators, 
and the trace is implicit in the discussions.}
\beq
x=\Phi_1,~y=\Phi_2,~z=\Phi_3.
\eeq

\begin{figure}[!htp]
\begin{center}
\psfrag{1}[cc][][0.8]{1}
\psfrag{a}[cc][][0.7]{$\Phi_2$}
\psfrag{c}[cc][][0.7]{$\Phi_2$}
\psfrag{b}[cc][][0.7]{$\Phi_3$}
\psfrag{d}[cc][][0.7]{$\Phi_3$}
\psfrag{e}[cc][][0.7]{$\Phi_1$}
\psfrag{4}[cc][][0.7]{a}
\psfrag{5}[cc][][0.7]{b}
\psfrag{6}[cc][][0.7]{c}
\psfrag{7}[cc][][0.7]{d}
\includegraphics[scale=1]{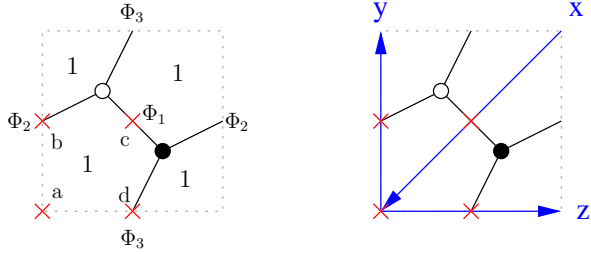}
\caption{Left: The dimer of $\mathbb C^3$ and the four fixed points under the point reflection. Right: The three fundamental 
mesons (edges crossed by the blue lines) assocciated to coordinates of $\C^3$.}
\label{Cfig1}
\end{center}
\end{figure}

The fundamental dimer cell and the fixed points under the involution are shown in Figure \ref{Cfig1}.

As described in \S\ref{C3}, the gauge group and three adjoints are
mapped to themselves. The possible orientifold 
sign choices and the resulting spectra are summarized in the first
three columns of Table \ref{Ctab1}, 
where signs for the four orientifold planes are ordered in a clockwise
fashion (abcd). Henceforth this 
ordering convention will be used throughout the paper. The action on
the mesons is given in the fourth 
column of Table \ref{Ctab1}. The fifth column denotes the number of
supersymmetry of the theory at each row.

\begin{table}[!htp]
\begin{center}
\begin{tabular}{cccccc}
Sign choice &Gauge group&Matter&$\inv(x,y,z)$&$\N$\\
\hline
$(-+++)$&$Sp$&$3~\symm$&$(-x,-y,-z)$&$4$\\
$(+---)$&$SO$&$3~\asymm$&$(-x,-y,-z)$&$4$\\
$(-+--)$&$Sp$&$1~\symm+2~\asymm$&$(+x,-y,+z)$&$2$\\
$(+-++)$&$SO$&$2~\symm+1~\asymm$&$(+x,-y,+z)$&$2$
\end{tabular}
\caption{Some possible orientifolds of the $\C^3$ theory. Additional (but obviously isomorphic) examples are obtained by 
permuting the signs of the $b,c,d$ orientifolds and the geometric actions on $x,y,z$, giving one of the gauge theories 
in this table.}
\label{Ctab1}
\end{center}
\end{table}

As expected the mesons are mapped to themselves under the $\IZ_2$
action but with possible signs. We propose the following 
rule to determine the sign of a meson under the orientifold action:

\medskip
\bigskip
\noindent{\bf \underline{Rule 1} :}
{\it
A meson passing through two orientifold planes, picks up a sign equal
to the product of the signs of these 
orientifold planes: It is even (odd) if the two orientifold planes have equal (opposite) sign.}
\bigskip
\medskip

This rule produces the geometric actions shown in the fourth column of
Table \ref{Ctab1}. This agrees with 
the geometric actions of O3-planes (the ${\cal N}=4$ theories) and the
geometric actions of O7-planes (${\cal N}=2$ theories), 
which lead to the orientifold field theories with spectra given in the second and third column.

The holomorphic 3-form which is preferred by the D3-branes is given by
\beq
\Omega=dx\wedge dy\wedge dz.
\eeq
Indeed the condition for unbroken supersymmetry in equation
\eqref{geneq1} is consistent with the geometric actions 
in Table \ref{Ctab1}. Furthermore, configurations with an even number
of orientifold planes of equal sign would lead 
to geometric actions where two coordinates flip sign, hence 
leaving $\Omega$ invariant. Such orientifolds (which have either
O9-planes or O5-planes) do not preserve any common 
supersymmetry with the D3-branes. Hence, as indicated above, we see a close relation between the sign rule and supersymmetry.

In fact $\Omega$ transforms as the meson $xyz$ or the meson $xzy$,
both are odd under the $\IZ_2$ action. 
These mesons are special, since they are terms in the
superpotential. Such mesons always exist in more general theories. 
It is a general fact that in order to have a supersymmetric
orientifold, mesons in the superpotential must be odd under 
the orientifold action. Formally, this follows from the fact that $W$
is a complex quantity which is determined 
by $\Omega$ in some string theory
constructions.\footnote{\label{orientsupo} In the mirror viewpoint the
  coefficients 
of the superpotential are given by the integral of $e^{(J+iB)}$ over a
disk worldsheet instanton. Mirror symmetry 
exchanges $e^{(J+iB)}$ and $\Omega$ and hence the connection between
terms in $W$ and the holomorphic form $\Omega$.} 
Furthermore, mesons in the superpotential can be written as the
product of all the GLSM fields in the toric 
description \cite{Franco:2006gc}. This product transforms under the
orientifold in the same way as the 
holomorphic 3-form, as one can check directly in many examples. This
is because $\Omega$ can be expressed 
in the GLSM description as the contraction of the holomorphic $n$-form
$\Omega_n=dp_1\wedge\cdots\wedge dp_n$ 
with the $(n-3)$ Killing vectors $i_{K_a}$ generating the $(\IC^*)^{n-3}$ holomorphic quotient symmetries
\beqa
\Omega= i_{K_1}\cdots i_{K_{n-3}} \Omega_n .
\eeqa

We can promote these observations to a new rule:

\medskip
\bigskip
\noindent{\bf \underline{Rule 2} :}
{\it A mesonic operator appearing as a superpotential term
  (surrounding a node in the parent dimer diagram) 
is odd under the orientifold action.}
\bigskip
\medskip

More generally, a mesonic operator defined by a homologically trivial loop in the dimer diagram picks 
up a sign given by $(-1)^k$, where $k$ is the number of nodes it
encloses. Since the total number of nodes 
in the dimer is even, this sign is independent of the choice of the enclosed region.

%===============================================================================
\subsubsection{Example: Orientifolds of the conifold}
\label{conifoldO}
%===============================================================================

\begin{figure}[!htp]
\begin{center}
\psfrag{1}[cc][][0.8]{1}
\psfrag{2}[cc][][0.8]{2}
\psfrag{a}[cc][][0.7]{$X^{(1)}_{12}$}
\psfrag{b}[cc][][0.7]{$X^{(2)}_{21}$}
\psfrag{c}[cc][][0.7]{$X^{(1)}_{21}$}
\psfrag{d}[cc][][0.7]{$X^{(1)}_{12}$}
\psfrag{e}[cc][][0.7]{$X^{(2)}_{21}$}
\psfrag{f}[cc][][0.7]{$X^{(1)}_{21}$}
\psfrag{g}[cc][][0.7]{$X^{(2)}_{12}$}
\includegraphics[scale=1]{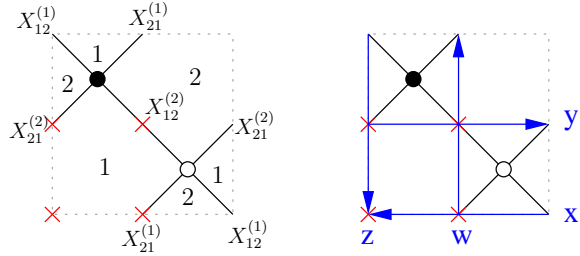}
\hspace{1cm}
\caption{Left: The fundamental cell of Figure 1 for the conifold with
  fixed-points of the orientifold action. 
Right: The mesonic operators. }
\label{conifig1}
\end{center}
\end{figure}

As the next example, let us consider the conifold singularity. The corresponding dimer diagram is illustrated in Figure 
\ref{conifig1}.

As for the $\IC^3$ example, the number of superpotential terms is 2 and therefore, using the sign rule, for this configuration the total number of orientifold planes with equal sign is odd. Using the Higgs mechanism one can show that the sign rule for both these theories should be the same. This is discussed in \S\ref{Higgsing}. Some possible orientifolds are shown in Table \ref{conitab1}. Other sign choices lead to isomorphic models related by a relabeling of fields and mesons and are not shown in the table.
\begin{table}[!htp]
\begin{center}
\begin{tabular}{cccc}
Charges&Gauge group&Matter&$\inv(x,y,z,w)$\\
\hline
$(+-++)$&$U$&$ \asymm+\symm+2~\antisymm+8~\fund$&$(+x,-y,-z,+w)$\\
$(-+--)$&$U$&$\symm+\asymm+2~\antiasymm+8~\antifund$&$(+x,-y,-z,+w)$\\
\end{tabular}
\caption{Two examples of orientifolds of the conifold theory. Other sign choices lead to isomorphic models 
related by permutations of $x,y,z,w$. We have added eight chiral multiplets in the fundamental or respectively 
anti-fundamental representation to make the theory anomaly free.}
\label{conitab1}
\end{center}
\end{table}

In both cases, a set of 8 (anti)fundamentals are added to cancel the chiral anomaly associated with 
the $\symm+\antiasymm$ ($\antisymm+\asymm$) combination. In a string theory construction with D3 branes on the conifold this follows from uncanceled contributions of the orientifold planes to tadpoles
of RR fields localized at the singularity. They are cancelled by the introduction of additional  D7-branes, which contribute additional fields from the D3-D7 sector. We come back to this point in the mirror perspective in \S\ref{tadsec}. 

% zzz changed sentence
We can write down the superpotential following the discussion in \S\ref{extraflavors}. Let us denote the different fields by $A, S, \bar S_1, \bar S_2, Q_i$, $i=1\ldots8$. Then the superpotential of the first model takes the form
\beq
W = S \bar S_1 A \bar S_2 -  \bar S_1 Q_i Q_i .
\eeq
where we have chosen to introduce the flavors by using the D7-brane naturally associated to $S_1$ (there exist other consistent choices).
A similar superpotential exists for the second model.

In order to obtain the geometric action, let us consider the basic mesonic operators in the parent theory, given by
\beq\label{conieq0}
x=X^{(1)}_{12}X^{(1)}_{21},~y=X^{(2)}_{12}X^{(2)}_{21},~z=X^{(1)}_{12}X^{(2)}_{21},~w=X^{(2)}_{12}X^{(1)}_{21},
\eeq
which in this case are in one-to-one correspondence with zigzag paths of the dimer and fulfill the usual relation $xy=wz$.

Using rule 1 we can infer the geometric actions on $x,y,z,w$, as shown
in the last column of Table \ref{conitab1}. 
The holomorphic 3-form can be locally written as
\beqa\label{conieq1}
\Omega=\frac{dx\wedge dy\wedge dz}{z} .
\eeqa
Hence it is odd under the orientifold action, as expected from a
supersymmetric orientifold quotient. 
One can show that this directly follows from the sign rule (global constraint on sign parity).

Orientifolds of the conifold can be studied using the T-dual HW setup,
which involves D4-branes suspended between one NS and one
NS$'$-brane. 
It is possible to orientifold this configuration by introducing
O6-planes: one is on top of the NS-brane (leading to an ${\cal N}=2$
subsector), 
namely two chiral multiplets in conjugate two-index representations ($\symm$ or $\asymm$
depending on the O6-plane charge); the second is on top of the NS$'$-brane, which cuts it in two
halves with opposite signs (leading to fields in $\symm+{\antiasymm}$ or ${\antisymm}+\asymm$).
The latter configuration has been dubbed a `fork'
\cite{Brunner:1998jr} due to its shape as a musical fork and was used
in \cite{Park:1999ep}. 
Cancellation of world-volume tadpoles on the NS$'$ requires the
introductions of eight half-D6 branes \cite{Hanany:1997}, 
which lead to additional flavors. Notice that any orientifold quotient
of this configuration necessarily contains one 
(and only one) fork. The geometric action of the orientifold quotient in these models were determined in \cite{Park:1999ep}, using the partial
resolution method.

Our construction in this section reproduces the orientifold field
theories obtained from the HW dual and which contain a fork
configuration. 
The appearance of just one fork follows from the global constraint on
sign parity. The geometric action on mesons as in Table
\ref{conitab1}, 
obtained from the orientifold, is in agreement with the geometric action in the dual HW setup \cite{Park:1999ep}.

%===============================================================================
\subsubsection{Example: $\C^2/\Z_2 \times \C$}
%===============================================================================

\label{c2z2sec}

\begin{figure}[!htp]
\begin{center}
\psfrag{x121}[cc][][0.65]{$X^{(1)}_{21}$}
\psfrag{x122}[cc][][0.65]{$X^{(2)}_{21}$}
\psfrag{x211}[cc][][0.65]{$X^{(1)}_{12}$}
\psfrag{x212}[cc][][0.65]{$X^{(2)}_{12}$}
\psfrag{x11}[cc][][0.65]{$\Phi_{1}$}
\psfrag{x22}[cc][][0.65]{$\Phi_{2}$}
\psfrag{1}[cc][][0.8]{1}
\psfrag{2}[cc][][0.8]{2}
\includegraphics[scale=1.05]{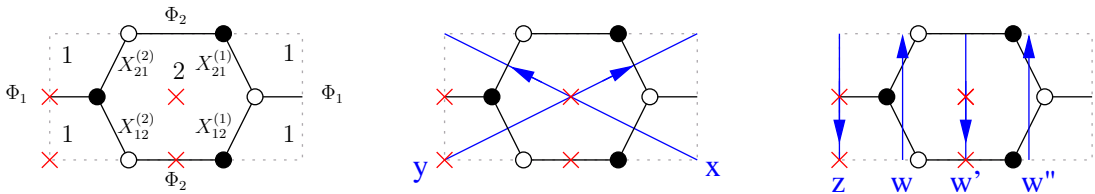}
\caption{Left: $\C^2/\Z_2 \times \C$ dimer with fixed-points. Center and right: The mesonic operators. 
We show two equivalent representations $w$ and $w''$ of the same meson, and an intermediate path $w'$.}
\label{mesonC2Z2}
\end{center}
\end{figure}

 Let us consider orientifolds of the $\C^2/\Z_2 \times \C$ orbifold. The dimer diagram with the
orientifold planes and the basic mesons are shown in \fref{mesonC2Z2}. They satisfy the relation
$xy=w^2$, with $z$ free. There is also a meson $u$ corresponding to a closed loop around a node,
which is not explicitly shown since it is not basic. It can be expressed in terms of the others as $u=z w$.

This example illustrates an additional situation that we might
encounter. Our rules above allow to read the orientifold action on
mesons 
that pass over orientifold planes and hence are mapped to themselves,
such as $x$, $y$ and $z$, or the meson corresponding to 
a superpotential term $u$. In this example, there is a new type of
meson $w$ that is not mapped to itself but rather to 
another path that is labeled $w''$.

%%%%%%%%%%%%%%%%%%%%%%%
%%%%%%%%%%%%%%%%%%%%%%%

The path $w$ can be transformed into its image $w''$ using some number
of F-term relations (in this case one). At some intermediate stage(s), 
labeled $w'$ in this example, the path can go over two orientifold planes. This pair of orientifold planes
can be in any of the combinations contemplated by rule 1. The total sign picked by $w$ is the product of signs dictated 
by rule 1
for each pair of nodes that are transversed at intermediate steps, times one minus sign for each F-term equation used 
coming
from rule 2. Another way of interpreting this procedure is that we can deform both $w$ and $w''$ into $w'$ by moving 
them to the right and left, respectively.
Both paths become identical and the sign can be determined using rule 1. The combined displacements of $w$ and $w''$ 
to an intermediate path can be viewed as
the total displacement from $w$ to $w''$, which involves one F-term equation. This reasoning applies to any other theory. 
We can summarize our conclusions as:

\medskip
\bigskip
\noindent{\bf \underline{Rule 3} :}
 {\it Consider a mesonic operator corresponding to a path which is not invariant under the orientifold action, but mapped to 
 another path representing the same mesonic operator. The two paths define a strip\footnote{In fact, they define two stripes. 
 Either of them can be used. In the next section we discuss how this freedom can be used to constrain sign parity.} 
 enclosing $2k$ nodes, and orientifold planes, of charges $q_i=\pm 1$. Then, the sign picked up by the operator 
 is $\prod_i q_i\, (-1)^{k} $.}
\bigskip
\medskip

Note that rule 3 is not an additional independent rule, but a combination of rules 1 and 2 in a practical form to 
deal with mesons that are not mapped to themselves.

The sign rule on orientifold charges implies in this case that there
is an even number of orientifold planes with equal charge. 
Different orientifold models of this system are shown in Table
\ref{Ctab2}. Other choices amount to overall sign flips, 
corresponding to overall flips of the Chan-Paton actions.

\begin{table}[!htp]
\begin{center}
\begin{tabular}{cccc}
Charges&Gauge group& 2-index tensors &$\inv(x,y,w,z)$ \\
\hline
$(-+-+)$&$Sp \times Sp$& $\symm_1+\symm_2$  &$(+x,+y,+w,-z)$ \\
$(-++-)$&$Sp \times SO$&  $\symm_1+\asymm_2$  &$(-x,-y,+w,-z)$ \\
$(----)$&$Sp \times Sp$&  $\asymm_1+\asymm_2$  &$(+x,+y,-w,+z)$ \\
$(--++)$&$Sp \times SO$& $\asymm_1+\symm_2$ &$(-x,-y,-w,+z)$
\end{tabular}
\caption{Some possible orientifolds of the $\C^2/\Z_2 \times \C$
  theory. Other choices amount to overall sign flips. 
The indices 1 and 2 refer to the adjoint fields of the parent
theory. Only the two-index tensor representations 
in the matter sector are listed. There are 2 additional bi-fundamental fields.}
\label{Ctab2}
\end{center}
\end{table}

It is straightforward to check that $sign(u)=sign(z)sign(w)=-1$ in all cases, in agreement with rule 2.

To write down the superpotential e.g. for the first model let us
introduce the notation $S_1, S_2, X^{(1,2)}$ for the 
two symmetric fields and the two bi-fundamental fields, and note that
the four terms in the parent theory are mapped to 
two in the orientifolded theory. We get
\beq
W=S_1 X^{(1)} X^{(2)} - S_2 X^{(2)} X^{(1)} .
\eeq
Similar superpotentials can be written for the other three models.
D3-branes at $\IC^2/\IZ_2\times \IC$ can be T-dualized to a Type IIA configuration with D4-branes suspended 
between two NS-branes. 
The introduction of orientifold planes in these configurations was described in
\cite{Uranga:1998uj}, and the geometric actions in the Type IIB geometry have been constructed 
in \cite{Park:1998zh,Park:1999eb} using CFT techniques. Hence we can compare our results with these constructions.

In fact the  four models in Table \ref{Ctab2} correspond to orientifolds in these references, 
where in Type IIA language the different projections follow from the relative orientation of the O-planes 
and the two NS-branes. The four models correspond to: two O6-planes with positive charge, two O6-planes 
with opposite charges, two O6$'$-planes with opposite charges and two O6$'$-planes with positive charge, respectively. 
Also, the geometric actions in Table \ref{Ctab2}, obtained using our rules for orientifold action on mesons,  
agree with the ones previously derived using other methods \cite{Park:1998zh,Park:1999eb}. Note that in configurations 
with O6$'$-planes, the vector multiplet and the adjoint chiral multiplet for each group have opposite projections. 
This is nicely reproduced by the dimer, and is ultimately implied by the global constraint on orientifold signs.

%===================================================================================
\subsection*{Other orientifolds of $\IC^2/\IZ_2 \times \IC$}
%===================================================================================

\begin{figure}[!htp]
\begin{center}
\psfrag{x121}[cc][][0.65]{$X^{(1)}_{21}$}
\psfrag{x122}[cc][][0.65]{$X^{(2)}_{21}$}
\psfrag{x211}[cc][][0.65]{$X^{(1)}_{12}$}
\psfrag{x212}[cc][][0.65]{$X^{(2)}_{12}$}
\psfrag{x11}[cc][][0.65]{$\Phi_{1}$}
\psfrag{x22}[cc][][0.65]{$\Phi_{2}$}
\psfrag{1}[cc][][0.8]{1}
\psfrag{2}[cc][][0.8]{2}
\includegraphics[scale=1.2]{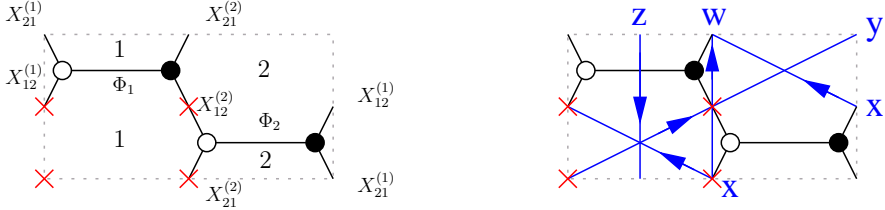}
\caption{Left: $\IC^2/\Z_2 \times \IC$ dimer with fixed-points for a different orientifold action. Right: 
The mesonic operators.}
\label{mesonC2Z2bis}
\end{center}
\end{figure}

We can consider other possible orientifolds of
$\IC^2/\IZ_2\times \IC$ which are obtained by a different embedding of the dimer diagram into the $\IT^2$. 
\fref{mesonC2Z2bis} shows the dimer diagram with the orientifold planes and the basic mesons. Recall that we 
also have the meson $u$ corresponding to a closed path around a node.

%%%%%%%%%%%%%%%%%%%%%%%
%%%%%%%%%%%%%%%%%%%%%%%

Our rules above allow to read the orientifold action on mesons. We use rule 3 for $z$. Table \ref{C2/Z2_tab_2} shows 
the spectrum and geometric action for some choices of signs.

\begin{table}[!htp]
\begin{center}
\begin{tabular}{cccc}
Charges&Gauge group& 2-index tensors &$\inv(x,y,w,z)$ \\
\hline
$(++++)$&$ U $& $ 1 \, \mbox{Adj.} + 2 \symm + 2 \antisymm$  &$(+x,+y,+w,-z)$ \\
$(++--)$&$ U $&  $ 1 \, \mbox{Adj.} + \symm + \antisymm + \asymm + \antiasymm  $ &$(-x,-y,+w,-z)$ \\
$(+-+-)$&$ U $&  $ 1 \, \mbox{Adj.} + \symm + \antisymm + \asymm + \antiasymm $ & $(+x,+y,-w,+z)$ \\
$(+--+)$&$ U $& $ 1 \, \mbox{Adj.} + 2 \left( \asymm + \antisymm \right)+16 \fund $ &$(-x,-y,-w,+z)$
\end{tabular}
\caption{Some possible orientifolds of the $\C^2/\Z_2 \times \C$
theory.}
\label{C2/Z2_tab_2}
\end{center}
\end{table}

Once again, we can immediately verify that $sign(u)=sign(z)sign(w)=-1$ in all cases. Other choices amount to overall flip of the signs, and thus produce the same geometric actions.
As in the conifold example, we have added fundamental fields when necessary to cancel the chiral anomaly. This corresponds to the addition of non-compact D7-branes.

Once again, these models can be T-dualized to Hanany-Witten Type IIA setups. In this case, the NS-branes are on top of O6-planes. These theories have been studied from that perspective in \cite{Park:1998zh,Park:1999eb,Park:1999ep} and correspond to having two O6-planes with positive charge, two O6-planes with opposite charges, two forks and two oppositely oriented forks, respectively.

%zzz 
The superpotential for the first model can be written as follows: Denote the fields in the adjoint and two-index tensors and conjugates as $\Phi$, $S_{1,2}$, $\bar S_{1,2}$. Then we have
\beq
W = \Phi S_1 \bar S_1 - \Phi S_2 \bar S_2 ,
\eeq
with similar terms for the other model that has no fork. The superpotential for models with extra fundamental flavors can be calculated as described in \S\ref{extraflavors}, giving contributions of the form
\beqa
W= \bar S_1 Q_iQ_i + \bar S_2 Q'_i Q'_i,
\eeqa
where we have chosen to introduce the flavors by using 8 D7-branes
associated to $\bar S_1$ and 8 to $\bar S_2$. Other consistent choices are possible, whose superpotentials are obtained similarly.
%================================================================
\subsection{The global sign rule and Higgsing}
%================================================================
\label{Higgsing}
It is illustrative to consider how the above construction of
orientifold theories interplays with the Higgs mechanism. 
We argue below that using the Higgs mechanism one can fix the global
constraint on the number of orientifold planes 
of equal sign in arbitrary dimer diagrams. Another simpler derivation for the prescription is described in \S\ref{section_constraint_signs}.

\begin{figure}[!htp]
\begin{center}
\psfrag{1}[cc][][0.7]{1}
\psfrag{2}[cc][][0.7]{2}
\psfrag{a}[cc][][1]{a)}
\psfrag{b}[cc][][1]{b)}
\includegraphics[scale=0.6]{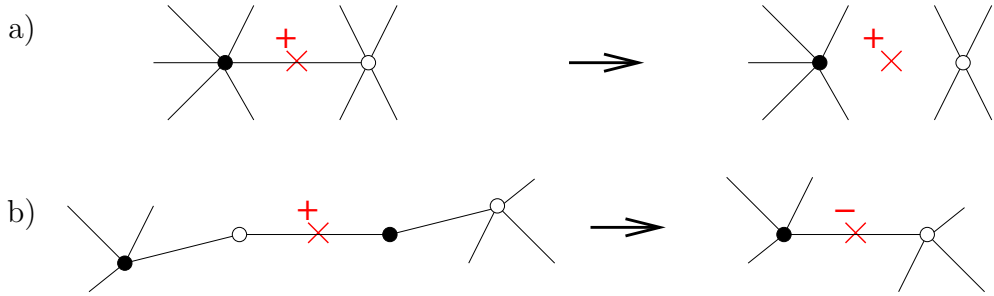}
\caption{a) Behavior of orientifold plane (red cross) upon removal of edges. b) Behavior of orientifold plane upon 
integrating out matter.}
\label{higgs}
\end{center}
\end{figure}

Higgsing in the parent theory corresponds to removal of edges
associated with the fields acquiring vevs in the dimer 
diagram \cite{Franco:2005rj}. In the presence of orientifold
quotients, removal of an edge should be accompanied by 
the removal of its orientifold image, since both correspond to a
single field in the orientifold theory. There are 
two important points to consider in this procedure:
\begin{itemize}

\item If the removed edge sits on top of an orientifold plane, the
  orientifold plane preserves its sign, and ends up 
on top of a face of the dimer, see Figure \ref{higgs}a. This is
  consistent with the orientifold rules 
in \S\ref{section_generalities} since a $U(N)$ gauge factor breaks
  down to an $SO(N)$ by Higgsing with 
a chiral multiplet in the $\symm$ representation, and to $Sp(N/2)$ with a field in the $\asymm$ representation.

\item After the edge removal, one may get bi-valent nodes involving an
  edge on top of an orientifold plane 
as shown in \fref{higgs}b. Bi-valent nodes represent quadratic terms (i.e. mass terms) in the 
superpotential. Given the $\IZ_2$ symmetry of the parent theory, the bi-valent nodes come in pairs.
Integrating out the massive fields corresponds to collapsing the bi-valent nodes \cite{Franco:2005rj} but the orientifold 
plane flips sign. This is consistent with the field theory result, since the edges not on top of the orientifold plane 
correspond to $\asymm+\symm$, and only one of these combinations becomes massive with the two-index tensor on top of 
the orientifold plane.

\end{itemize}

%%%%%%%%%%%%%%%%%%%%%%%
%%%%%%%%%%%%%%%%%%%%%%%%%

An example of the first point appears in Higgsing the orientifolds of
the conifold to orientifolds of $\IC^3$. 
An example of the second point appears in Higgsing the orientifolds of
$\IC^3/(\IZ_2\times \IZ_2)$ to 
orientifolds of the SPP theory.\footnote{For the field theory discussion we refer to \cite{Park:1999ep}.}

The above rules allow to fix the global constraint on the number of orientifold planes of equal sign in arbitrary dimer 
diagrams by Higgsing them down to a known dimer diagram. Some examples are discussed in following sections.

%===============================================================================
\subsection{The global constraint for the fixed point signs revisited}
\label{section_constraint_signs}
%===============================================================================

In all the models considered in the previous sections, we are able to determine which parity of the fixed point signs  leads to supersymmetric orientifolds. This is consistent with (un)Higgsing to known theories. It is in principle possible, although generally much more involved, to use one of these approaches for a generic toric singularity.

The transformation rules of mesonic operators under the orientifold
action, derived in the previous section, 
offer a simpler method for fixing the signs. For an arbitrary dimer,
the parity of orientifold plane signs 
can be determined by requiring consistency among the sign assignments that follow from the three rules.

With a reasoning similar to the one used to arrive at rule 3, we can derive a compact principle fixing the sign parity. 
Take any mesonic operator and move it once around the torus until it comes back to itself. It picks up a sign given by
\beq
(-1)^{SP}(-1)^{\# F}=(-1)^{SP}(-1)^{N_W/2},
\label{sign_parity}
\eeq
where $SP$ denotes the overall sign parity and $F$ the number of F-term relations required to move the path once around 
the torus. In detail, when moving around the torus, the path representing the meson passes over all the orientifold planes 
and pick their signs. This gives rise to the $(-1)^{SP}$ factor in (\ref{sign_parity}). The $(-1)^{\# F}$ factor comes from 
the F-term relations required to move the path around the torus and return to its original position. All the $N_W$ 
superpotential terms are used in this process. Since every field in the quiver is present in exactly two superpotential 
terms, we take care of all of them by using the F-term equations for $N_W/2$ fields. Since the meson should not pick up 
a sign in this process, (\ref{sign_parity}) has to be equal to one. Hence, we conclude that the sign parity is equal to 
the parity of $N_W/2$. This is precisely the sign rule announced in \S\ref{section_generalities}.

%%%%%%%%%%%%%%%%%%%%%%%%%%%%%%%%%%%%%%%%%%%%%%%%%%%%%%%%%%%%%%%%%%%%%%%%%%%%%%%%
\subsection{Further examples}
\label{moremodels}
%%%%%%%%%%%%%%%%%%%%%%%%%%%%%%%%%%%%%%
%===============================================================================
\subsubsection{Orientifolds of $\C^3/\Z_3$}
\begin{figure}[!htp]
\begin{center}
\psfrag{a}[cc][][0.65]{$X^{(2)}_{23}$}
\psfrag{b}[cc][][0.65]{$X^{(1)}_{31}$}
\psfrag{c}[cc][][0.65]{$X^{(3)}_{23}$}
\psfrag{d}[cc][][0.65]{$X^{(2)}_{12}$}
\psfrag{e}[cc][][0.65]{$X^{(1)}_{23}$}
\psfrag{f}[cc][][0.65]{$X^{(3)}_{31}$}
\psfrag{g}[cc][][0.65]{$X^{(2)}_{23}$}
\psfrag{h}[cc][][0.65]{$X^{(1)}_{12}$}
\psfrag{i}[cc][][0.65]{$X^{(3)}_{23}$}
\psfrag{j}[cc][][0.65]{$X^{(2)}_{31}$}
\psfrag{k}[cc][][0.65]{$X^{(1)}_{23}$}
\psfrag{l}[cc][][0.65]{$X^{(3)}_{12}$}
\psfrag{1}[cc][][0.8]{1}
\psfrag{2}[cc][][0.8]{2}
\psfrag{3}[cc][][0.8]{3}
\includegraphics[scale=1]{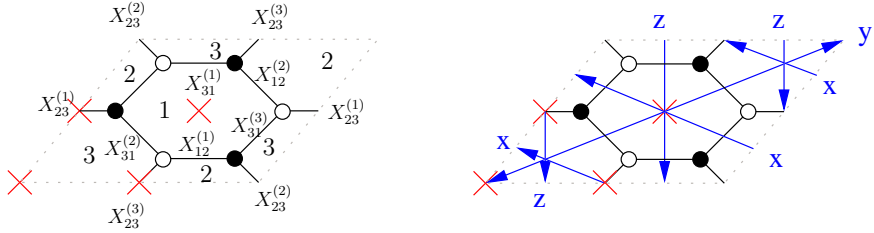}
\caption{Left: The dimer of the orbifold $\C^3/\Z_3$ with fixed-points. Right: The mesonic operators.}
\label{C3Z3fig1}
\end{center}
\end{figure}

%===============================================================================
\label{C3Z3sec}
Let us consider $\C^3/\Z_3$ which can be also seen as the orbifold point of $\O_{\P^2}(-3)$, i.e. the complex cone over $dP_0$. The dimer and the fixed-points under the point-reflection involution as well as the mesonic operators corresponding to the coordinates of the geometry are depicted in Figure \ref{C3Z3fig1}.
Under the involution one gauge group stays fixed while two are identified. Three bi-fundamentals are mapped to themselves while the remaining ones are identified. The holomorphic 3-form survives from the parent $\C^3$ theory. The mesonic operators show identical behavior as in $\C^3$, i.e. they have only one common fixed-point. Hence, the same considerations apply, i.e. only an odd sign parity leads to an $\mathcal N=1$ supersymmetric theory. Some possible sign choices and the resulting theories are listed in Table \ref{C3Z3tab1}.
\begin{table}[!htp]
\begin{center}
\begin{tabular}{ccc}
Charges&Gauge group&Matter\\
\hline
$(++-+)$&$Sp\times U$&$3~\symm_2+3~(\fund,\antifund)$\\
$(--+-)$&$SO\times U$&$3~\asymm_2+3~(\fund,\antifund)$\\
$(+---)$&$Sp\times U$&$1~\symm_2+2~\asymm_2+3~(\fund,\antifund)$\\
$(-+++)$&$SO\times U$&$2~\symm_2+1~\asymm_2+3~(\fund,\antifund)$\\
\end{tabular}
\caption{Some possible $\N=1$ orientifolds of $\C^3/\Z_3$.}
\label{C3Z3tab1}
\end{center}
\end{table}
The orientifolds of the first and second model of Table \ref{C3Z3tab1} are closely related to the orientifold projection introduced in \cite{Angelantonj:1996uy}, see e.g. \cite{Lykken:1997gy,Lykken:1997ub,Aldazabal:2000sa,Uranga:2002pg} for its construction using D3-branes.

All the spectra shown in Table \ref{C3Z3tab1} are anomalous unless the ranks of the gauge groups are restricted 
or additional
(anti)fundamental matter is added. It is instructive to discuss this example in more detail, since similar 
considerations apply to the rest of the models
in the paper. Contrary to what happens for the models we have discussed so far, there is more than one way to 
render them consistent. For example, for the first model in Table \ref{C3Z3tab1} with gauge group $Sp(n_1)\times U(n_2)$, the most 
general anomaly-free spectrum reads

\beq
3~\symm_2+3~(\fund,\antifund)+3~(4-n_1+n_2)\antifund,
\eeq
where a negative number of $\antifund$ is understood as a positive number of $\fund$. All these choices translate to branes wrapping compact or non-compact cycles. It would be interesting to reproduce this constraint along the lines of \cite{Imamura:2006ub}.

%===============================================================================
\subsubsection{Orientifolds of the Conifold$/\Z_{N}$}
%===============================================================================
In this section we consider orientifolds of certain orbifolds of the conifold, introduced in \cite{Uranga:1998vf}, and corresponding to $L^{aaa}$ geometries with $a=N$. The tools should be familiar by now, so our discussion is sketchy.

\begin{figure}[!htp]
\begin{center}
\psfrag{1}[cc][][1]{1}
\psfrag{2}[cc][][1]{2}
\psfrag{3}[cc][][1]{3}
\psfrag{4}[cc][][1]{4}
\includegraphics[scale=1]{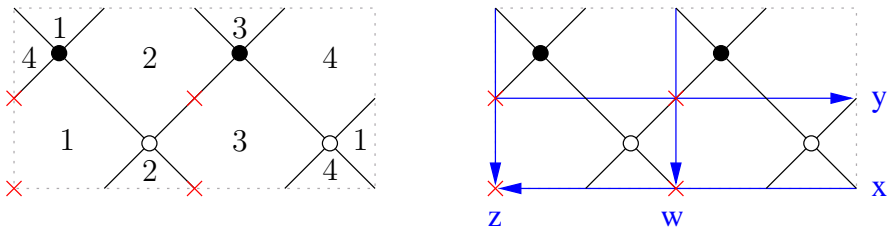}
\hspace{1cm}
\caption{Left: Dimer of the conifold$/\Z_2$. Right: The mesonic operators.}
\label{coniZ2fig1}
\end{center}
\end{figure}

Let us start with the $\Z_2$ orbifold of the conifold obtained by the following geometric action:
\beq
x\rightarrow x,~y\rightarrow y,~z\rightarrow -z,~w\rightarrow-w.
\eeq
This orbifold has been discussed in detail in \cite{Argurio:2007qk}. The holomorphic 3-form $\Omega$ of the conifold is invariant under this orbifold action as can be inferred from (\ref{conieq1}). The dimer and the mesonic operators are drawn in Figure \ref{coniZ2fig1}.
The basic structure of mesonic operators and their relation with the orientifold points is similar to that of the conifold. The sign parity of the model is positive, though. Hence the number of negative orientifold 
planes is even.

The structure of the possible orientifold models is as follows.
The resulting gauge group is $U(N_1)\times U(N_2)$ and there are two antisymmetric and two symmetric representations (two among the four are conjugated). The chiral combinations require additional flavors, as in fork configurations. This can be obtained upon introduction of extra non-compact branes in the system. The models can be described using a T-dual HW setup with two NS and two NS$'$-branes. The non-chiral models correspond to HW models with e.g. the NS-branes on top of O6-planes (of different possible signs) and the NS$'$-branes mapped to each other by the orientifold action. The chiral models correspond to HW models with the NS$'$-branes on top of O6-planes. These O6 planes are split in halves, and lead to a chiral matter content as in a fork configuration. The NS-branes are mapped to each other.

\begin{figure}[!htp]
\begin{center}
\psfrag{1}[cc][][0.7]{1}
\psfrag{2}[cc][][0.7]{2}
\psfrag{3}[cc][][0.7]{3}
\psfrag{N}[cc][][0.7]{N}
\psfrag{N+1}[cc][][0.7]{N+1}
\psfrag{N+2}[cc][][0.7]{N+2}
\psfrag{N-1}[cc][][0.7]{N-1}
\psfrag{N-2}[cc][][0.7]{N-2}
\psfrag{2N}[cc][][0.7]{2N}
\psfrag{2N-1}[cc][][0.7]{2N-1}
\psfrag{2N-2}[cc][][0.7]{2N-2}
\includegraphics[scale=1]{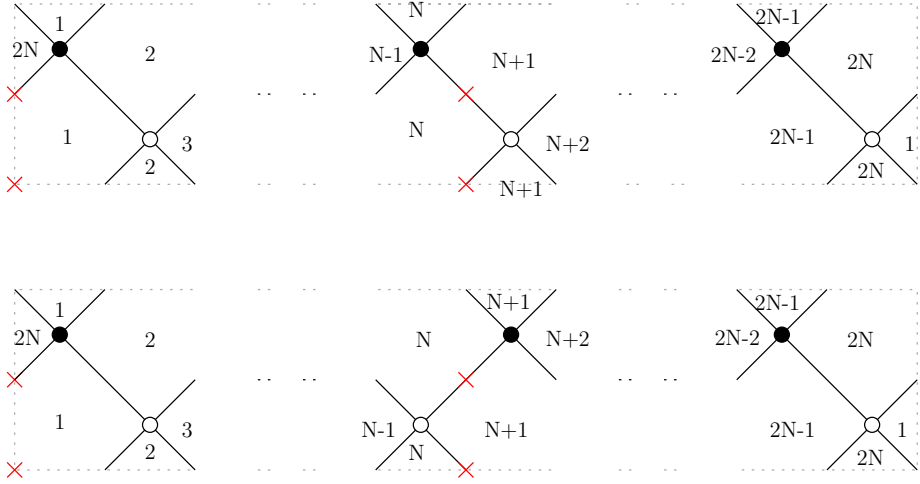}
\hspace{1cm}
\caption{Top: Dimer of the conifold$/\Z_N$ for odd $N$. Bottom: For even $N$.}
\label{coniZNfig2}
\end{center}
\end{figure}

It is easy to generalize this to any $\Z_N$ orbifold of the conifold. The dimers for odd as well as for even $N$ are given in \fref{coniZNfig2}.
Some basic possible orientifolded theories are given in Table
\ref{coniZ2tab1}. For all these models we have $N$ factors of gauge group $U(N_i)$ with $i=1...N$ and bi-fundamentals $\sum_{i=1}^{N-1} (\fund_i, \antifund_{i+1})+(\antifund_i, \fund_{i+1})$. The remaining matters are specified in Table \ref{coniZ2tab1}. The different geometric actions are 
straightforwardly obtained from our rules.

% zzz
\begin{table}[!htp]
\begin{center}
\begin{tabular}{ccc}
Charges&G. g.&Matter\\
\hline
$(+++-)$&$\prod_{i=1}^{N}U(N_i)$&$\symm_1+\antisymm_1+\symm_N+\antiasymm_N+8\antifund_N$\\
$(+---)$&$\prod_{i=1}^{N}U(N_i)$&$\symm_1+\antiasymm_1+\asymm_N+\antiasymm_N+8\antifund_1$\\
\hline
 $(++++)$&$\prod_{i=1}^{N}U(N_i)$&$\symm_1+\antisymm_1+\symm_N+\antisymm_N$\\
 $(+-+-)$&$\prod_{i=1}^{N}U(N_i)$&$\symm_1+\antiasymm_1+\symm_N+ \antiasymm_N+8\antifund_1+8\antifund_N$\\
\end{tabular}
\caption{Basic $\N=1$ orientifolds of the conifold$/\Z_N$. The first two rows correspond to orbifolds with odd $N$, 
while the last two rows correspond to orbifolds with even $N$.}
\label{coniZ2tab1}
\end{center}
\end{table}

%==================================================================
\subsubsection{Orientifolds of SPP}
%==================================================================

\begin{figure}[!htp]
\begin{center}
\psfrag{a}[cc][][0.65]{$\Phi_3$}
\psfrag{b}[cc][][0.65]{$X_{31}$}
\psfrag{c}[cc][][0.65]{$X_{13}$}
\psfrag{d}[cc][][0.65]{$X_{23}$}
\psfrag{e}[cc][][0.65]{$X_{32}$}
\psfrag{f}[cc][][0.65]{$X_{21}$}
\psfrag{g}[cc][][0.65]{$X_{12}$}

\psfrag{1}[cc][][0.7]{1}
\psfrag{2}[cc][][0.7]{2}
\psfrag{3}[cc][][0.7]{3}
\includegraphics[scale=1]{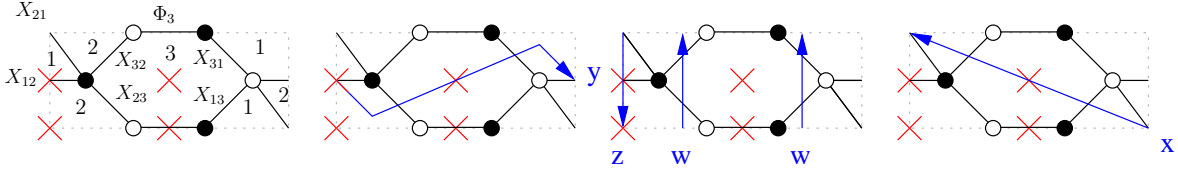}
\hspace{1cm}
\caption{Dimer of SPP and the mesonic operators.}
\label{SPPdimer}
\end{center}
\end{figure}

As another example, consider the real cone over the first member of the $L^{aba}$ family, $L^{1,2,1}$. This is also known as the complex cone over the suspended pinch point, for short SPP. The dimer and the mesonic operators corresponding to the coordinates of the geometry are shown in Figure \ref{SPPdimer}.

Using the rules one can easily extract the spectra and action on coordinates for various sign choices as listed in Table \ref{SPPtab1}. In addition to the fields listed, all the models contain a $(\fund_1,\antifund_2)+(\fund_1,\fund_2)$ pair. 
The last two models also require
the addition of antifundamentals to cancel gauge anomalies.

\begin{table}[!htp]
\begin{center}
\begin{tabular}{cccc}
Charges&Gauge group& Matter &$\inv(x,y,z,w)$\\
\hline
$(++++)$&$SO\times U$&$\symm_1+\symm_2+\antisymm_2$&$(+x,+y,+z,-w)$\\
$(----)$&$Sp\times U$&$\asymm_1+\asymm_2+\antiasymm_2$&$(+x,+y,+z,-w)$\\
$(--++)$&$SO\times U$&$\symm_1+\asymm_2+\antiasymm_2$&$(-x,-y,+z,-w)$\\
$(++--)$&$Sp\times U$&$\asymm_1+\symm_2+\antisymm_2$&$(-x,-y,+z,-w)$\\
$(+-+-)$&$SO\times U$&$\asymm_1+\symm_2+\antiasymm_2+8~\antifund_2$&$(+x,-y,-z,+w)$\\
$(-+-+)$&$Sp\times U$&$\symm_1+\symm_2+\antiasymm_2+8~\antifund_2$&$(+x,-y,-z,+w)$\\
\end{tabular}
\caption{Some possible orientifolds of the SPP.}
\label{SPPtab1}
\end{center}
\end{table}

Hence, we easily recover the orientifolds derived via field-theory techniques in \cite{Park:1999ep}.

\bigskip
It is straightforward to use our rules to systematically construct orientifolds of general
$L^{aba}$ theories. Some technical details of these theories can be found in Appendix C. Since
the basic ingredients have already been discussed, we refrain from a detailed classification and
instead proceed to the construction of new examples. 
%==================================================================
\subsubsection{Orientifolds of $L^{1,5,2}$}
%==================================================================

\begin{figure}[!htp]
\begin{center}
\psfrag{1}[cc][][0.7]{1}
\psfrag{2}[cc][][0.7]{2}
\psfrag{3}[cc][][0.7]{3}
\psfrag{4}[cc][][0.7]{4}
\psfrag{5}[cc][][0.7]{5}
\psfrag{6}[cc][][0.7]{6}
\includegraphics[scale=.8]{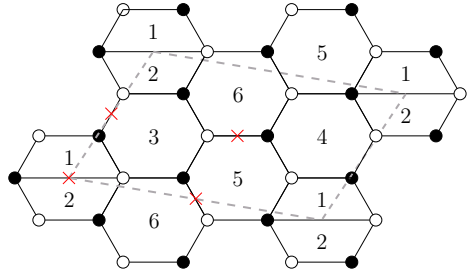}
\hspace{1cm}
\caption{Dimer diagram for $L^{1,5,2}$.}
\label{L152dimer}
\end{center}
\end{figure}

Finally, in order to illustrate the power of our methods, we close this section with an example
that was not amenable to the techniques previously available in the literature. \fref{L152dimer}
shows the dimer diagram for $L^{1,5,2}$ \cite{Franco:2005sm}.

The parent theory has 10 superpotential terms. Hence, the sign rule determines that the sign
parity must be odd. The resulting gauge group is $U(n_1)\times U(n_3) \times U(n_5)$, with
matter content
\beq
(\fund_1,\antifund_3)+(\antifund_1,\antifund_3)+(\fund_1,\antifund_5)+2~(\fund_3,\fund_5)+(\antifund_3,\fund_5)
+\bar{R}_1+S_3+\bar{T}_5+\bar{V}_5.
\eeq
$\bar{R}_1$, $S_3$, $\bar{T}_5$ and $\bar{V}_5$ are two-index tensor representations determined by the signs 
of the orientifold planes.

\begin{table}[!htp]
\begin{center}
\begin{tabular}{cc}
Charges& 2-index tensors \\
\hline
$(+++-)$& $\antisymm_1 + \symm_3 + \antisymm_5 + \antiasymm_5$ \\
$(++-+)$& $\antisymm_1 + \symm_3 + \antisymm_5 + \antiasymm_5$ \\
$(+-++)$& $\antisymm_1 + \asymm_3 + 2~ \antisymm_5$ \\
$(-+++)$& $\antiasymm_1 + \symm_3 + 2~ \antisymm_5$
\end{tabular}
\caption{Two-index representations for some possible orientifolds of the $L^{1,5,2}$
theory. Other choices amount to overall sign flips.}
\label{tabdP3III}
\end{center}
\end{table}

As in previous examples, gauge anomalies are canceled by constraining the ranks of gauge groups and adding (anti)fundamental fields.

%===============================================================================
\section{Orientifolds from dimers with fixed lines}
\label{lines}
%===============================================================================

After discussing in detail the large class of involutions preserving
the mesonic $U(1)^2$ flavor symmetries, in this section 
we consider a different type of involution preserving a linear
combination of them. As pointed in \S\ref{dimersec}, they are 
described in the dimer diagram as symmetries which leave fixed lines.

%======================================================================
\subsection{Generalities}
\label{Generalities}
%======================================================================

\begin{figure}[!htp]
\begin{center}
\psfrag{a}[cc][][1]{a)}
\psfrag{b}[cc][][1]{b)}
\psfrag{c}[cc][][1]{c)}
\includegraphics[scale=0.7]{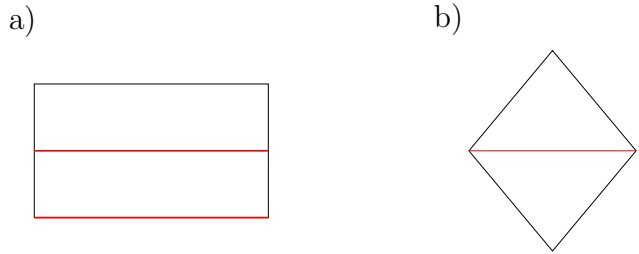}
\caption{The two possible unit cell geometries for orientifolds with fixed lines: rectangle and rhombus.}
\label{twounitcells}
\end{center}
\end{figure}

Orientifolds with fixed lines are obtained for dimer diagrams which
are invariant under a line reflection.\footnote{An additional 
constraint is that the symmetry should map black to black nodes and white to white nodes.} 
Symmetries with fixed lines exist for dimer diagrams with unit cell of two possible kinds, as shown in Figure 
\ref{twounitcells}. As in Figure \ref{twounitcells}a, the unit cell can be a `rectangle', in which case it has two 
independent orientifold lines. Or as in Figure \ref{twounitcells}b it can be a `rhombus', in which case there is a single
orientifold line.\footnote{In the orientifold literature it is well known that tori with line symmetries 
exist only for two possible complex structures (rectangular and rhombus). This issue has been discussed in detail
e.g. in \cite{Angelantonj:1999xf,Blumenhagen:2000ea}. It corresponds to the fact that in Type I the orientifold flips the 
B-field so only two choices are left: $B=0$ and $B=1/2$ (dual to rectangular and rhombus unit cell, respectively)}
Using the interpretation of the dimer diagram as a brane tiling of
NS-branes with suspended D5-branes \cite{Franco:2005rj, Franco:2005sm}, 
such orientifold lines are physical orientifold planes. Hence, for the case in Figure \ref{twounitcells}a we can consider 
independent sign choices (charges) for the two orientifold lines. 
The structure of the unit cell depends only on the dimer diagram under consideration.

For each choice of orientifold signs, the rules to read out the spectrum are similar to the case of fixed points, 
see \S\ref{section_generalities}. Our description here is thus more sketchy.
\begin{itemize}
\item Faces mapped to themselves by the orientifold action have their gauge factors projected down to $SO$ or $Sp$ for positive or negative orientifold line charge, respectively. Faces paired up by the orientifold action are identified, and give rise to a single unitary gauge factor.
\item Edges mapped to themselves correspond to chiral multiplets in the two-index symmetric or antisymmetric tensor representations for positive or negative orientifold line charge, respectively. Edges paired up by the orientifold action are identified, and give rise to a single bi-fundamental field.
\item Nodes mapped to themselves give rise to interaction terms in the superpotential.
\end{itemize}

%==================================================================
\subsection{Few examples and the geometric action}
%==================================================================

In this section we consider several examples of orientifold with fixed lines. We also describe the geometric action on mesons, which can be obtained using rules similar to those of fixed point orientifolds.

\subsubsection{Line orientifolds of $\mathbb C^3$}
\label{Cfixed}
\begin{figure}[!htp]
\begin{center}
\psfrag{1}[cc][][0.8]{1}
\psfrag{a}[cc][][0.7]{$\Phi_2$}
\psfrag{c}[cc][][0.7]{$\Phi_2$}
\psfrag{b}[cc][][0.7]{$\Phi_3$}
\psfrag{d}[cc][][0.7]{$\Phi_3$}
\psfrag{e}[cc][][0.7]{$\Phi_1$}
\psfrag{4}[cc][][0.7]{a}
\psfrag{5}[cc][][0.7]{b}
\psfrag{6}[cc][][0.7]{c}
\psfrag{7}[cc][][0.7]{d}
\includegraphics[scale=1]{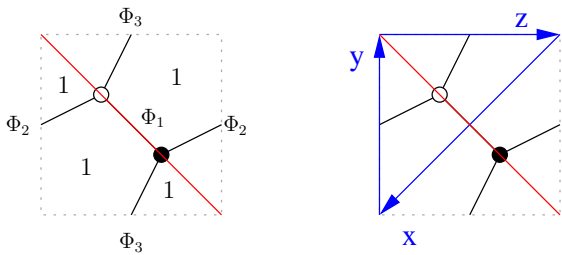}
\caption{Left: The dimer of $\mathbb C^3$ with line reflection. Right: The three fundamental mesons.}
\label{Cfixedline}
\end{center}
\end{figure}

As a little warmup, let us come back to $\C^3$. But this time, let us consider the line-reflection symmetry as 
depicted in Figure \ref{Cfixedline}.

% zzz

Clearly, The orientifold line passes through the face and depending on
the sign of the single O-plane, we obtain a $SO/Sp$ gauge theory. One
of the fields is mapped to itself and projects down to the adjoint of
the SO/Sp gauge group, which is denoted by $\Phi$. The other two fields
are exchanged and 
lead to one two-index symmetric and one antisymmetric representation,
denoted by $S, A$ respectively.  

The superpotential can be obtained from the parent theory by simply
replacing the original fields 
by the components surviving the projection:
\beq
W = \Phi A S - \Phi S A.
\eeq
The superpotential in coming examples can be found similarly.

Concerning the geometric action, we anticipate that
the line reflection exchanges the mesons $y$ and $z$ and keeps the meson $x$ fixed:
\beq
y\leftrightarrow z,~x\rightarrow x.
\eq
% zzz
This agrees with the fact that the model corresponds to the introduction of an O7-plane in a system of D3-branes in flat space.
The sign transformation of the mesons will be more systematically discussed in \S\ref{fixedlinerules}.

\subsubsection{$\IC^2/\IZ_N\times \IC$, even $N$}

Let us consider the example of $\IC^2/\IZ_N\times \IC$, distinguishing the even and odd $N$ cases. The geometry 
has a T-dual realization as a HW setup with $N$ parallel NS5-branes. Since these models are orbifolds, we can also 
construct the orientifolds we are to discuss using CFT techniques. Both descriptions can be found 
in \cite{Uranga:1999mb,Feng:2001rh}. Thus, the models provide a good starting point to derive the rules for the 
construction of the orientifolds in terms of dimer diagrams with fixed lines.

\begin{figure}[!htp]
\begin{center}
\psfrag{a}[cc][][1]{a)}
\psfrag{b}[cc][][1]{b)}
\psfrag{1}[cc][][0.7]{1}
\psfrag{2}[cc][][0.7]{2}
\psfrag{3}[cc][][0.7]{3}
\psfrag{4}[cc][][0.7]{4}
\includegraphics[scale=0.5]{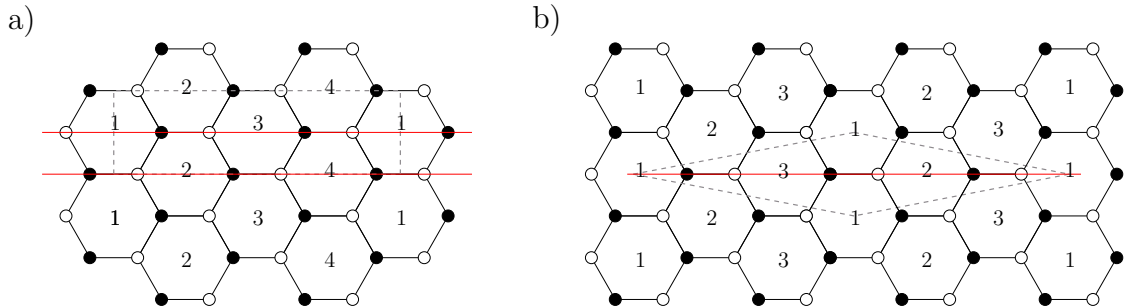}
\caption{Prototypical examples of orientifolds of the dimer of $\IC^2/\IZ_n\times \IC$
with orientifold fixed lines. Figure a) illustrates the even $n=4$ case, while the odd $n=3$ case is illustrated 
in Figure b). They correspond to rectangular and rhombus unit cells, respectively.}
\label{C2ZNline}
\end{center}
\end{figure}

Consider the even $N$ case, for example $\IC^2/\IZ_4\times \IC$ shown in Figure \ref{C2ZNline}a.
The unit cell has a geometry such that the two orientifold lines are independent, so their signs can be chosen 
independently. This leads to the four possible orientifold models whose gauge group and two-index tensor structure is given in Table \ref{C2ZNlinetab}. Here and in the 
following examples we take as convention that we label the fixed lines from bottom to top, i.e. $(+-)$ means that 
the lower fixed line has positive charge while the upper line has negative charge. Note that all models presented 
in Table \ref{C2ZNlinetab} possess in addition four bi-fundamentals:
\beq
(\fund_1,\fund_2) +(\fund_2,\fund_3) +(\fund_3,\fund_4)+(\fund_4,\fund_1).
\eq
% zzz
Denoting the two-index tensors $T_i$, and the bi-fundamentals $X_{i,i+1}$, the superpotential for the different models has the structure
\beqa
W\, =\, T_i X_{i,i+1}X_{i,i+1} - T_{i+1} X_{i,i+1} X_{i,i+1}.
\label{supoline}
\eeqa

\begin{table}[!htp]
\begin{center}
\begin{tabular}{ccc}
Charges&Gauge group&2-index tensors \\
\hline
$(+-)$&$Sp \times SO \times Sp \times SO$&$\symm_1 + \asymm_2 + \symm_3 + \asymm_4$ \\
$(-+)$&$SO \times Sp \times SO \times Sp$&$\asymm_1 + \symm_2 + \asymm_3 + \symm_4$ \\
$(++)$&$SO \times SO \times SO \times SO$&$\symm_1 + \symm_2 + \symm_3 + \symm_4$ \\
$(--)$&$Sp \times Sp \times Sp \times Sp$&$\asymm_1 + \asymm_2 + \asymm_3 + \asymm_4$ \\
\end{tabular}
\caption{The fixed line orientifolds of $\IC^2/\IZ_4\times \IC$.}
\label{C2ZNlinetab}
\end{center}
\end{table}

For instance, the model presented in the first row of Table \ref{C2ZNlinetab}
has $\N=2$ supersymmetry and reproduces the HW T-dual with an O4-plane. Notice the interesting feature that the alternating 
structure of projections on the gauge factors and two-index tensors, which in the HW is reproduced by an alternating 
charge of the O4 \cite{Evans:1997hk}, here comes about simply from the fact that different faces feel different 
orientifold planes (and similarly for edges).

Clearly, a similar model with completely opposite projections is obtained by considering the lower orientifold 
line to be negative and the upper to be positive, as given in the second row of Table \ref{C2ZNlinetab}.

Let us use the parameterization $xy=w^N$ for $\IC^2/\IZ_N$, and a coordinate $z$ for the additional $\IC$. 
Regarding the coordinates as mesons, we have an exchange of $x$ and $y$. The meson $z$ crosses two orientifolds 
of opposite signs, suggesting it is odd, the same holds for $w$. 

Let us consider the case of both orientifold lines with positive sign. The corresponding models can be found in the third 
and fourth row of Table \ref{C2ZNlinetab}.

The theory has $\N=1$ supersymmetry and reproduces the HW T-dual with an O8-plane, described in \cite{Feng:2001rh}.
One can also infer that the geometric action exchanges the mesons $x$ and $y$ while the meson $z$ crosses two 
orientifolds of same sign, suggesting it is even, similar for $w$.

%===================================================================
\subsubsection{$\IC^2/\IZ_N\times \IC$, odd $N$}
%===================================================================

Let us consider the odd $N$ case, for instance the case $\IC^2/\IZ_3\times \IC$ shown in Figure \ref{C2ZNline}b. 
The unit cell has a geometry such that there is only one orientifold line. 
Notice that this automatically forbids the analogs of the $\N=2$ orientifolds described above for the even $N$ case. 
This is in complete agreement with the impossibility to introduce O4-planes in HW configurations with an odd number 
of NS-branes.

The two possible models are given in Table \ref{C2ZNlinetab2}, where we show the gauge group and two-index tensor structure. There we use the notation $(++)$, $(--)$ for positive and negative orientifold lines (to keep the analogy with the previous case).

\begin{table}[!htp]
\begin{center}
\begin{tabular}{ccc}
Charges&Gauge group&2-index tensors \\
\hline
$(++)$&$SO \times SO \times SO $&$\symm_1 + \symm_2 + \symm_3$ \\
$(--)$&$Sp \times Sp \times Sp $&$\asymm_1 + \asymm_2 + \asymm_3$ \\
\end{tabular}
\caption{The fixed line orientifolds of $\IC^2/\IZ_3\times \IC$.}
\label{C2ZNlinetab2}
\end{center}
\end{table}
In addition, both models possess three bi-fundamentals
\beq
(\fund_1,\fund_2) +(\fund_2,\fund_3) +(\fund_3,\fund_1).
\eq
% zzz
The structure of the superpotential is as in (\ref{supoline})

The models correspond to $\N=1$ theories, in agreement with the T-dual HW setup with O8-planes described 
in \cite{Feng:2001rh}. Notice that in these HW models the O8-plane does not have any alternating structure, 
so they really exist for odd $N$ as well.

\subsubsection{The geometric action}
\label{fixedlinerules}
As for fixed point orientifolds, the simplest way to specify the geometric action of the orientifold quotient is 
by providing rules for the action on the gauge invariant mesonic operators. These can be easily obtained from 
simple examples (for which the geometric action is known via other techniques) and extended to general validity. 
We simply state them as follows 

\medskip
\bigskip
\noindent{\bf \underline{Rule 1} :}
{\it
A meson mapped to itself under the orientifold action picks a sign $(-1)^k$ where $k$ is the intersection number 
(counted with orientation) with negative orientifold lines.
}
\bigskip

\noindent{\bf \underline{Rule 2} :}
{\it
Two mesons mapped to each other under the orientifold action are related without any relative sign.
}
\\\\
A simple consequence of these rules is that mesons corresponding to superpotential terms associated to nodes on top of 
orientifold lines are even under the orientifold action. This does not contradict the fact that the complete superpotential 
is odd (cf. footnote \ref{orientsupo}), since in general different terms in the superpotential are swapped under the 
orientifold action.\footnote{The explanation in the mirror picture is as follows: These involutions switch the 
orientation of paths bounding the faces in the dimer. Hence, the corresponding 1-cycles in the mirror Riemannian 
surface do so. Thus, the superpotential terms are mapped to themselves, but an additional sign flip comes in due to 
the orientation change (for illustration compare Figures \ref{trivinvfig0} and \ref{coniinv}).} 

For instance, consider $\IC^3$ with the orientifold action $y\leftrightarrow z$, $x\to x$ as discussed in \S\ref{Cfixed}. 
The operators $t_1=tr(\Phi_1\Phi_2\Phi_3)$ and $t_2=tr(\Phi_2\Phi_1\Phi_3)$ are exchanged, so the meson (modulo F-terms) 
$t\equiv t_1\equiv t_2$ is even. Still, the complete superpotential $W=t_1-t_2$ is odd.

It is straightforward to apply the above rules to obtain the geometric action for the orientifolds constructed in the 
previous subsections. Let us use the parameterization, familiar from previous sections,  $xy=w^N$ for $\IC^2/\IZ_N$, 
and a coordinate $z$ for the additional $\IC$. Regarding the coordinates as mesons, we have an exchange of $x$ and $y$. 
The meson $z$ is mapped to itself and crosses two orientifold lines. Finally, the same holds for $w$. From the above rules 
we obtain the action
\beq
x\leftrightarrow y,~z\to \varepsilon z,~w\to \varepsilon w.
\eq
With $\varepsilon=-1$ for the models with one positive and one
negative orientifold line (and $\varepsilon=+1$ for the other 
examples). This precisely agrees with the geometric actions proposed in \cite{Uranga:1999mb} for the T-dual of the HW with 
O4-planes, and in \cite{Feng:2001rh} for the T-dual of the HW with O8-plane.
Notice that in all cases, the meson associated to the superpotential terms (which can be regarded as $xw$) is even.

\medskip

We hope that this examples suffice to illustrate the rules to construct orientifolds of dimers with fixed lines. In the next 
section we apply these rules to construct further families of examples.

%===============================================================%================================================================
\subsection{Further examples}
%================================================================

%================================================================
\subsubsection{General $L^{aba}$ theories}
%================================================================

It is straightforward to construct orientifolds with fixed lines for the entire $L^{aba}$ family. In this section we 
illustrate it with a few examples.

\begin{figure}[!htp]
\begin{center}
\psfrag{a}[cc][][1]{a)}
\psfrag{b}[cc][][1]{b)}
\psfrag{1}[cc][][0.7]{1}
\psfrag{2}[cc][][0.7]{2}
\psfrag{3}[cc][][0.7]{3}
\psfrag{4}[cc][][0.7]{4}
\includegraphics[scale=0.6]{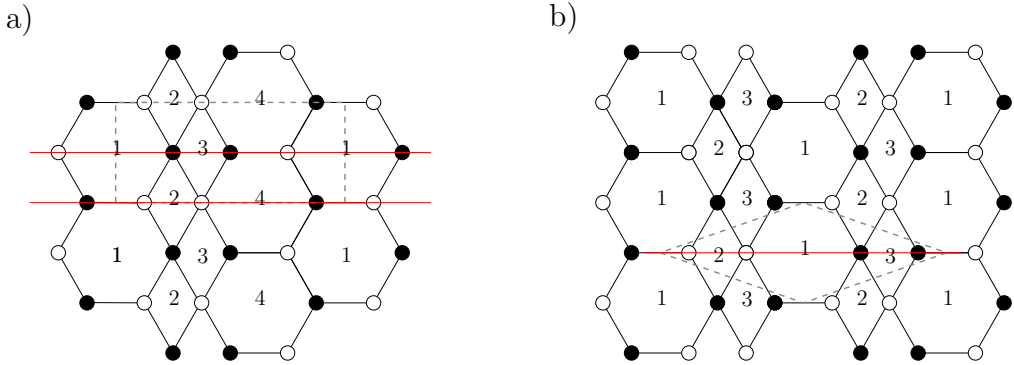}
\caption{Orientifold of the dimer of two $L^{aba}$ theories, corresponding to rectangular and rhombus unit cells, respectively. Left: $L^{131}$, Right: $L^{121}$.
}
\label{Labaline}
\end{center}
\end{figure}

Consider for instance the dimer shown in Figure \ref{Labaline}a, corresponding to the $L^{131}$ geometry.  
There are two independent orientifold lines, whose signs can be chosen independently. The gauge group and spectrum 
for the different choices are easily obtained using our rules.

This geometry is T-dual to a HW setup with 3 NS-branes and 1 NS$'$-brane.
The orientifold models we have described have not been discussed in the literature, but it is possible to use this 
T-dual picture to double-check our results. Indeed it is easy to show (and we skip its discussion) that models with 
opposite signs for the orientifold lines correspond to HW setups with an O4-plane, while choices with equal signs 
correspond to HW setups with an O8-plane. There is full agreement between the gauge group and matter spectra obtained 
using both constructions.

Consider now the model in Figure \ref{Labaline}b, corresponding to the $L^{1,2,1}$ geometry, namely the SPP theory.  
There is only one independent orientifold line, so there is only one sign to choose.
Again, these are new models, which have not been previously discussed in the literature. Since the SPP geometry is 
T-dual to a HW setup with 2 NS-branes and 1 NS$'$-brane, one can use the T-dual picture to verify our results. 
Indeed the models obtained agree with the introduction of an O8-plane in the T-dual HW setup.
Note also that the impossibility to introduce an O4-plane in the HW setup (due to the odd number of NS5-branes), 
agrees nicely with the geometry of the unit cell, which does not allow for other orientifold models.

\medskip

The generalization to the general $L^{aba}$ theory is clear. In particular, one can exploit the T-duality with a 
HW setup with $a$ NS$'$-branes and $b$ NS-branes along the circle direction to check results. The construction of the 
dimer goes as follows. One goes along the circle of the HW picture, and for each interval between two parallel branes 
one introduces a column of hexagons in the dimer, and for each interval between two rotated branes one introduces 
a column of rhombi. The complete picture is consistent, and admits the fixed line orientifolds we have discussed. 
For $a+b$ even, the corresponding dimer has a unit cell leading to two independent orientifold lines. The four possible choices of sign reproduce the different orientifolds, which are T-dual of models with O4-planes and O8-planes. For $a+b$ odd, there is only one orientifold line. The two possible choices of sign reproduce the different orientifolds, which are T-dual to HW models with an O8-plane.

\subsubsection{Orbifolds of the conifold}
\label{otherconisec}

\begin{figure}[!htp]
\begin{center}
\psfrag{a}[cc][][1]{a)}
\psfrag{b}[cc][][1]{b)}
\psfrag{1}[cc][][0.9]{1}
\psfrag{2}[cc][][0.9]{2}
\psfrag{3}[cc][][0.9]{3}
\psfrag{4}[cc][][0.9]{4}
\includegraphics[scale=1]{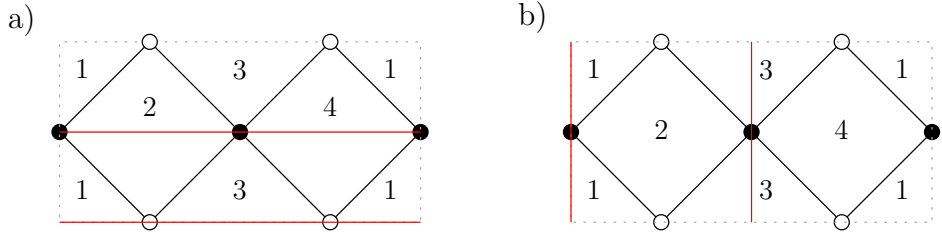}
\caption{Orientifolds with fixed lines for the dimer diagram of the $\IZ_2$ orbifold of the conifold.}
\label{coniZ2line}
\end{center}
\end{figure}

As a further set of examples, we would like to consider certain orbifolds of the conifold, introduced in 
\cite{Uranga:1998vf} (also known as generalized conifolds \cite{Aganagic:1999fe}) . They are also known as $L^{aaa}$ theories, hence some of their orientifolds fall in the above description. However given 
some additional symmetries of the dimer in this case, there are some new orientifold examples. To be more specific, 
in Figure \ref{coniZ2line} we show the possible orientifolds of the dimer diagram for the  particular case of the 
$\IZ_2$ orbifold of the conifold, which is illustrative enough for our purposes.

Let us consider the configuration in Figure \ref{coniZ2line}a. Its discussion fits in the general framework of the 
$L^{aba}$ theories mentioned above, so we simply sketch their description. The unit cell is such that there are 
two independent orientifold lines. The different choices lead, using our construction, to orientifold models in 
nice agreement with possible orientifold in the T-dual HW setups with two NS-branes and two NS$'$-branes. Specifically, 
dimer diagrams with orientifold lines of different sign give models which are T-dual to the introduction of an O4-plane  
in the HW dual. Also, dimer diagrams with orientifolds lines of the same sign yield models T-dual to HW setups with an O8-plane.

% zzz

The gauge theory has a structure $G_1\times H_2\times G_3\times H_4$, with $G,H$ being orthogonal or symplectic depending on the sign of the orientifold line passing though the corresponding face. The matter content is given by chiral bi-fundamentals $X_{i,i+1}$ in the $(\fund_i,\fund_{i+1})$, and the superpotential is
\beqa
W= X_{12}^2 X_{23}^2 - X_{23}^2 X_{34}^2 + X_{34}^2 X_{41}^2 - X_{41}^2 X_{12}^2.
\eeqa

% zzz

Describing the geometry as $xy=z^nw^n$, with the basic mesons defined as in Figure \ref{coniZ2fig1}, the geometric actions correspond to
\beq
x\leftrightarrow y,~ z\to \varepsilon z,~w\to\varepsilon w,
\eq
with here and later on $\varepsilon$ is given by the product of orientifold line signs 
(negative for opposite signs, positive for equal sign). 

Consider now the configuration in Figure \ref{coniZ2line}b. Again there are two independent orientifold lines, 
whose signs can be chosen independently. The structure of the orientifold theory on $\IZ_N$ orbifold of the conifold 
is as follows. The gauge group reads
\beqa
G_1\times U(n_2) \times \ldots \times U(n_{N}) \times G_{N+1},
\eeqa
with $G_i$ given by $SO/Sp$ for $O^{+/-}$ respectively. 

For the $N=2$ case in the figure, we have $G_1\times U(n_2)\times G_3$, and the matter content is given by
%zzz
\beqa
(\fund_1,\antifund_2)+(\fund_1,\fund_2) +(\fund_2,\fund_3) +(\antifund_2,\fund_3),
\eeqa
with superpotential
\beqa
W\, =\, X_{12}^2 X_{21}^2  - X_{12}X_{23}X_{32}X_{21} + X_{23}^2 X_{32}^2.
\eeqa

Using our techniques, the geometric action of the orientifold can be obtained as
\beq
z\leftrightarrow w,~ x\to \varepsilon x,~ y\to \varepsilon y.
\eq

The above orientifold corresponds under T-duality to a HW model with O6-planes away from the NS- and NS$'$-branes, 
and mapping them to each other (thus the orientifold planes are at 45 degrees with the NS-branes in the 48 and 59 directions). 
This orientifold has been briefly considered in \cite{Park:1999ep}.

\smallskip

Notice that for the particular case of the conifold the two orientifolds described above (with horizontal or vertical 
fixed lines) are isomorphic (up to a relabeling of coordinates). This is related to the enhanced global symmetries 
of the conifold, which lead to a very symmetric dimer diagram.

Thus we have two possible HW T-dual interpretations for the orientifold of $xy=zw$ by the orientifold 
$\omega \sigma (-1)^{F_L}$, with $\sigma$ acting as
\beq
x\leftrightarrow y,~z\to \varepsilon z,~ w\to \varepsilon w.
\eq
This is in fact correct, as one can see as follows: One can regard the conifold as a $\IC^*$ , parameterized by $x,y$, 
fibered over $z,w$ and degenerating at $z=0$ an $w=0$. T-dualizing along the circle in this $\IC^*$ 
(namely the orbit of $x\to e^{i\theta}x, y\to e^{-i\theta}y$) gives one NS along $z$ and one NS$'$ along $w$. 
In this T-dual picture the orientifold action reproduces an O4-plane for $\varepsilon=-1$ and an O8-plane 
for $\varepsilon=+1$.  On the other hand, one can regard the same equation as describing a $\IC^*$ fiber 
parameterized by $z,w$ fibered over $x,y$. T-dualizing along the circle in this fibration, one gets one 
NS along $x$ and one NS$'$ along $y$. In this T-dual, the orientifold action introduces and O6-plane at 
45 degrees with the NS- and NS$'$-branes. Hence the two HW configurations are correct T-dual descriptions 
of the orientifold model. It is amusing that the dimer representation is clever enough to realize this 
and make manifest that the two orientifolds are actually the same!

%%%%%%%%%%%%%%%%%%%%%%%%%%%%%%%%%%%%%%%%%%%%%%%%%%%%%%%%%%%%%%%%%%%%%%%%%%%%%%%
\section{The mirror perspective}
%%%%%%%%%%%%%%%%%%%%%%%%%%%%%%%%%%%%%%%%%%%%%%%%%%%%%%%%%%%%%%%%%%%%%%%%%%%%%%%
\label{mirrorsec}

The physical realization of the dimer diagram has been shown to be quite manifest upon use of mirror 
symmetry \cite{Feng:2005gw}. The different gauge factors arise from D6-branes wrapped on 3-cycles 
in the mirror geometry, with the bi-fundamental multiplets arising from their intersection and the 
superpotential couplings arising from open string disk worldsheet instantons. The dimer diagram is 
moreover recovered from the mirror geometry by a certain projection on a two-torus.

We may therefore expect that there is a natural interpretation of the orientifold action in the 
mirror geometry, and that some aspects of the previous discussion receive a new light from this viewpoint. 
In this section we provide a description of orientifolding in the mirror geometries, and its interplay with 
our previous dimer diagram description. Although we do not achieve a systematic understanding of all 
aspects of orientifolding, we recover many interesting results.

\subsection{Review of the mirror picture }

\noindent{\bf The mirror configuration}
\\\\
Under local mirror symmetry, a system of D3-branes located at the singular point of a non-compact 
toric Calabi-Yau 3-fold $\mf$ are mapped to a set of D6-branes wrapping supersymmetric 3-cycles 
in the mirror Calabi-Yau $\mmf$ \cite{Hanany:2001py,Feng:2002kk}. The mirror manifold $\mmf$ can be 
seen as a double fibration over the complex plane \cite{Hori:2000ck,Hori:2000kt}:
\beq
\begin{split}\label{mgeoeq1}
z&=uv, \nonumber \\
z&=P(x_1,x_2),\\
\end{split}
\eq
where $x_1,x_2\in\C^\times$, $u,v,z\in\C$ and $P(x_1,x_2)$ is the Newton polynomial of the toric diagram of  
the singularity.  Note that one can read off $P$ directly from the corresponding toric diagram 
\cite{Leung:1997tw,Hanany:2001py}.
More specifically, for a singularity whose toric diagram is given by a set of points $\{ (n_i,m_i)\}$, we have
\beqa
P(x_1,x_2)\, =\, \sum_i\, x_1^{n_i}\, x_2^{m_i}.
\eeqa
For instance, for the conifold we have $P=1+x_1+x_2+x_1x_2$.
Note that the $SL(2,\IZ)$ freedom in choosing the basis of the toric lattice corresponds to a redefinition of 
coordinates in the mirror geometry.

The first fiber defines a $\IC^*$ fibration, degenerating at the origin of the base $z=0$, where 
the $\IS^1$ in $\IC^*$, denoted $\IS^1_c$, shrinks to zero. The second fiber describes a punctured 
Riemannian surface, which is denoted as $\Sigma$.  Its genus
$g$ equals the number of internal points of the toric diagram. The fiber
over $z=0$, denoted $\Sigma_0$, is important for our purposes.
It corresponds to a smooth Riemann surface which can be thought of as a
thickening of the web diagram dual to the toric diagram \cite{Aharony:1997ju,Aharony:1997bh,Leung:1997tw}. 
At certain critical points $z^*_i$ on the base, certain 1-cycles $[c_i]$ of $\Sigma$ pinch off. In detail, 
the critical points $z^*_i$ can be obtained by solving
\beq\label{cpointseq}
\partial_{x_1}P=0,~\partial_{x_2}P=0,
\eq
for $x_1,x_2$ and substituting back into the fiber defining $\Sigma$.

The geometry has a non-trivial compact 3-homology. We can define a set of 3-cycles $[C_i]$ by taking the 
segment $[0,z^*_i]$ on the base and fibering the two-torus $\IS^1_c\times [c_i]$ over it. Since $\IS^1_c$ 
shrinks to zero at $z=0$ and $[c_i]$ shrinks to zero at $z_i^*$, this defines a 3-cycle with the topology of $\IS^3$. 
The mirror image of the D3-branes at the singularity correspond to D6-branes wrapped on these 3-cycles. 
Namely, the fractional D3-brane giving rise to the $i^{th}$ gauge factor is mapped to a D6-brane wrapped 
on the 3-cycle $[C_i]$. The bi-fundamental chiral multiplets $(\fund_i,\antifund_j)$ arise from open strings 
stretching between the $i^{th}$ and $j^{th}$ D6-branes, at their intersections. That is, there are 
$I_{ij}=[C_i]\cdot [C_j]$ bi-fundamentals $(\fund_i,\antifund_j)$ in the gauge theory. Notice from the 
above description that the 3-spheres intersect only on the fibers above $z=0$. Hence, the geometry of 
the D6-brane cycles and their intersections is encoded in $\Sigma_0$.

\begin{figure}[!htp]
\begin{center}
\psfrag{1}[cc][][0.7]{1}
\psfrag{2}[cc][][0.7]{2}
\includegraphics[scale=1.2]{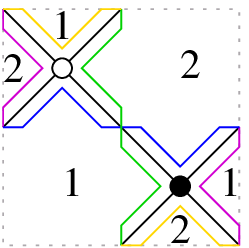}
\caption{Dimer of the conifold with the corresponding four zigzag paths (for simplicity we
omitted the orientation of the zigzag paths).} \label{conidimer_zigzag}
\end{center}
\end{figure}

The close relation between $\Sigma_0$ and the gauge theory works in the reverse way, and in fact
it is easy to recover $\Sigma_0$ and the geometry of the D6-branes in terms of the dimer
diagram. Given a dimer diagram, one can define zigzag paths (these, along with the related
rhombi paths, were introduced in the mathematical literature on dimers in
\cite{Kenyon2003,Kenyon:2003uj}, and applied to the quiver gauge theory context in
\cite{Hanany:2005ss}), as paths composed of edges, and which turn maximally to the right at e.g.
black nodes and maximally to the left at white nodes. They can be conveniently shown as oriented
lines that cross once at each edge and turn at each vertex, as shown in Figure
\ref{conidimer_zigzag}. We denote the colorful diagram corresponding
to the set of such zigzag-lines as the 
Harlequin diagram.\footnote{Any graph has an associated Harlequin diagram consisting of
  its zigzag paths. In short, we will introduce 
a tiling of the $\Sigma_0$
that we denote shiver diagram. Both the dimer and shiver diagrams have associated Harlequins. We are confident that the reader can 
distinguish between the two constructions based on the context.}
Notice that at each edge two zigzag paths must have opposite orientations. For dimer models
describing toric gauge theories, these zigzag paths never self-intersect and form closed loops
wrapping $(p,q)$ cycles on the $\IT^2$. This is shown for the conifold in Figure
\ref{conidimer_zigzag}.

\begin{figure}[!htp]
\begin{center}
\epsfxsize=10cm
\hspace*{0in}\vspace*{.2in}
\psfrag{A}[cc][][0.8]{A}
\psfrag{B}[cc][][0.8]{B}
\psfrag{C}[cc][][0.8]{C}
\psfrag{D}[cc][][0.8]{D}
\psfrag{a}[cc][][0.65]{$X^{(1)}_{12}$}
\psfrag{b}[cc][][0.65]{$X^{(1)}_{21}$}
\psfrag{c}[cc][][0.65]{$X^{(2)}_{12}$}
\psfrag{d}[cc][][0.65]{$X^{(2)}_{21}$}
\psfrag{p}[cc][][0.65]{$\P^1$}
\psfrag{i}[cc][][1]{a)}
\psfrag{ii}[cc][][1]{b)}
\includegraphics[scale=0.6]{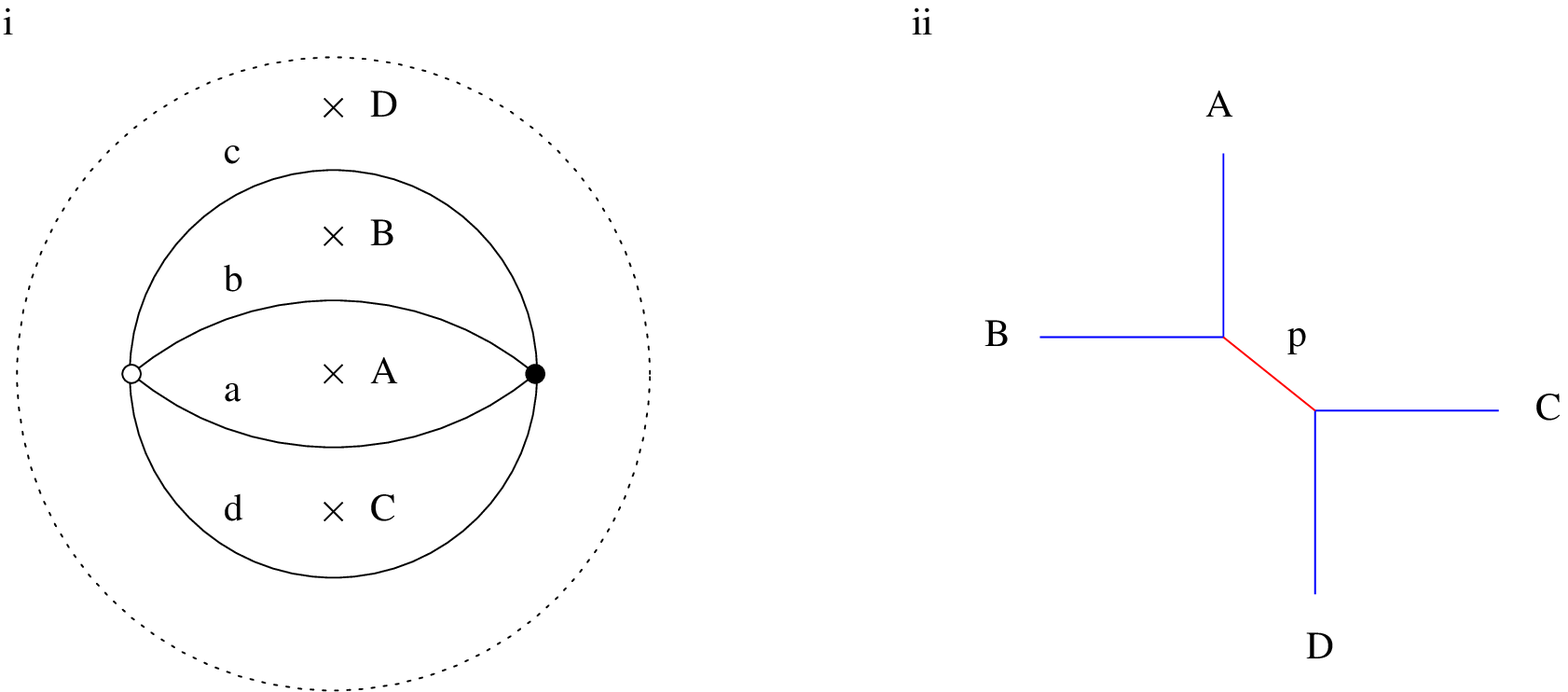}
\caption{\small a) Shiver of the conifold b) The corresponding web diagram which provides a
skeleton of $\Sigma_0$ with asymptotic legs
corresponding to the punctures.}
\label{coniriem}
\end{center}
\end{figure}

\begin{figure}[!htp]
\begin{center}
\psfrag{A}[cc][][0.8]{A}
\psfrag{B}[cc][][0.8]{B}
\psfrag{C}[cc][][0.8]{C}
\psfrag{D}[cc][][0.8]{D}
\includegraphics[scale=0.8]{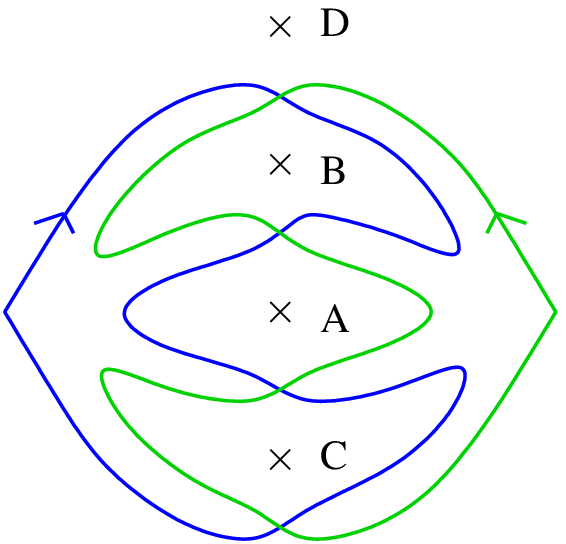}
\caption{Harlequin diagram of the conifold shiver. The two zigzag paths of the conifold shiver
corresponding to the 1-cycles in $\Sigma_0$ the D6-branes are wrapped which give rise to the two
gauge groups.} \label{conisurf}
\end{center}
\end{figure}

As shown in \cite{Feng:2005gw}, the dimer diagram can be associated to a tiling of the Riemann
surface $\Sigma_0$ in the mirror geometry, which we denote the `shiver diagram'. Specifically,
each zigzag path of the dimer encloses a face of the shiver, and the $(p,q)$ charge of the
associated leg in the web diagram is the $(p,q)$ homology charge of the zigzag path in the dimer
$\IT^2$. The tiling of $\Sigma_0$ for the conifold is shown in Figure \ref{coniriem}a, while the
corresponding web diagram is shown in Figure \ref{coniriem}b.
The dimer diagram moreover encodes the 1-cycles in the mirror Riemannian surface, associated to
the different gauge factors in the gauge theory. Consider a gauge factor associated to a face in
the dimer diagram. One can consider the ordered sequence of zigzag path pieces that appear on
the interior side of the edges enclosing this face. By following these pieces in the tiling of
$\Sigma_0$ one obtains a non-trivial 1-cycle in $\Sigma_0$ which corresponds precisely to that
used to define the 3-cycle wrapped by the mirror D6-branes carrying that gauge factor. Using
this map, it is possible to verify all dimer diagram rules (edges are bi-fundamentals, nodes are
superpotential terms) mentioned at the beginning. An amusing feature is that these non-trivial
1-cycles in $\Sigma_0$ are given by zigzag paths of the tiling of $\Sigma_0$. The non-trivial
1-cycles in the mirror Riemann surface for the case of the conifold are shown in Figure
\ref{conisurf}.

\begin{figure}[!htp]
\begin{center}
\includegraphics[scale=0.6]{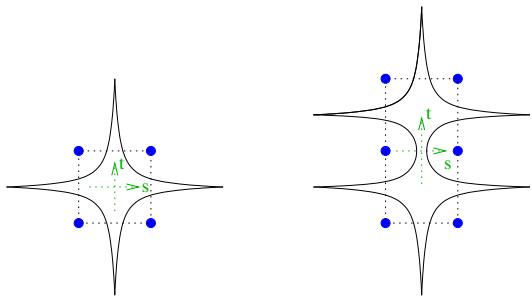}
\caption{Left: Amoebae of the conifold. Right: Amoebae of conifold/$\Z_2$.}
\label{algafig1}
\end{center}
\end{figure}

A direct construction of the shiver diagram from the knowledge of the toric gauge theory and its 
superpotential is explained in detail in \S\ref{Schains}.
\\\\
{\bf The amoeba and alga maps}
\\\\
There exists certain projections of the $\Sigma$ fibration which turn out to be useful in relating 
the mirror configuration to the dimer diagram \cite{Feng:2005gw}. Consider the amoeba projection, 
given by the map $\Sigma\rightarrow \R^2$:
\beq\label{algaeq1}
 (x_1,x_2)\in\Sigma\rightarrow (s=\log(|x_1|),t=\log(|x_2|))\in\R^2,
\eq
It has been widely studied in mathematics, and it has the useful property that the punctures of 
$\Sigma$ map to "tentacles" perpendicular to the convex polyhedron associated to the toric variety 
of $\mf$ \cite{mikh2001}. Namely, it provides a smooth version of the web diagram for the toric singularity. 
For example, the amoeba of the conifold and a $\Z_2$ orbifold thereof are illustrated in Figure \ref{algafig1}.

Let us consider a different projection. Namely, the projection of $\Sigma$ onto the phases of 
$x_1=|x_1|e^{i\Phi}$ and $x_2=|x_2|e^{i\Theta}$. This projection was dubbed alga map in \cite{Feng:2005gw}, 
and provides a natural map
 $\Sigma\rightarrow \IT^2$:
\beq
\label{algaeq0}
(x_1,x_2)\in\Sigma\rightarrow (\Phi,\Theta)\in \IT^2,
\eq
Moreover, at least for certain cases, it is possible to identify this $\IT^2$ with the dimer 
$\IT^2$ and recover the dimer diagram from the image of the alga \cite{Feng:2005gw}. 
As is discussed in \S\ref{classes}, this map turns out to be useful for translating the 
orientifold action from the dimer to an involution on $\Sigma$.\\\\
 \noindent {\bf Tadpole cancellation}
\\\\
An important consistency condition of systems of D-branes at singularities is cancellation of 
RR tadpoles of compact support. This condition is manifest in the mirror geometry, as cancellation of 
compact homology charge of the wrapped 3-cycles, see \cite{Hanany:2001py,Uranga:2002pg}. 
Given that the set of 3-cycles $[C_i]$ in the mirror geometry provides a basis of the compact 3-homology, 
an equivalent definition is as follows: the total homology class of the system must have zero intersection 
number with any of the basic 3-cycles $[C_i]$. As usual, this implies cancellation of gauge anomalies for 
the four-dimensional field theory.

The simplest solution to these constraint is to consider an equal number of D-branes corresponding to each gauge factor 
(the conformal case of D3-branes at a singularity). In the mirror geometry, it is possible to see that the set of 3-cycles 
$[C_i]$ can recombine into a single 3-cycle with the topology of $\IT^3$. This is the mirror of the statement that 
in this situation the different fractional branes can form a bound state corresponding to a dynamical D3-brane able 
to move off the singular point.

In the presence of orientifold planes, the orientifold plane has a contribution to RR charge, proportional to the 
compact 3-homology class of the 3-cycle it wraps. This implies that the set of D6-branes required to cancel RR tadpoles 
is different from those in unorientifolded system (and in particular that no recombination into a $\IT^3$ cycle is possible).
 An example is discussed in \S\ref{tadsec}.
\\\\
\noindent{\bf Supersymmetry and calibrations}
\\\\
In order that the D6-branes and O6-planes in the mirror geometry preserve a common supersymmetry, they need to wrap 
special Lagrangian submanifolds. This is made more explicit in \S\ref{calisec}, where a (local) calibration condition 
is given for the embedding of the D6-branes and O6-planes to fulfill.
\\\\
\noindent{\bf A subtlety for genus 0}
\\\\
\label{genuspuz}
We would like to conclude this section with a clarification regarding the genus
0 case. As already mentioned in \cite{Feng:2005gw}, for $\Sigma$ of genus 0 it may happen that
the mirror fibration has fewer critical points than needed to explain the number of gauge
factors. For instance for the mirror of $\C^3$ there are no critical points, and for the
conifold there is only a single critical point.

Looking at these examples in detail one however finds no paradox. For instance, the mirror of the D3-branes in $\IC^3$ 
is given by a D6-brane wrapped on a smooth $\IT^3$ in the mirror geometry (given by fibering $\IS^1_c$ times a 1-cycle 
in $\Sigma$ over a 1-cycle in the base encircling the origin).

Similarly, a single critical point for the conifold is not necessarily a problem. The $\Sigma$ fibration in this case 
is given by\footnote{To avoid confusion, note that we switched on a non-trivial B-field. As is clear, without B-field 
the mirror interpretation breaks down at the conifold point.}
\beq
1+x_1+x_2-x_1x_2-z=0,
\eq
Clearly, there is only a single critical point at $z=2$, above which the fiber is
\beq
(1-x_1)(1-x_2)=0.
\eq

This describes two complex planes given by $x_1=1$, $x_2$ arbitrary and $x_2=1$, $x_1$
arbitrary. Under the amoeba projection (\ref{algaeq1}) they reproduce the web diagram of the
conifold (cf. Figure \ref{algafig1}), with the interior shrunk to a point. This implies that the
two 1-cycles associated to the two gauge factors shrink at this point, and can be used to
construct two 3-spheres. This fits with the picture in  \cite{Hanany:2001py}.

\begin{figure}[!htp]
\begin{center}
\psfrag{a}[cc][][1]{a)}
\psfrag{b}[cc][][1]{b)}
\includegraphics[scale=0.5]{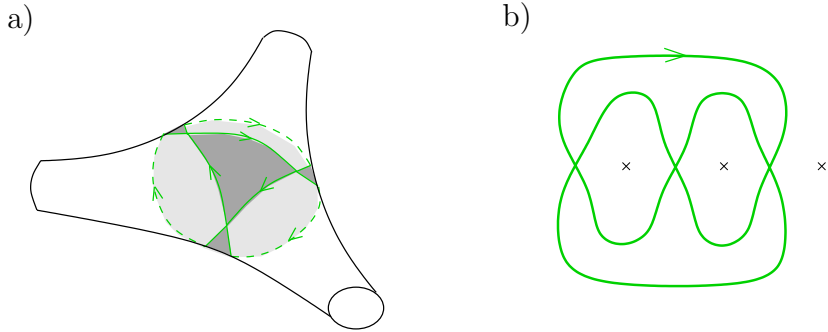}
\caption{a) The $\IC^3$ shiver Harlequin on the Riemann sphere with three punctures. The darkly shaded faces mark the 
world-sheet disk instantons. b) Heuristic picture of the $\IC^3$ shiver Harlequin diagram. There is one self-intersecting 
cycle, with three intersection points giving the three adjoint fields of $\mathcal{N}=4$ SYM.}
\label{c3shiver}
\end{center}
\end{figure}

We conclude this review section by noting that one can nicely visualize many of the properties of the mirror geometry 
by drawing the mirror surface $\Sigma_0$ with the 1-cycles wrapped by the D6-branes as illustrated in Figure \ref{c3shiver}a.

\subsection{Orientifolds in the mirror system}

In this section we would like to consider the introduction of orientifold quotients in the mirror configurations. 
As described above the mirror configuration is given by a set of D6-branes wrapped on special lagrangian 3-cycles 
in the mirror Calabi-Yau geometry. The natural orientifolding operation in this system is the introduction of O6-planes 
on the 3-cycles fixed under an antiholomorphic action, that is modding out by $\omega {\tilde \sigma} (-1)^{F_L}$, where $\omega$ is 
worldsheet parity and ${\tilde \sigma}$ is an antiholomorphic action on the mirror geometry (hence acting as $\Omega\to {\ov \Omega}$ 
on the holomorphic 3-form).\footnote{Systems of D6-branes in the presence of O6-planes have been extensively studied 
in compact setups in the intersecting brane model literature, 
see e.g. \cite{Uranga:2003pz,Blumenhagen:2005mu,Blumenhagen:2006ci,Marchesano:2007de} for reviews.} 
Notice that this implies that the orientifold preserves a common supersymmetry with the D6-brane system. 
Different choices of the orientifold geometry corresponds to different orientifolds of the D-brane systems 
(as is studied in Type IIB using the dimer diagram). Since on the mirror side one has to deal with the 
(in general complicated) geometry of $\Sigma$, a systematic classification of possible orientifolds is beyond 
the scope of this work. Our aim is rather to present how different ingredients of the orientifold arise 
in the mirror picture.

\begin{figure}[!htp]
\begin{center}
\includegraphics[scale=0.4]{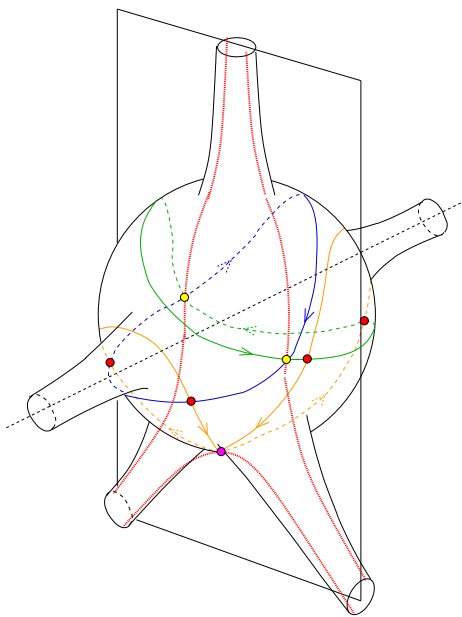}
\caption{Orientifold action on the mirror Riemann surface $\Sigma_0$ of the SPP singularity. 
The red lines are the non-compact 1-cycles wrapped by the O6-planes, while the other colored lines are 
the 1-cycles wrapped by the D6-branes.
}
\label{trivinvfig2}
\end{center}
\end{figure}

It is illustrative to describe at this stage some general features of the orientifold mirror geometries.
The antiholomorphic action on the geometry (\ref{mgeoeq1}) can be defined by specifying antiholomorphic actions on 
the base and fibers. Namely, the O6-planes in the geometry span a real dimension on the base, and on each of the fibers.
Clearly, given that essentially all the information of the gauge theory is encoded on the Riemann surface $\Sigma_0$ 
over $z=0$, much of the information of the orientifold is encoded in the antiholomorphic action on $\Sigma_0$. 
In Figure \ref{trivinvfig2} we show a prototypical example of such involution, for the case of the mirror of 
the SPP singularity.

\begin{figure}[!htp]
\begin{center}
\psfrag{a}[cc][][0.65]{$X_{21}$}
\psfrag{b}[cc][][0.65]{$X_{32}$}
\psfrag{c}[cc][][0.65]{$X_{13}$}
\psfrag{d}[cc][][0.65]{$\Phi_3$}
\psfrag{e}[cc][][0.65]{$X_{31}$}
\psfrag{f}[cc][][0.65]{$X_{12}$}
\psfrag{g}[cc][][0.65]{$X_{23}$}
\includegraphics[scale=0.6]{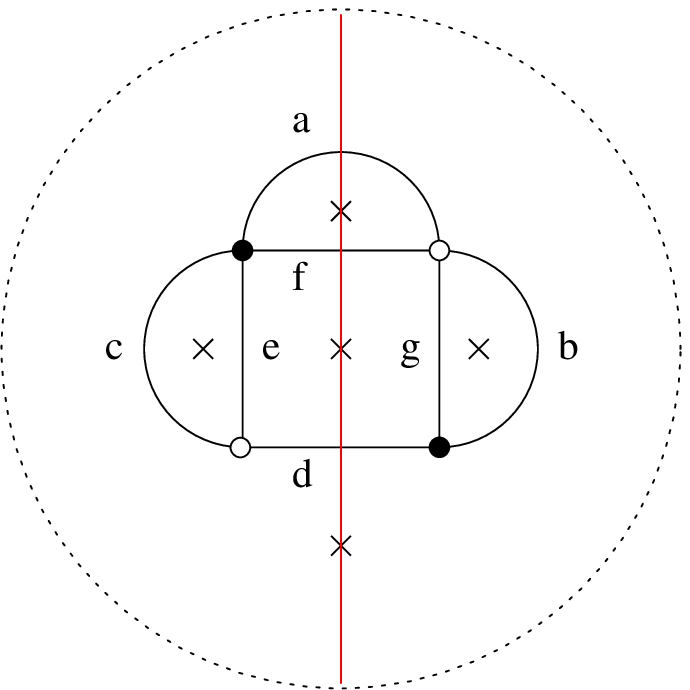}
\hspace{2cm}
\includegraphics[scale=0.4]{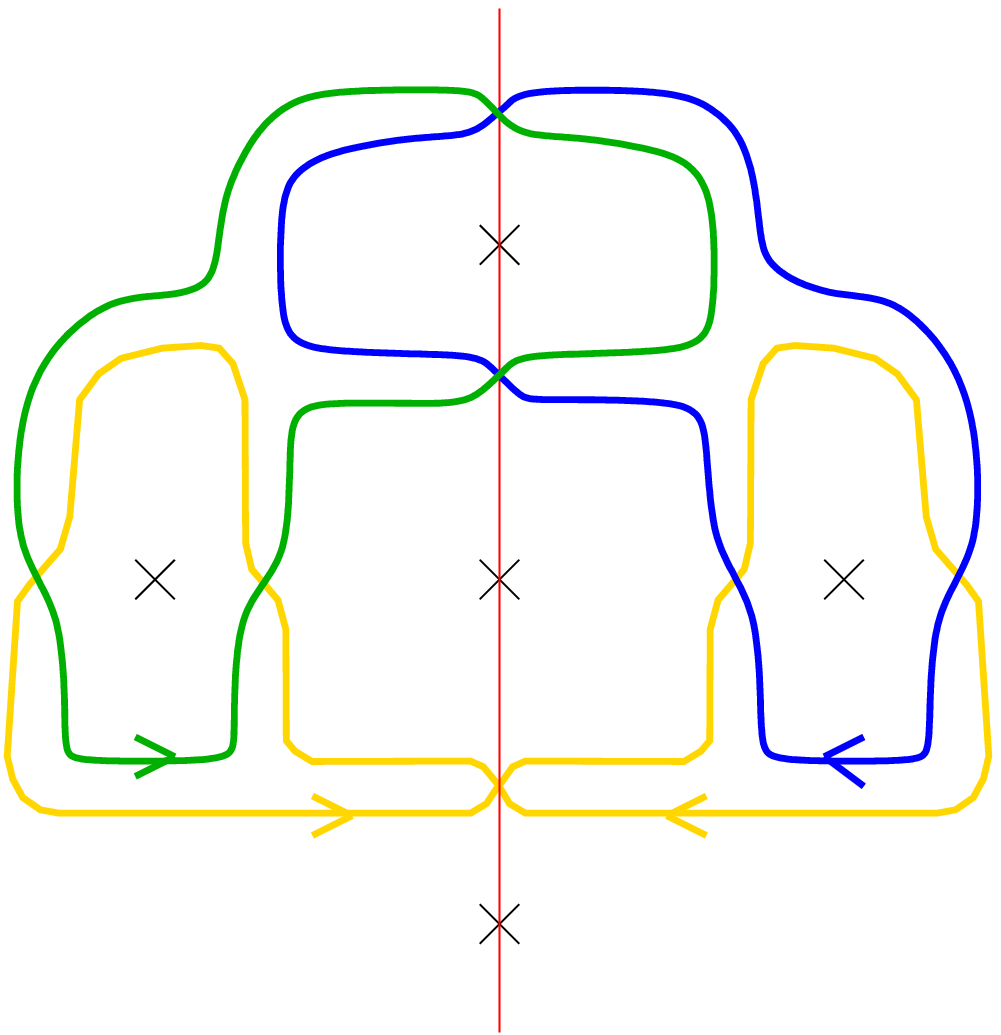}
\caption{Left: Antiholomorphic involution acting on the shiver of SPP. Right: Action on the
corresponding Harlequin diagram.} \label{trivinvfig0}
\end{center}
\end{figure}

The anti-holomorphic action on $\Sigma_0$ can more practically be visualized on the shiver diagram of the system. 
This allows a direct interpretation of the action of the orientifold on the gauge theory data. 
The action in Figure \ref{trivinvfig2} is shown on the shiver of the SPP in Figure \ref{trivinvfig0}.

Note that the fixed set contains a set of non-compact branches escaping to infinity along punctures. 
In principle these orientifold branches correspond to disconnected O6-planes in the geometry, and 
hence can have different RR charges and thus imply different orientifold projections. We expect that 
the freedom in choosing these charges (along with other discrete choices on how the involutions act on 
the rest of the mirror geometry) reproduce the different orientifolds of a given system. However we have not achieved 
such a complete understanding, and here simply restrict to well-understood aspects of the mirror orientifold.

In either of these pictures it is easy to read out the effect of the orientifold on the gauge theory.
For instance, the gold D6-branes are mapped to themselves, while the green and blue are mapped to each other.\footnote{Of course this statement is valid for an appropriate action of the orientifold on the $z$-base, 
which is easy to construct and whose details we skip for the moment.} This implies that the gauge group of 
the orientifold field theory is
\beqa
G_1\times U(N_2),
\eeqa
with $G_1$ corresponding to $SO/Sp$ depending on the sign of the orientifold branch crossing the gold D6-brane. 
Concerning the matter content, the orientifold action on the intersections implies that it is given by
\beqa
(\fund,\antifund) + (1,R) + (1,{\ov S}) + (T,1),
\eeqa
where $R,S,T$ are two-index tensor representations determined by the RR charge of the O6-plane branches passing 
through the corresponding intersection (and possibly the relative geometry of the D6-branes and O6-plane 
in the complete geometry).

These pictures of the mirror construction provide a good understanding of the orientifold action on the gauge theory 
ingredients. However, we have not found a proper general explanation of detailed projections of the gauge groups 
to orthogonal and symplectic factors, and of the chiral multiplets mapped to themselves to two-index symmetric and 
antisymmetric representations.
Leaving this more detailed understanding for future work, we proceed with the description of several classes of 
involutions and their effect on the gauge theory.

\subsection{Classes of orientifold involutions}
\label{classes}
It is possible to use the alga projection to find the involutions of the mirror geometries that correspond to 
the orientifold of the dimer diagrams studied in \S\ref{dimersec}.  The main idea to achieve this is to start 
from the action on the coordinates $(\Phi,\Theta)$ of the dimer $\IT^2$, and infer an action on the variables 
$x_1$ and $x_2$ of the curve $\Sigma_0$. By demanding invariance of $\Sigma$ under the action, one can infer 
an extension to the whole double fibration given in (\ref{mgeoeq1}), i.e. infer an action on $u,v$ and $z$.\footnote{It is implicitly understood that we are at special points in moduli space such that the involution exists.}

\subsubsection{Involutions mirror to orientifold dimers with fixed points}
\label{mirrorpoint}

Let us start with the orientifolds that act as a point reflection on the dimer diagram. Using the alga map (\ref{algaeq0}) 
it follows that point reflection in the dimer diagram corresponds to complex conjugation of $x_1$ and $x_2$.

This can be extended to an antiholomorphic involution of the $\Sigma$ fibration by specifying the additional action 
$z\rightarrow\cc z$. Note that this maps the fiber over $z$ to the fiber over ${\ov z}$. Note also that this in general 
exchanges the degeneration points $z_i^*$ of the fibration.

In most examples (see later for examples), supersymmetry of the O6-plane with respect to the D6-branes fixes that the 
latter should span the $\IS^1_c$ circle in the $\IC^*$ fibration. This implies that the anti-holomorphic action on the 
$\IC^*$ fibration is
\beq
(u,v)\rightarrow (\cc v,\cc u).
\eq
Note that the phase in this action is fixed by requiring that the $\IS^1_c$ is mapped point-wise to itself (so that the 
orientifold planes really exist).

Given the above action on the geometry, one can obtain the action on the  3-spheres the D6-branes are wrapped on, 
namely $C_i\rightarrow C_{\cc i}$, where $\cc i$ denotes the complex conjugate of the critical point. This allows to 
obtain the effect of the orientifold on the gauge theory, as illustrated in the following concrete example:

Consider the orientifolds of $\IC^3/\IZ_3$, studied in \S\ref{C3Z3sec} from the viewpoint of the dimer diagram. 
The mirror of the geometry and the system of D-branes has been discussed in detail in 
\cite{Hori:2000ck,Cachazo:2001sg,Hanany:2001py}. The mirror geometry
is given by
\beqa
z & = & uv, \nonumber \\
z & = & P(x_1,x_2)\, =\, x_1+x_2+{x_1^{\, 2}x_2^{\, 2}},
\eeqa

The degenerations of the $\Sigma$ fibration can be obtained by solving (\ref{cpointseq}). They are located at
\beqa
z_k^* \, =\, r \, e^{2\pi i\, k/3},
\eeqa
with $r$ a constant not relevant to the discussion below.

\begin{figure}[!htp]
\begin{center}
\psfrag{1}[cc][][0.7]{1}
\psfrag{2}[cc][][0.7]{2}
\psfrag{3}[cc][][0.7]{3}
\psfrag{z}[cc][][0.7]{z}
\includegraphics[scale=1]{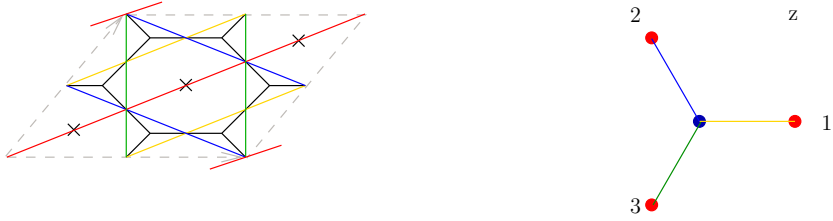}
\caption{The mirror $\Sigma$ fibration for the $dP_0$ theory. On the left we show the elliptic fiber $\Sigma_0$, 
with the shiver and with punctures shown as crosses. On the right the red dots correspond to the critical points 
of the elliptic fibration, while the blue dot correspond to the critical point of the circular fibration over the 
base $z$.
The blue, green and gold lines correspond to the cycles wrapped by the D6-branes, while the red line corresponds 
to the orientifold plane (which splits into three branches).}
\label{dp0mirror}
\end{center}
\end{figure}

The structure of the $\Sigma$ fibration is drawn in Figure \ref{dp0mirror}. It allows to immediately read off the 
structure of the orientifold field theory. We see that $\IS^3_1$ is mapped to itself while $\IS^3_2$ and $\IS^3_3$ are 
exchanged under the involution, yielding the gauge group
\beqa
G_1\times U(n_2),
\eeqa
with $G_1$ an orthogonal or symplectic factor.
 Concerning the matter content, one can directly read out the intersections mapped to themselves or exchanged by 
 the orientifold action and obtain the spectrum
 \beqa
 3\, [\, (\fund_1,\antifund_2) + (1,R_2)\, ],
 \eeqa
 where $R_2$ is a two-index tensor. This agrees with the structure  obtained in \S\ref{C3Z3sec}.

For later discussion it is convenient to focus on a concrete choice. Consider the three orientifold branches to 
have negative RR charge. This corresponds to the representation $R_2$ corresponding to the two-index antisymmetric 
representation, and should correspond to $G_1$ being an orthogonal $SO(n_1)$ factor. Other choices should reproduce 
the different orientifolds classified in \S\ref{C3Z3sec}.

Clearly many other examples can be worked out similarly. We leave them as exercises for the interested reader.

\subsubsection{Involutions mirror to orientifold dimers with fixed lines}
\label{mirrorline}

Here we would like to consider other kinds of involutions, that correspond to orientifolds of the dimer with 
fixed lines, which exist for particular geometries. For concreteness, we consider two examples. Namely, 
the non-trivial orientifolds of $\C^3$ and of the conifold as discussed in \S\ref{Cfixed} and \S\ref{otherconisec}.
 Generalization to other examples should be clear.

Let us start with the mirror geometry of $\C^3$, where the $\Sigma$ fibration is given by
\beq
z=1+x_1+x_2.
\eq
Via the alga map, we can translate the action on the dimer $\IT^2$ to an involution on $\Sigma$:
\beq
x_1\rightarrow \cc x_2,~ x_2\rightarrow \cc x_1.
\eq
Hence, the base and the circular fibration transform as in \S\ref{mirrorpoint}.

\begin{figure}[!htp]
\begin{center}
\psfrag{a}[cc][][1]{a)}
\psfrag{b}[cc][][1]{b)}
\includegraphics[scale=0.5]{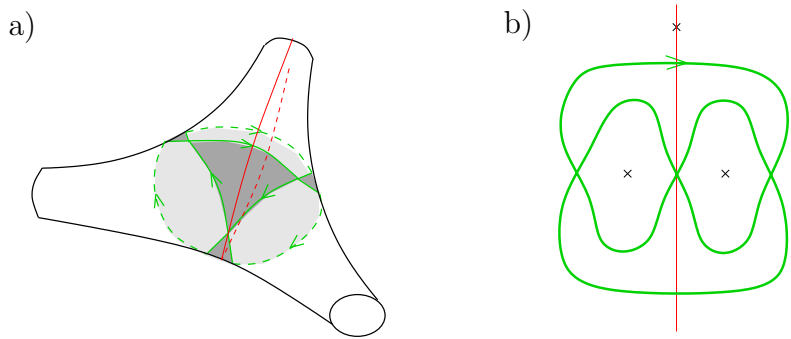}
\caption{a) $\Sigma_0$ of $\IC^3$ with the fixed-line involution (red line). b) Action on the Harlequin diagram.}
\label{c3shiverfixed}
\end{center}
\end{figure}

It is easy to deduce that the corresponding fixed-point set corresponds to a 1-cycle that runs from a puncture 
back to the same puncture, as illustrated in Figure \ref{c3shiverfixed}a.
We infer that we obtain a $SO/Sp$ gauge group with a single adjoint, one symmetric and one anti-symmetric tensor, as expected from \S\ref{Cfixed}.

As another example, let us consider the mirror of the conifold.
The action with fixed lines in the dimer diagram of the conifold corresponds to a reflection of one 
circular coordinate leaving the other invariant. Recalling that the alga map interprets these circles 
as phases of coordinates in the mirror geometry, we infer the following action on the variables $x_1$ and $x_2$:
\beq\label{otherinveq1}
\minv(x_1)=\cc x_1,~\minv(x_2)=\frac{1}{\cc x_2}.
\eq
This can indeed be extended to a symmetry of the $\Sigma$ fibration by requiring that the coordinate $z$ on the base 
transforms as
\beqa
z\to \frac{\ov z}{\cc x_2}.
\eeqa
Notice that this is indeed an involution, given the specific transformation of $x_2$ in (\ref{otherinveq1}). 
Similarly, it can be extended to an involution of the $\IC^*$ fibration (with the circle on its fixed point set)
 by introducing the action
\beqa
u\to {\cc v} \quad ; \quad v\to \frac{\cc u}{\cc x_2}.
\eeqa
The fixed point set is given by two components $x_1\in \R$, $x_2=+1$, $z\in \R$, $u={\cc v}$
and $x_1\in \R$, $x_2=-1$, $iz\in\R$, $u=-{\cc v}$. This suggests that the fixed set in the Riemann surface $\Sigma_0$ 
passes through two punctures. Indeed, the effect of this involution on $\Sigma_0$ can be captured by using the amoeba 
map  (\ref{algaeq1}). In detail, the action on the amoeba image, namely the web diagram, is given by
\beq
\minv(s,t)=(s,-t).
\eq
It corresponds to a $\Z_2$ reflection keeping two punctures ($t=0$, $s\to \pm \infty$) fixed, while mapping the other 
two ($s=0$, $t\to \pm \infty$) to each other.

The O6-plane has two branches, which can in principle have either sign.
\begin{figure}[!htp]
\begin{center}
\psfrag{a}[cc][][0.65]{$X^{(1)}_{12}$}
\psfrag{b}[cc][][0.65]{$X^{(1)}_{21}$}
\psfrag{c}[cc][][0.65]{$X^{(2)}_{12}$}
\psfrag{d}[cc][][0.65]{$X^{(2)}_{21}$}
\includegraphics[scale=0.6]{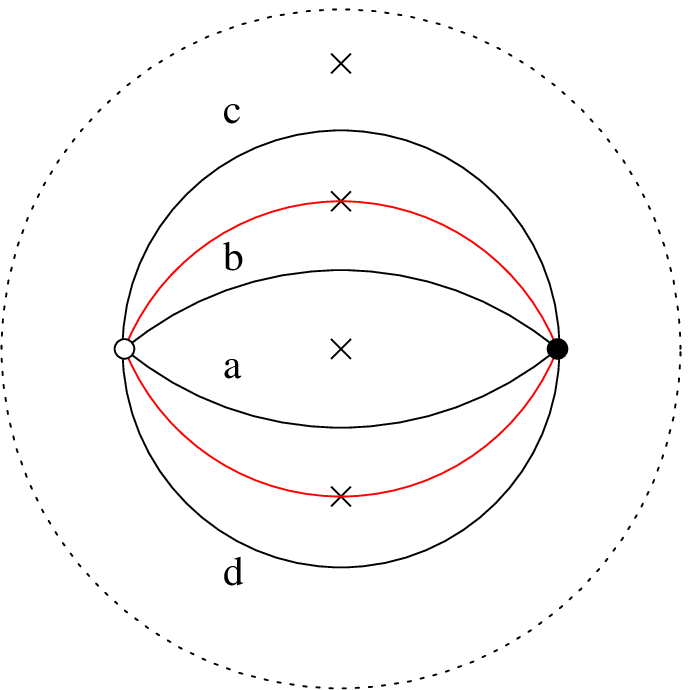}
\hspace{2cm}
\includegraphics[scale=0.8]{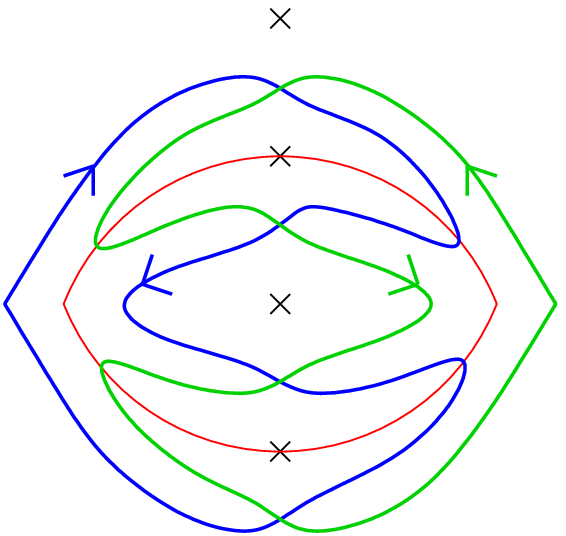}
\caption{Left: The action of the involution on the shiver of the conifold. Right: Action on the
corresponding Harlequin diagram.} \label{coniinv}
\end{center}
\end{figure}

In Figure \ref{coniinv} we see the action on the conifold shiver diagram. One can immediately
obtain the effect of the orientifold on the gauge theory. Each of the 1-cycles associated to the
D6-branes are mapped to themselves under the orientifold action, hence the gauge group is
\beqa
G_1\times G_2,
\eeqa
with $G_i$ and orthogonal or symplectic gauge factor, determined by the sign of the orientifold branch crossing the 
corresponding 1-cycle. Concerning the matter content, no intersection is invariant under the orientifold action, 
rather any intersection is mapped to an image. Hence, the matter content is
\beqa
2~(\fund_1,\fund_2).
\eeqa
The resulting theories are in precise agreement with the construction in terms of dimer diagrams in \S\ref{otherconisec}.

\subsection{Tadpole cancellation}
\label{tadsec}

An advantage of the mirror description of these systems is that it makes it manifest the RR charges carried by the 
D6-branes and O6-planes in terms of the 3-homology classes of the 3-cycles they wrap. In this language, 
cancellation of RR tadpoles corresponds to the condition that the compactly supported homology classes of the 
different objects add up to zero. Since this point has already been discussed in the literature 
(see e.g. \cite{Hanany:2001py,Uranga:2002pg}), we simply sketch the general idea and provide a simple example.

Consider first the situation in the absence of orientifold planes. We have stacks of $N_a$ D6-branes wrapped on 
3-cycles $[\Pi_a]$. The total homology charge of the system should cancel in compact homology. This is equivalent 
to demanding the vanishing of its intersection with the elements $[C_b]$ of a basis of compact homology:
\beqa
\sum_a N_a [\Pi_a]\cdot [C_b] = 0 \quad \forall b.
\eeqa
For branes at singularities, the 3-cycles wrapped by the branes are themselves elements of the basis. 
The conditions $\sum_i N_a[C_a]\cdot[C_b]=0$ are then equivalent to the conditions of cancellation of 
non-abelian gauge anomalies on the D6-brane wrapped on $[C_b]$. We thus recover the equivalence between 
cancellation of compact RR tadpoles and cancellation of gauge anomalies.

As a simple illustrative example, consider the $dP_0$ theory, whose mirror geometry is described in  
\S\ref{mirrorpoint}.  The gauge theory on the D6-brane system is $\prod_i U(N_i)$ 
with matter $3~\sum_i (\fund_i,\antifund_{i+1})$. The condition of cancellation of anomalies requires 
$N_1=N_2=N_3$. This is in agreement with the fact that the total homology class
$N (\, [C_1]  + [C_2] + [C_3]\, )$ has zero intersection number with any of the basis 3-cycles $[C_a]$ 
(and hence is zero in compact homology).

A similar result holds in the presence of orientifold planes, as has been described in the intersecting brane 
literature. Focusing to the case of D6-branes wrapping basis cycles, we consider a set of $N_a$ D6-branes on 
3-cycles $[C_a]$ and their orientifold images on $[C_{a'}]$. The O6$^{\pm}$-planes wrap 3-cycles with total 
compact homology class $\pm 4 [\Pi_{O6^{\pm}}]$, where the factor of 4 comes from the O6-plane charge. 
Cancellation of compact homology charge requires
\beqa
[\Pi_{\rm tot}]\, =\, \sum_a \, N_a \, [C_a] \, + \, \sum_{a'}\,  N_{a'}\, [C_{a'}] \, +\, 4~\left([\Pi_{O6^+}]
-[\Pi_{O6^-}]\right) =0.
\eeqa
Equivalently, one may require $[\Pi_{\rm tot}]\cdot [C_a]=0$ for all basis 3-cycles:
\beqa
\sum_{b} N_b [C_b]\cdot[C_a]+\sum_{b'} N_{b'}[C_{b'}]\cdot[C_a]+4~\left([\Pi_{O6^+}]-[\Pi_{O6^-}]\right)\cdot[C_a]=0.
\eeqa
These conditions are equivalent to the cancellation of anomalies on the D6-brane wrapped on $[C_a]$. 
This is easily shown by noticing that the field content on such brane is as follows:
\begin{itemize}
\item $[C_a]\cdot[C_b]$ chiral multiplets in the $(\fund_a,\antifund_b)$,
\item $[C_a]\cdot [C_{b'}]$ in the $(\fund_a,\fund_b)$,
\item $[C_a]\cdot [C_{O6^{\pm}}]$ fields in the $\symm_a$, respectively $\asymm_a$ representation from $aa'$ 
intersections on top of orientifold planes,
\item $\frac 12 [C_a]\cdot ( [C_{a'}] - [C_{O6^+}] - [C_{O6^-}])$ fields in the $\symm_a+\asymm_a$ from $aa'$ 
intersections away from orientifold planes.
\end{itemize}

Hence we conclude that cancellation of RR tadpoles is equivalent to cancellation of gauge anomalies on D6-branes 
on all possible basis 3-cycles. It is important that in configuration where some multiplicity $N_a$ is zero, 
one still needs to impose $[\Pi_{\rm tot}]\cdot [C_a]=0$. In this sense, one needs to require cancellation of anomalies, 
even for gauge factors absent in the specific model at hand. In fact this condition is not satisfied by some models 
in the literature.

Cancellation of RR tadpoles in coming examples are carried out by checking cancellation of anomalies on the D6-branes. 
To illustrate with an example, let us again consider the orientifold of the $dP_0$ theory in \S\ref{mirrorpoint}. 
The geometry of the D6-branes and O6-planes is shown in Figure \ref{dp0mirror}. As discussed, in the case where all 
orientifold plane branches carry negative charge it leads to a gauge theory with gauge group $SO(n_1)\times U(n_2)$ 
and chiral multiplets $3(\fund_1,\antifund_2)+3(1,\asymm_2)$. Here $n_1$ correspond to the number of D6-branes in the 
3-cycle $[C_1]$, and $n_2$ is the number of D6-branes on $[C_2]$ and $[C_3]$ (which must be equal due to the orientifold 
symmetry). Cancellation of tadpoles corresponds to cancellation of $SU(n_2)$ anomalies (since $SO(n_1)$ is automatically 
anomaly free), and leads to the condition $n_1+4=n_2$.

Other examples can be discussed similarly. Notice that in general it may not be possible to satisfy the RR tadpoles with 
compact D6-branes. In such situations, it may be possible to satisfy them by introducing non-compact D6-branes. 
These are mirror to D7-branes in the Type IIB picture.

\subsection{Calibration}
\label{calisec}

In the introduction of O6-planes in the system, we have implicitly used a constraint that we would like to make explicit 
at this point. In order that the various D6-branes and O6-planes preserve a common supersymmetry, they need to wrap 
special lagrangian 3-cycles, i.e. the phase of $\Omega$ must be constant on the volume of the wrapped 3-cycle. This 
calibration condition for the geometries at hand can be made explicit as follows (cf. \cite{Uranga:2002pg}). Let us 
introduce a local coordinate $w$ on $\Sigma_0$ near an intersection locus of D6-branes and an O6-plane. Then the 
holomorphic 3-form is locally given by
\begin{equation}
\Omega=dz\wedge dw\wedge\frac{du}{u}.
\end{equation}
Parameterizing the 3-spheres $[C_i]$ as
\beq\label{calieq0}
z=|z^*_i|e^{i\theta_i},~u=\sqrt{|z^*_i|}e^{i\nu},~v=\sqrt{|z^*_i|}e^{-i\nu}e^{i\theta_i},~w=|w|e^{i\phi},
\eq
we obtain that $\Omega$ restricted to the D6-brane volume is given by
\beq
\Omega|_{D6}=ie^{i(\theta+\phi)}d|z|\wedge d|w|\wedge d\nu.
\eq
We thus conclude that the 3-cycles are special lagrangian for
\beq\label{calieq1}
\theta=-\phi.
\eq
In other words, the angle in the base must be aligned with the angle in $\Sigma_0$.

The non-compact O6-planes (as well as possible) non-compact D6-branes can be parameterized similarly. Namely, 
considering they wrap the $\IS^1_c$ circle in the $\IC^*$ fibration, their angles in the elliptic fibration 
should obey (\ref{calieq1}). This kind of constraint has been implicitly used in \S\ref{mirrorline} to require 
that the O6-plane spans the $\IS^1_c$ in the $\IC^*$ fiber. Namely, the angle between the O6-plane and the 
D6-branes in $\Sigma_0$ can be argued to be $\pi/2$, while the angle on the $z$-plane is $-\pi/2$; hence it is 
calibrated only if it wraps $\IS^1_c$.

\section{Applications}
\label{appl}

In this section we discuss several possible physical applications of models of the kind studied in this paper. 
They correspond to diverse systems where the presence of orientifold planes is crucial in the physics.

\subsection{Dynamical supersymmetry breaking}

One of the most interesting applications of systems of D-branes at singularities is to study the dynamics of gauge 
theories on fractional branes. This can lead to the strong dynamics phenomena of confinement 
\cite{Klebanov:2000hb,Franco:2005fd} or to supersymmetry breaking \cite{Berenstein:2005xa,Franco:2005zu,Bertolini:2005di} 
with runaway to infinity \cite{Franco:2005zu,Intriligator:2005aw,Forcella:2006}. Upon the addition of flavor D7-branes, 
this can also lead to metastable vacua \cite{Franco:2006es,Garcia-Etxebarria:2007vh}.\footnote{For other realizations 
of metastable vacua using D-branes at singularities see \cite{Argurio:2006ny,Argurio:2007qk}.}

One also expects rich strong dynamical phenomena on systems of D-branes at orientifold singularities. 
A full classification and study is beyond the scope of the present paper, but we would like to mention the 
existence of configurations of D-branes leading to dynamical supersymmetry breaking in the vacuum.

\begin{figure}[!htp]
\begin{center}
\psfrag{1}[cc][][0.7]{1}
\psfrag{2}[cc][][0.7]{2}
\psfrag{3}[cc][][0.7]{3}
\psfrag{4}[cc][][0.7]{4}
\psfrag{5}[cc][][0.7]{5}
\psfrag{0}[cc][][0.7]{0}
\psfrag{a}[cc][][0.5]{a}
\psfrag{b}[cc][][0.5]{d}
\psfrag{c}[cc][][0.5]{c}
\psfrag{d}[cc][][0.5]{b}

\includegraphics[scale=0.6]{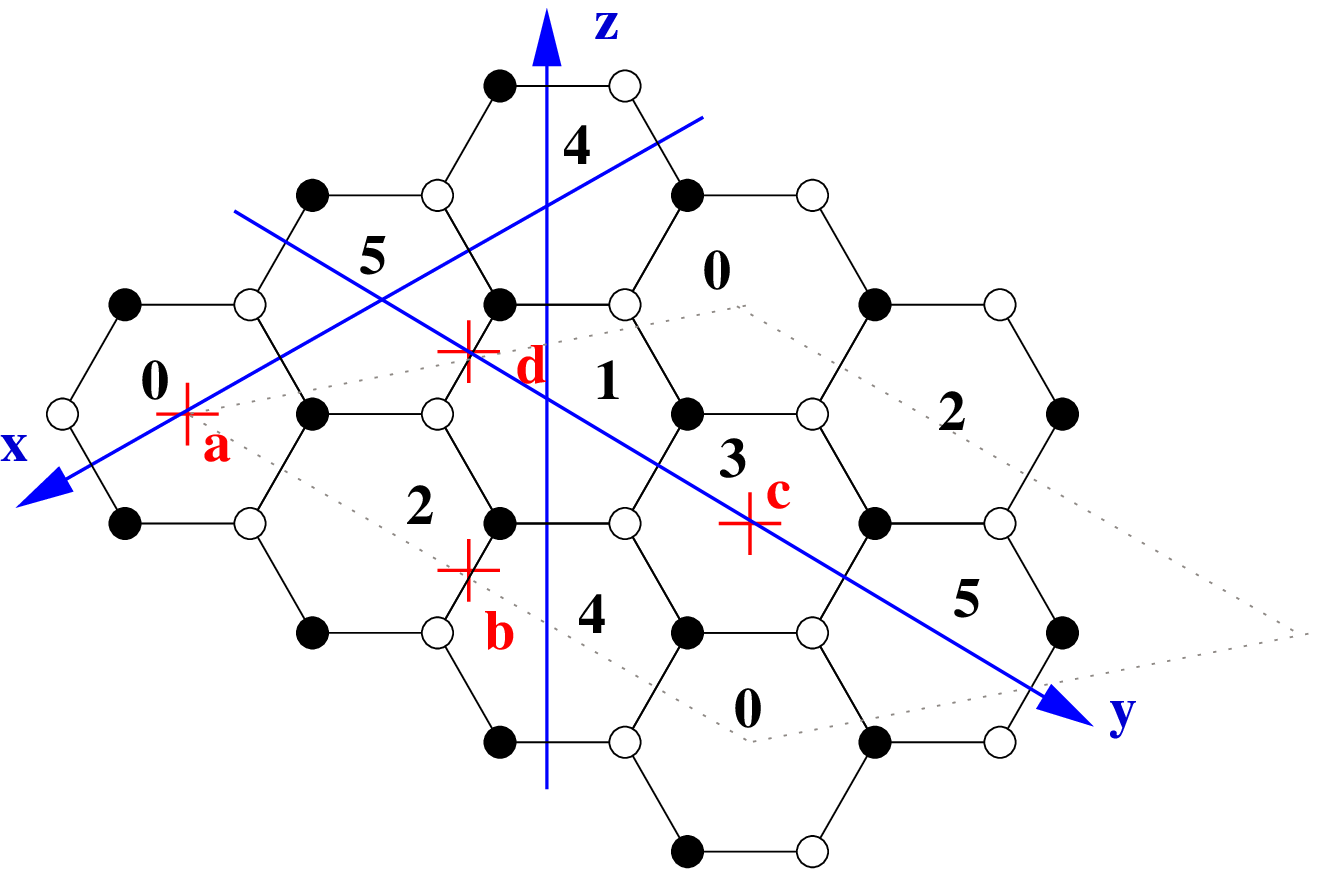}
\caption{Dimer diagram for an orientifold of the $\IC^3/\IZ_6'$ theory.}
\label{dsbZ6}
\end{center}
\end{figure}

The model we like to present is based on the $\IC^3/\IZ_6'$ geometry, where the $\IZ_6'$ is generated by
\beq
z_i \to e^{2i\pi v_i} z_i,
\eq
with $v=(1,2,-3)/6$. The orientifolded dimer diagram is shown in Figure \ref{dsbZ6}, where we also show some of 
the basic mesons, which are related to the $\IC^3$ coordinates by
\beq
x=z_1^{\, 6}, ~y=z_2^{\,3}, ~z=z_3^{\, 2}.
\eq

As extensively discussed in \S\ref{points}, one can obtain different models by choosing different signs for the 
orientifold points. Each choice of sign setup corresponds to a different orientifold action. Consider the choice of 
orientifold signs $(a,b,c,d)=(++--)$. Using our rules, the gauge theory for general ranks can be obtained as
\beqa
&SO(n_0)\times U(n_1)\times U(n_2)\times Sp(n_3) & \nonumber \\
& (\fund_0,\antifund_1) + (\fund_1,\antifund_2) + (\fund_2,\antifund_3)+& \nonumber \\
& + (\fund_0,\antifund_2) + (\fund_1,\fund_3) + \asymm_2+\antisymm_1 +& \nonumber \\
&+  [\, (\fund_0,\fund_3) + (\fund_1,\fund_2) + (\antifund_1,\antifund_2) \,]&.
\eeqa
As described in \S\ref{tadsec}, the requirement of cancellation of non-abelian gauge anomalies is equivalent to the 
requirement of cancellation of compact RR tadpoles. The condition reads
\beqa
-n_0 + n_2+n_3-n_1-4=0.
\eeqa
This condition can be solved by e.g. $n_1=n_3=0$, $n_0=N$, $n_2=N+4$. The gauge theory becomes $SO(N)\times U(N+4)$ with 
matter $(\fund,\antifund)+(1,\asymm)$. The $U(1)$ gauge factor is anomalous and becomes massive by coupling to a suitable 
RR field, hence disappears from the massless spectrum.
For the particular case of $\N=1$, one indeed obtains a fully consistent configuration, leading to a $SU(5)$ theory 
with chiral multiplets in the $10+{\ov 5}$ and no superpotential. This theory has been argued to show dynamical 
supersymmetry breaking \cite{Affleck:1983mk}.

The above model corresponds to the geometric action
\beq
x\to -x,~ y\to -y, ~z\to -z.
\eq
Equivalently, the orientifold group is $(1+\theta+\ldots +\theta^5) (1+\Omega \alpha (-1)^{F_L})$, where $\alpha$ acts as
\beq
(z_1,z_2,z_3)\to (e^{2i\pi/12}, e^{4i\pi /12}, e^{-6i\pi /12}).
\eq
In fact, it is possible (but lengthy) to directly construct this orientifold model using CFT techniques.

\begin{figure}[!htp]
\begin{center}
\psfrag{1}[cc][][0.8]{1}
\psfrag{2}[cc][][0.8]{2}
\psfrag{3}[cc][][0.8]{3}
\psfrag{4}[cc][][0.8]{4}
\psfrag{5}[cc][][0.8]{5}
\psfrag{6}[cc][][0.8]{6}
\psfrag{7}[cc][][0.8]{7}

\includegraphics[scale=0.6]{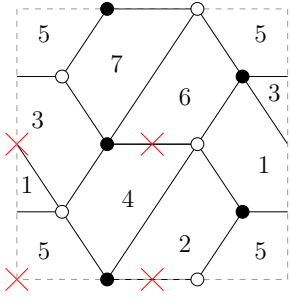}
\caption{Dimer diagram for an orientifold of the $PdP_4$ theory.}
\label{PdP4}
\end{center}
\end{figure}

It is easy to use our tools to construct other examples realizing the same gauge theory, thus leading to dynamical 
supersymmetry breaking vacua. As a further example to illustrate our tools, consider Model I of 
pseudo del Pezzo 4 ($PdP_4$) \cite{Feng:2002fv,Butti:2006hc}, whose
dimer is shown 
in Figure \ref{PdP4}.\footnote{$PdP_4$ is a non-generic, toric blow-up of $dP_3$. It was first introduced in \cite{Feng:2002fv}, 
where the corresponding gauge theory was also investigated.}

The structure of the gauge theory is
\beqa
& G_5\times U(n_1)\times U(n_2)\times U(n_4) & \nonumber \\
& (\fund_5,\antifund_1) + (\fund_5, \fund_4) + (\fund_5,\fund_2) + T_1 + &\nonumber \\
& + (\antifund_1,\antifund_4) + (\fund_1,\antifund_2) + (\fund_2,
\antifund_4) + {\ov R}_2 + S_4 &.
\eeqa
The signs of the orientifold points
determine the orientifold projections on the gauge factor $G_5$ and the two-index tensor representations 
$T_1$, $R_2$ and $S_4$, respectively. The dimer diagram contains 8 superpotential couplings, hence the sign parity
is even.

For the choice of signs $(+-+-)$, we have $G_5=SO(n_5)$, $T_1=\asymm_1$, ${\ov R}_2=\antiasymm_2$ and 
$S_4=\symm_4$. The tadpole/anomaly cancellation conditions for the theory are simply
\beq
n_5-n_1-n_2+n_4+4=0,
\eq
which we can satisfy by taking $n_2=n_4=0$, $n_5=N$, $n_1=N+4$. The resulting theory is $SO(N)\times U(N+4)$ 
with matter $(\fund,\antifund)+(1,\asymm)$. Thus for $\N=1$ it reproduces the dynamical supersymmetry breaking 
theory of interest.

\medskip

Note that there have been several proposals of orientifold models in the literature which claim to realize 
the above $SU(5)$ gauge theory \cite{Wijnholt:2007vn,Antebi:2007xw}. Although insightful in suggesting orientifold 
models to reproduce dynamical supersymmetry breaking, the proposals show technical problems in the specific models 
presented, as one can show using our new tools.

The model in \cite{Wijnholt:2007vn} is based on an orientifold of $\IC^3/(\IZ_3\times \IZ_3)$ corresponding 
to a dimer with fixed lines. The orientifold projection on gauge factors and two-index tensors in the reference 
is not compatible with the existence of a single orientifold fixed line for this model.\footnote{Note that 
other models in this reference, for instance the 3-2 model realized on $dP_5$ are non-toric and we cannot use 
our tools to check their consistency.}

The orientifold model proposed in \cite{Antebi:2007xw} is based on $\IC^3/\IZ_6'$ with an orientifold action 
$(z_1,z_2,z_3)\to (z_1,z_2,-z_3)$. This corresponds to the action
\beq
x\to x, ~y\to y, ~z\to z,
\eq
and can be described with the dimer diagram with fixed points given above. The corresponding sign setup leading to 
this geometric action is $(++++)$ or $(----)$. Thus, the resulting spectra do not agree with the theory of interest 
since there is no model with orthogonal gauge factor and matter in two-index antisymmetric tensor representations.

Since these models are based on orbifolds, we have also double-checked our results by direct computations using 
CFT techniques.

It is remarkable that our new tools facilitate the construction of orientifolds in a simple and systematic way. 
We expect them to be extremely useful in the study of infrared dynamics of systems of D-branes at orientifold 
singularities.

It is almost suggestive that one can find whole families of orientifold models yielding dynamical supersymmetry 
breaking. We leave a detailed investigation and classification of these models for future work.

%==================================================================
\subsection{D-brane instantons}
%==================================================================

In Type II compactification, many important non-perturbative effects arise from Euclidean D-brane instantons. 
In configurations with 4d spacetime filling D-branes leading to a gauge sector, such D-brane instanton can induce 
superpotentials corresponding to non-trivial field theory operators.
This can correspond to a field theory instanton effect (when the structure of the Euclidean D-brane instanton 
in the Calabi-Yau geometry is the same as that of one of the 4d spacetime filling branes),  
see \cite{Haack:2006cy,Florea:2006si,Akerblom:2006hx,Bianchi:2007wy} for recent discussions, 
or to a D-brane instanton without such field theory interpretation (stringy D-brane instantons), 
see \cite{Blumenhagen:2006xt,Ibanez:2006da,Florea:2006si,Cvetic:2007ku,Argurio:2007vq,Ibanez:2007rs} 
for recent discussions. Focusing in the latter case,  the D-brane instanton typically has fermion zero modes  
$\alpha_{a}$, $\beta_b$ transforming in the $\antifund_a$, $\fund_b$ under the 4d gauge factors $a$, $b$. 
If there is a coupling $\alpha_a {\cal O}_{ab} \beta_b$ where ${\cal O}_{ab}$ is a 4d chiral operator in the 
$(\fund_a,\antifund_b)$, integration over the fermion zero modes leads to insertions of ${\cal O}$ in the superpotential 
\cite{Ganor:1996pe}.

As discussed in general in \cite{Ibanez:2007rs} and in particular cases in 
\cite{Bianchi:2007fx,Bianchi:2007wy,Argurio:2007vq}, stringy D-brane instantons in perturbative Type II compactifications
 without fluxes generically have four uncharged fermion zero modes, and cannot generate superpotential couplings. 
 In the presence of orientifold planes, the orientifold projection on instantons with $O(1)$ world-volume gauge group 
 removes two fermion zero modes, thus allowing the possibility of non-perturbative superpotentials.

This observations make orientifolds of D-branes at singularities a natural setup to study this non-perturbative effects. 
In fact, some concrete examples have already appeared in \cite{Argurio:2007qk,Argurio:2007vq}.
It is clear that the techniques developed in this paper provide many additional examples of D-brane instantons 
in this kind of systems. Namely, one can consider D-brane instantons with the internal structure corresponding to 
one of the faces of the dimer diagram (and on top of an $O^-$-plane to
have $O(1)$ symmetry).\footnote{Note the sign flip in the 
orientifold action on the gauge group of D-brane instantons as compared with 
4d spacefilling D-branes.} This instanton is stringy if there are no 4d spacefilling branes for the corresponding face. 
Edges between the instanton face and faces of 4d spacefilling branes correspond to instanton fermion zero modes, and 
nodes correspond to their interactions. Thus one easily obtains the structure of the induced superpotentials.

\begin{figure}[!htp]
\begin{center}
\psfrag{a}[cc][][1]{a)}
\psfrag{b}[cc][][1]{b)}
\psfrag{1}[cc][][0.7]{1}
\psfrag{2}[cc][][0.7]{2}
\psfrag{3}[cc][][0.7]{3}
\psfrag{4}[cc][][0.7]{4}
\psfrag{5}[cc][][0.7]{5}
\psfrag{6}[cc][][0.7]{6}
\psfrag{7}[cc][][0.7]{7}
\psfrag{8}[cc][][0.7]{8}
\psfrag{9}[cc][][0.7]{9}
\includegraphics[scale=0.6]{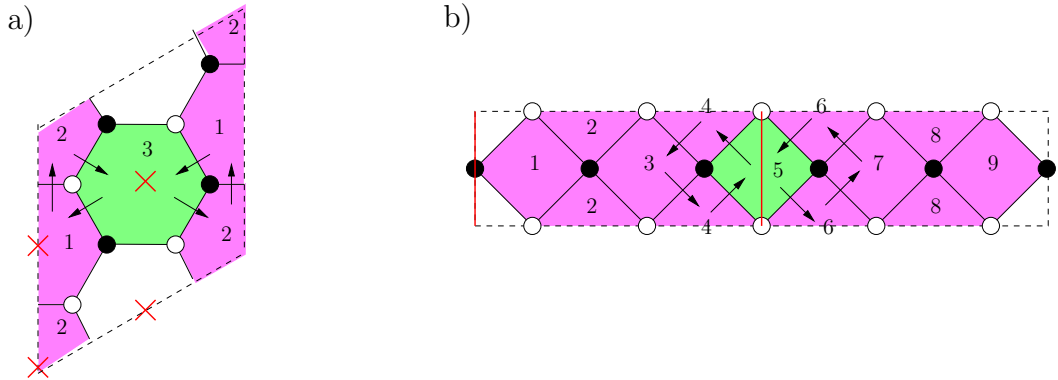}
\caption{D-brane instantons in systems of D-branes at orientifolded singular geometries. Figure a) shows an 
instanton (filling the green face) in an orientifold of $\IC^3/(\IZ_2\times\IZ_2)$. Figure b) shows an 
instanton in an orientifold of a generalized conifold.}
\label{instanton}
\end{center}
\end{figure}

In Figure \ref{instanton} we show several examples of stringy D-brane instantons (shown as shaded faces) generating 
non-perturbative superpotentials. Figure \ref{instanton}a shows a D-brane instanton in the orientifold of 
$\IC^3/(\IZ_2\times \IZ_2)$ studied in \cite{Argurio:2007vq}.
The magenta faces correspond to 4d spacefilling D-branes, while the green face corresponds to a D-brane instanton. 
Notice the fermion zero modes of the latter $\alpha_{31}$, $\beta_{23}$ (and their orientifold images) arising from 
edges between magenta and green faces, and their cubic coupling $\alpha \Phi_{12} \beta$ to the 4d chiral multiplet 
$\Phi_{12}$.
Figure \ref{instanton}b shows a D-brane instanton in the orientifold of the $L^{aba}$ geometry considered 
in \cite{Argurio:2007qk}. Notice the fermion zero modes $\alpha_{54}$, $\beta_{45}$, and the coupling 
$\alpha X_{43} X_{34} \beta$.

It is straightforward to use our techniques to obtain many other examples of stringy D-brane instantons 
generating superpotentials in systems of D-branes at singularities with orientifold actions. We leave a more 
detailed discussion for future work.

%=====================================================================
\section{Conclusions}
\label{conclu}
%=====================================================================

We have introduced dimer model techniques that allow an easy determination
of the gauge theory on D-branes over orientifolds of general toric 
singularities. As in the unorientifolded case, dimers provide the 
most comprehensive available classification of such theories.

% zzz
The rules have been presented as a heuristic extrapolation of some
simple families of examples. However, it is 
important to emphasize that they are automatically consistent with
partial resolution (and its inverse process 
of un-Higgssing) and therefore extend by induction to arbitrary toric singularities. This constitutes a proof of the rules we have presented.

Under the new light, the pre-existing examples based on worldsheet 
computations for orbifolds and theories with T-dual Hanany-Witten setups
become trivial.

We explained how to extract the geometric action of the orientifold
using mesonic operators. We also explored in detail the orientifolds from
the point of view of the mirror geometry.

To give a flavor of some possible applications of our ideas, we chose 
dynamical supersymmetry breaking and D-brane instantons. 
We are confident that our tools will permit a systematic investigation
of large families of models and fuel 
considerable progress in model building.

We also expect other exciting directions to profit from our work. To name one, 
it is natural to expect that they will help extending recent 
progress in the combinatorics of quiver gauge theories (counting of chiral 
BPS gauge invariant operators) to orientifold theories.

%=====================================================================
\acknowledgments
%=====================================================================

We thank R. Argurio, M. Bertolini, M. Buican, I. Garcia-Etxebaria, S. Kachru, J. McGreevy and M. Wijnholt for useful discussions. 
S.F. is supported by the DOE under contract DE-FG02-91ER-40671. 
D.K. likes to thank D. L\"ust for support and CERN-TH for hospitality during part of the project. A.U. thanks Ecole Polytechnique and SISSA for hospitality during part of this project. 
The work of J. P. is supported by the Science Research Center Program of KOSEF through the Center for Quantum
Spacetime (CQUeST) of Sogang University with the grant number R11-2005-021. J. P. appreciates the hospitality
of CERN-TH and of theoretical particle physics group of University of Pennsylvania during his stay. 
The work by A.U. has been supported by CICYT (project FPA-2003-02877), Comunidad de Madrid (project HEPACOS), the EU networks MRTN-CT-2004-503369, MRTN-CT-2004-005104, and partially by the European Union Excellence Grant MEXT-CT-2003-509661.
D.V. is supported by the
U.S. Department of Energy under cooperative research agreement \#DE-FC02-94ER40818.

%=====================================================================

\appendix

\section{From quivers to dimers and shivers}
\label{appQDS}
The authors of \cite{Feng:2005gw} constructed the dimer and shiver of a given toric quiver gauge theory via a 2- or 
respectively 3-step procedure. Firstly, they glued the superpotential terms appropriately together to obtain the 
periodic quiver, then they graph dualized the periodic quiver to obtain the dimer and finally performed an untwisting 
procedure to arrive at the shiver. In the following, we give a more practical algorithm to obtain the dimer as well as 
the shiver, which in principal follows the same steps but without the need to draw the intermediate graphs. Especially, 
the given algorithms are more suitable for a computer implementation. The main ingredient we use are D(imer)- and 
respectively S(hiver)-chains. These are chains of bi-fundamental fields which correspond to fundamental cells of 
the dimer or respectively of the shiver.\footnote{In fact, the fields need not necessarily be bi-fundamentals. 
For example, adjoint fields may occur in the superpotential as well. However, one can interpret such fields as 
bi-fundamentals of the same group and hence are sometimes referred to as bi-fundamentals.}

\subsection*{S-chains}
\label{Schains}
A S-chain is constructed in the following way:
Take an arbitrary bi-fundamental field contained in a monomial of the superpotential as the first field in the S-chain. 
The next field is the field right of it in the same monomial. If the field is the last one in the monomial, 
take the first field of the monomial as the next field. Then jump to the unique other monomial containing 
this field.\footnote{Since the quiver is assumed to be toric, each bi-fundamental can only occur twice and the 
jump to the next monomial is uniquely determined.} Take the right field of it as the next one in the S-chain and 
again jump to the unique monomial containing also this field. Iterate until the chain closes.\footnote{Since a 
S-chain encloses a puncture in the mirror Riemannian surface, the number of S-chains equals the number of punctures 
and further it can not intersect itself up to multiple windings.}

Actually, a S-chain is nothing else than a zigzag path of the corresponding dimer.

In order to see this, let us translate the S-chain defining algorithm into the language of the periodic quiver:
Each monomial of the superpotential corresponds to a set of nodes of the quiver connected by a closed path of 
arrows of one orientation (clockwise or anti-clockwise). Hence, each monomial term can be seen as a closed cell 
with an orientation given by the orientation of the arrows. Cells which are formed by a closed loop of clockwise 
arrows we assign a positive sign and call them positive cells, while cells which are formed by anti-clockwise 
arrows we assign a negative sign and denote them as negative cells. The periodic quiver is constructed by gluing 
these cells together along the edges, i.e. bi-fundamental fields, they have in common. Since each bi-fundamental 
field occurs in a toric quiver exactly twice, once in a monomial with plus, and once in one with minus sign, 
the gluing gives a totally periodic graph, there positive and negative cells alternate (the graph dual is the dimer).
The fact that our algorithm only visits two fields of a monomial, one while entering the monomial, and one to exit 
to the next monomial, corresponds in the periodic quiver to entering a cell and leaving it in the next 
step.\footnote{Actually, the meaning of entering and leaving a cell is a bit confusing since an edge always 
belongs to two different cells. In fact, we mean by leaving a cell that the next edge followed is in a different 
cell, while by entering we understand that the edge followed before belonged to another cell.} Leaving a positive 
cell corresponds to make a step right while leaving a negative cell corresponds to make a step left. Since always 
cells of different sign are glued together, our algorithm actually follows a zigzag path in the periodic quiver 
which under graph dualization goes over to a zigzag path in the dimer.

Recalling the untwisting procedure of \cite{Feng:2005gw}, we conclude that in order to obtain the shiver 
we simply need to glue the S-chains at there common fields appropriately together.

Further, since an S-chain encircles a puncture in the shiver, it can also be identified as the difference 
of two perfect matchings of the corresponding dimer (cf. \cite{Hanany:2005ss,Garcia-Etxebarria:2006aq}).

As an illustrative example of this algorithm, consider the conifold. The superpotential is given by
\begin{equation}
W=X^{(1)}_{12}X^{(1)}_{21}X^{(2)}_{12}X^{(2)}_{21}-X^{(1)}_{12}X^{(2)}_{21}X^{(2)}_{12}X^{(1)}_{21}.
\end{equation}
\begin{figure}[!htp]
\begin{center}
\psfrag{1}[cc]{$X^{(1)}_{12}$}
\psfrag{2}[cc]{$X^{(1)}_{21}$}
\psfrag{3}[cc]{$X^{(2)}_{12}$}
\psfrag{4}[cc]{$X^{(2)}_{21}$}
\psfrag{x}[cc]{$-$}
\psfrag{5}[cc]{$X^{(1)}_{12}$}
\psfrag{6}[cc]{$X^{(2)}_{21}$}
\psfrag{7}[cc]{$X^{(2)}_{12}$}
\psfrag{8}[cc]{$X^{(1)}_{21}$}
\psfrag{a}[cc]{1.}
\psfrag{b}[cc]{2.}
\psfrag{c}[cc]{3.}
\psfrag{d}[cc]{4.}
\includegraphics{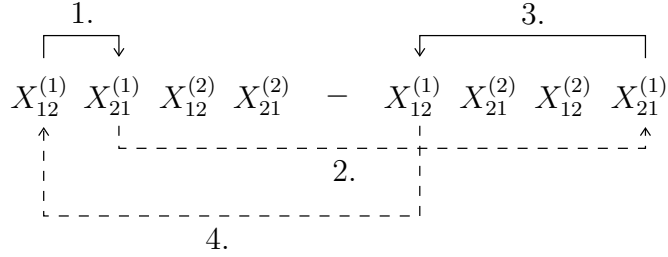}
\caption{Illustration of the algorithm giving the S-chains for the conifold.}
\label{ZIZZAGalgofig1}
\end{center}
\end{figure}

Using the S-chain algorithm, as depicted in Figure \ref{ZIZZAGalgofig1}, 
we obtain the following S-chains:
\begin{equation}
\begin{split}
A:~X^{(1)}_{12}-X^{(1)}_{21},\\
B:~X^{(1)}_{21}-X^{(2)}_{12},\\
C:~X^{(2)}_{12}-X^{(2)}_{21},\\
D:~X^{(2)}_{21}-X^{(1)}_{12}.
\end{split}
\end{equation}
Gluing them together yields the shiver illustrated in Figure \ref{coniriem}.

As another example, consider the Suspended Pinch Point (SPP) with superpotential given by \cite{Uranga:1998vf}
\begin{equation}
W=-X_{21}X_{12}X_{23}X_{32}+X_{23}\Phi_3 X_{32}-X_{13}\Phi_3 X_{31}+X_{13}X_{31}X_{12}X_{21},
\end{equation}
where $\Phi_3$ is an adjoint field.

We derive the following S-chains:
\begin{equation}
\begin{split}
A&:~X_{21}-X_{12},\\
B&:~X_{23}-X_{32},\\
C&:~X_{13}-X_{31},\\
D&:~X_{12}-X_{23}-\Phi_3-X_{31},\\
E&:~X_{32}-X_{21}-X_{13}-\Phi_3.\\
\end{split}
\end{equation}
Gluing yields the shiver given in Figure \ref{trivinvfig0}.

\subsection*{D-chains}
\label{Dchains}
A D-chain is defined as follows:
Take an arbitrary bi-fundamental field in a monomial of the superpotential as the first field 
in the D-chain. The next field is the field right of it in the same monomial. 
If the field is the last one in the monomial, take the first field of the monomial as 
the next field. Then jump to the unique other monomial containing this field. Take the field 
left of it as the next field. If the field is the first one in the monomial, take the last field 
of the monomial as the next field. Jump to the other monomial containing this field and iterate 
until the chain closes.\footnote{Care must be taken, since the D-chain may also possess 
self-intersections.}

A D-chain encloses a face of the dimer.

Again, this can be most easily seen in the language of the periodic quiver. In contrast to the S-chain algorithm, 
we change the orientation we follow once we enter a new cell, since the algorithm alternates between taking the 
left or right one as the next field. Hence, we simply collect the edges conneted to a single vertex in a clockwise or anti-clockwise fashion. Thus, under graph dualization the D-chain goes over to a fundamental cell of the dimer. 
Obviously, the number of D-chains equal the number of gauge groups.

The untwisting procedure then shows that the D-chain corresponds to a zigzag path in the shiver.

Note that the orientation of the zigzag paths is neither fixed in the dimer nor in the shiver. Both algorithms work 
fine as well if one exchanges left and right. This is as expected, since the orientation of zigzag paths in a bipartite 
graph is pure convention.

There is one potentially source of failure in both algorithms one must be aware of. Namely, it is crucial that the 
fields in the monomials of the superpotential are in correct order. This is clear since the fields are in general 
matrices which can not be permuted freely. Note that this is something not always being taken care of in the 
superpotentials which can be found in the literature.

\begin{figure}[!htp]
\begin{center}
\psfrag{1}[cc]{$X^{(1)}_{12}$}
\psfrag{2}[cc]{$X^{(1)}_{21}$}
\psfrag{3}[cc]{$X^{(2)}_{12}$}
\psfrag{4}[cc]{$X^{(2)}_{21}$}
\psfrag{x}[cc]{$-$}
\psfrag{5}[cc]{$X^{(1)}_{12}$}
\psfrag{6}[cc]{$X^{(2)}_{21}$}
\psfrag{7}[cc]{$X^{(2)}_{12}$}
\psfrag{8}[cc]{$X^{(1)}_{21}$}
\psfrag{a}[cc]{1.}
\psfrag{b}[cc]{2.}
\psfrag{c}[cc]{3.}
\psfrag{d}[cc]{4.}
\psfrag{e}[cc]{5.}
\psfrag{f}[cc]{6.}
\psfrag{g}[cc]{7.}
\psfrag{h}[cc]{8.}
\includegraphics{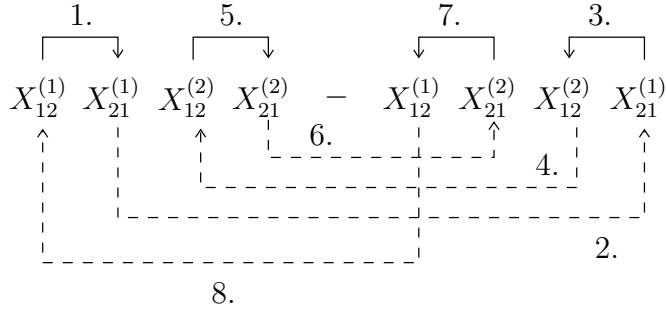}
\caption{Illustration of the algorithm defining the D-chains applied to the conifold.}
\label{ZIZZAGalgofig3}
\end{center}
\end{figure}

As an example for this algorithm, the D-chain algorithm applied to the superpotential of the conifold is 
depicted in Figure \ref{ZIZZAGalgofig3}.
We obtain the following D-chains:
\begin{equation}
\begin{split}
1:~X^{(1)}_{21}-X^{(2)}_{12}-X^{(2)}_{21}-X^{(1)}_{12},\\
2:~X^{(1)}_{12}-X^{(1)}_{21}-X^{(2)}_{12}-X^{(2)}_{21}.
\end{split}
\end{equation}
Gluing them together yields the well known dimer illustrated in Figure \ref{odimersfig1}.

As another example, let us apply the D-chain algorithm to the SPP:

We obtain the following D-chains:
\begin{equation}
\begin{split}
1:&~X_{21}-X_{12}-X_{31}-X_{13},\\
2:&~X_{12}-X_{23}-X_{32}-X_{21},\\
3:&~X_{23}-X_{32}-\Phi_3-X_{31}-X_{13}-\Phi_3.
\end{split}
\end{equation}
Note that one of the D-chains possesses a self-intersection, implying that the respective face in the dimer 
is glued to itself. The corresponding dimer is given in Figure \ref{SPPdimer}.

%----------------------------------------------------------------------------------------------------
%-----------------------------------------------------------------

\section{Mnemonics: Orientifolded Harlequin diagrams}

\label{section_harlequin}

The purpose of this Appendix is to introduce yet another practical point of view for orientifold quotients, namely orientifolded versions of the Harlequin diagrams of the dimer. We
focus on orientifolds with fixed points. The resulting Harlequin diagrams live on a
2-sphere with punctures.

Let us consider the $\IZ_2$ symmetry which acts on the dimer torus with four fixed points.
Modding-out by this action gives a sphere with four singular points. The zigzag paths of the
original dimer map to lines on this sphere. There are two possible types of zigzags in the dimer
graph:

\bigskip

\begin{itemize} 

\item Closed zigzags which avoid the singular points. They map to closed lines
on the sphere with nontrivial homotopy between the four punctures.

\item Open zigzags that pass through singular points. They occur when the
orientifold is lying on top of a bi-fundamental edge in the dimer diagram. In the vicinity of the orientifold
point a local complex $z$ coordinate can be chosen for which the $\IZ_2$ action is $z\mapsto
-z$. Modding-out this action eliminates half of the zigzag that passes through the origin. As a result 
such zigzags become open lines stretched between two singular points on the sphere.

\end{itemize}

\bigskip

%\subsection{$\IC^3$}

\bigskip

\begin{figure}[!htp]
\begin{center}
\psfrag{a}[cc]{(i)} \psfrag{b}[cc]{(ii)} \psfrag{c}[cc]{(iii)}
\includegraphics[scale=0.7]{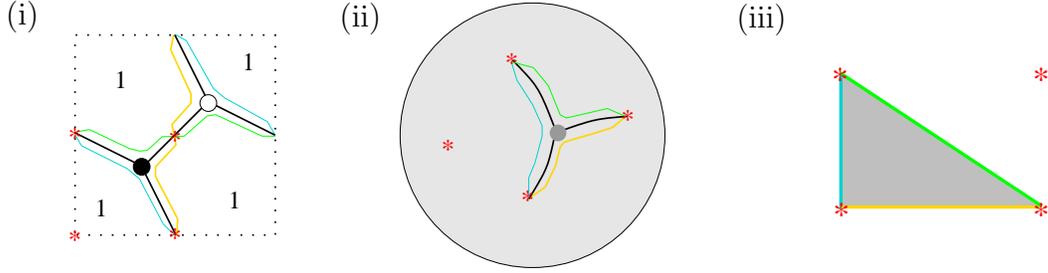}
\caption{(i) $\IC^3$ tiling with zigzags.  (ii) Orientifold tiling of the sphere.  (iii) $\IC^3$
orientifold Harlequin diagram. The three open zigzags create a single loop.} \label{har1}
\end{center}
\end{figure}

Let us apply these ideas to some concrete models. First, consider the simple example
of $\IC^3$. \fref{har1} shows the three
steps from the tiling to the sphere Harlequin diagram. 
$\IC^3$ has three zigzags, represented in blue,
green and yellow lines. All of them cross orientifold points, therefore they become open
zigzags. Modding-out the involution gives (ii), with the three zigzags ending on three singular
loci. The tiling here is drawn with a grey vertex to indicate that black and white nodes were
mapped to one another. One can consider this graph as a ``tiling'' of the sphere by one face which
contains the fourth orientifold point. The Harlequin diagram can be drawn on a plane, which
gives (iii). The grey triangle corresponds to the superpotential term in the orientifolded theory. 
The three lines are the open zigzags. The (infinite) white face corresponds to the gauge group. In this case, 
the group is $SO/Sp$ depending on the charge of the fixed point in it.

In general, the orientifolded Harlequin diagram will contain open and closed zigzags. Since zigzags come in pairs
for every fixed point, we are left with a set of closed loops after combining them. Whenever
an open zigzag crosses a singular point it changes color (this is for instance the case in \fref{har1},
where the loop touches three orientifold points). After drawing the zigzag lines, we can shade every
other face. The shaded faces correspond to superpotential terms. The remaining white faces are the gauge
groups. Whenever there is a fixed point in the middle of a white face, the corresponding
gauge group is $SO/Sp$, depending on the charge. Bifundamentals arise from intersections of
lines. (Anti)symmetric fields are present whenever two zigzags touch at a singular point.
Henceforth we follow the convention that the infinite face is white. One can always draw the
diagram in this way. The sign parity of the configuration is simple equal to the number of shaded faces modulo two 
(this is in agreement with the sign rule of \S\ref{C3}, which states that the sign parity is equal to the parity
of half the number of superpotential terms in the parent theory). Anomaly cancellation can be ensured by adding
extra chiral multiplets if necessary. 

We conclude by listing the orientifolded Harlequin diagrams for the examples in \S\ref{c2z2sec}.

\begin{figure}[!htp]
\begin{center}
\includegraphics[scale=0.8]{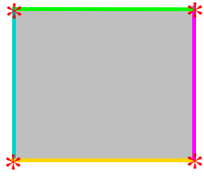}
\caption{The conifold. The (infinite) white face contains no orientifold, and thus it corresponds to a
$U(N)$ gauge group. The intersections of the four open zigzags give four (anti)symmetric fields
as listed in Table 2.} \label{harl2}
\end{center}
\end{figure}

\begin{figure}[!htp]
\begin{center}
\psfrag{a}[cc]{(i)} \psfrag{b}[cc]{(ii)}
\includegraphics[scale=0.7]{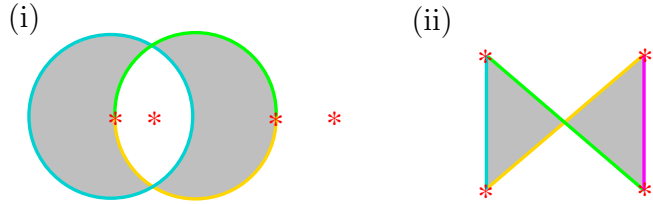}
\caption{(i)-(ii) Two different orientifolds of $\C^2/\Z_2 \times \C$. They correspond to
Figure 5 and Figure 6, respectively.} \label{harl3}
\end{center}
\end{figure}

\begin{figure}[!htp]
\begin{center}
\includegraphics[scale=0.9]{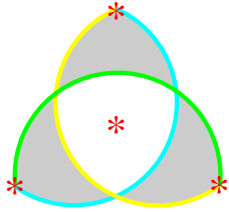}
\caption{$\C^3/\Z_3$. There are two white faces and one of them contains an orientifold. Thus,
the gauge group is $SO/Sp \times U$. The matter content can also be read off from the
diagram: three bi-fundamentals from the intersections and three (anti)symmetric fields from the
orientifolds as in Table 5.} \label{harl4}
\end{center}
\end{figure}

\begin{figure}[!htp]
\begin{center}
\includegraphics[scale=0.7]{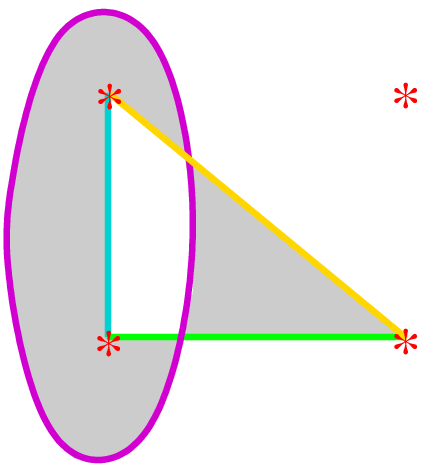}
\hspace{4cm}
\includegraphics[scale=0.7]{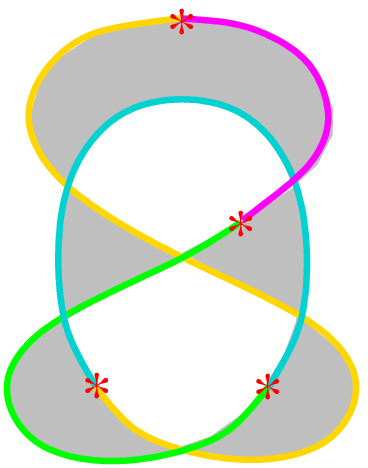}
\caption{Left: The Suspended Pinch Point theory. We simply add one more loop to the $\IC^3$ diagram. Right: The $L^{1,5,2}$ Harlequin diagram. There is a single loop which intersects itself six times and crosses the four orientifold points.}
\label{harl6}
\label{harl7}
\end{center}
\end{figure}

%------------------------------------------------------------------------------------------------------------
%---------------------------------------------------------
\clearpage

\section{$L^{aba}$ theories}

\label{section_Laba}

In this Appendix, we consider the generically singular $L^{aba}$ spaces.\footnote{$a$ and $b$
are positive integers with $a\le b$.} These are in the $L^{abc}$ family for which explicit
metrics are known \cite{metr1,metr2}. Probing a real cone over
$L^{aba}$ spaces with D3-branes gives interesting 
gauge theories \cite{Benvenuti:2005ja,
Franco:2005sm, Butti:2005sw}. The T-dual theory is an {\it elliptic model}, i.e. Type IIA
$\mathcal{N}=1$ Hanany-Witten setup \cite{Hanany:1996ie} which contains $a$ NS5-branes with
world-volume along 012345 and $b$ NS5'-branes along 012389. D4-branes (along 01236) are
suspended between the five-branes. There are $2\times(a+b)$ bi-fundamental fields. There are also
adjoints whenever two adjacent
fivebranes are parallel. We have referred to these Type IIA constructions extensively.
In this appendix we explain how they can be translated into harlequin and shiver diagrams, which can
then be orientifolded following the rules we have presented in the paper.

\begin{figure}[!htp]
\begin{center}
\psfrag{a}[cc]{(i)} \psfrag{b}[cc]{(ii)}
\includegraphics[scale=0.6]{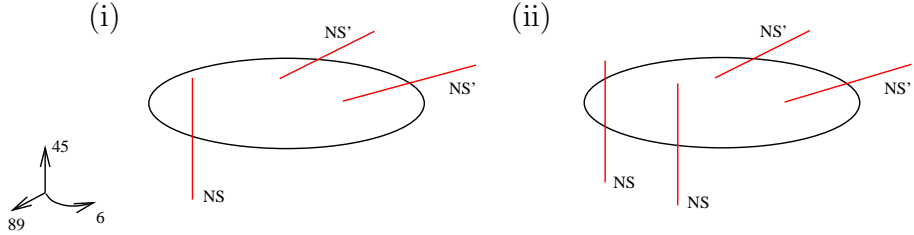}
\caption{(i) $L^{1,2,1}$ = the Suspended Pinch Point. (ii) $L^{2,2,2}$ = conifold$/\IZ_2$.}
\label{spp_elliptic}
\end{center}
\end{figure}

\fref{spp_elliptic} shows two simple examples.

Since we have been discussing orientifolds from the dimer point of view, the question
immediately arises: can this Type IIA picture be easily translated to the language of brane
tilings? The answer is yes. In order to show this, we work with Harlequin diagrams (of the dimer). 
The simple rules for constructing the diagrams are the following:

\bigskip
\begin{itemize}
\item{Start with an empty box for the fundamental cell of the tiling torus. In the following,
we heuristically identify the horizontal circular direction with $x^6$ of the elliptic models.}
\item{Draw a horizontal line. We choose the convention that it is directed to the left.}
\item{Draw vertical lines for the NS5-branes. The direction should be upward for NS and downward for NS$'$ branes.}
\item{Draw another horizontal line going to the right. It should go around the nontrivial vertical torus cycle whenever
two adjacent five-branes are parallel.}
\end{itemize}
\bigskip

\begin{figure}[ht]
\begin{center}
  \epsfxsize = 12cm
  \centerline{\epsfbox{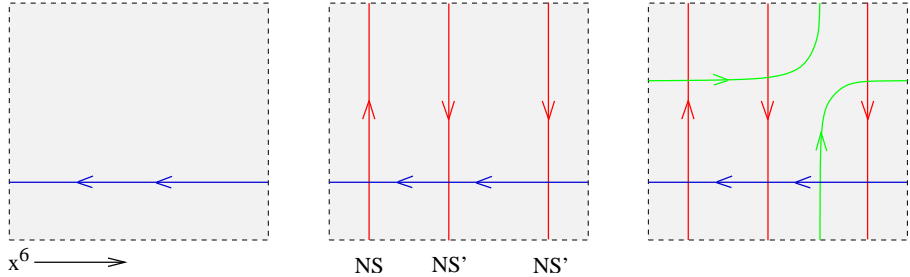}}
  \caption{Constructing the Harlequin diagram (of the dimer) of SPP. (i) First step: a horizontal blue line. (ii) Second step: 
  adding NS and NS$'$ branes.
  (iii) Last step: another ``horizontal'' green line which wraps the vertical cycle between two identical NS$'$-branes.}
  \label{spp_har}
\end{center}
\end{figure}

\begin{figure}[ht]
\begin{center}
  \epsfxsize = 4cm
  \centerline{\epsfbox{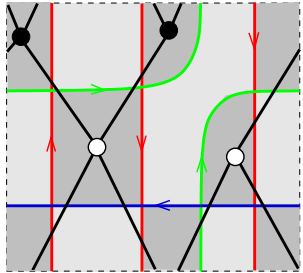}}
  \caption{The Harlequin diagram and the tiling of SPP.}
  \label{spp_har4}
\end{center}
\end{figure}

To illustrate our ideas, this is demonstrated for the SPP in \fref{spp_har}.
The resulting tiling can be seen in \fref{spp_har4}.

\begin{figure}[ht]
\begin{center}
\includegraphics[scale=0.7]{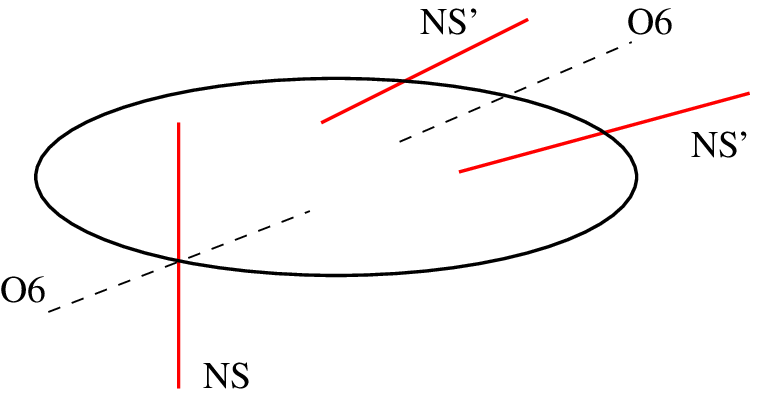}
\hspace{2cm}
\includegraphics[scale=0.7]{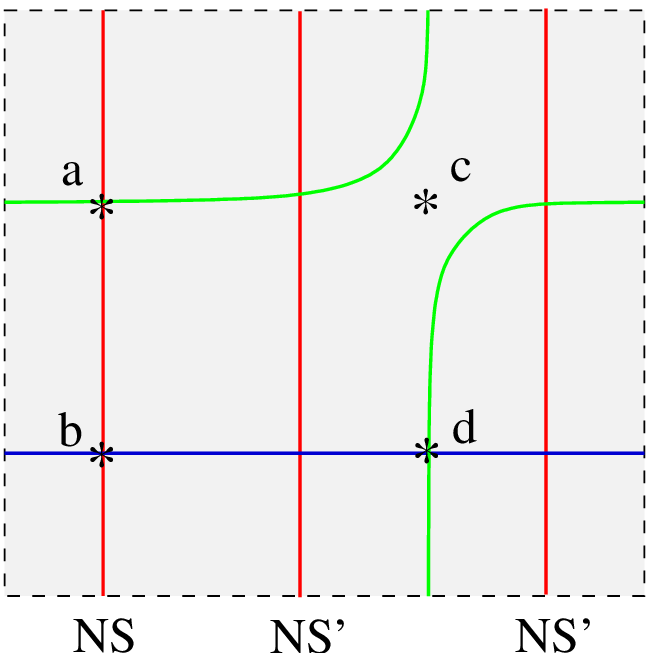}
 \caption{Left: Orientifolding SPP. Right: Four fixed points in the Harlequin diagram.}
  \label{spp_elliptic_ori}
  \label{spp_ori}
\end{center}
\end{figure}

Elliptic models can be orientifolded by including O6-planes. The $\IZ_2$ symmetry acts on
the $x^6$ direction and reduces it to an interval with two orientifold planes at the ends.
Several orientifolds from this class were constructed in \cite{Landsteiner:1997vd,
Landsteiner:1997ei, Uranga:1998uj, Park:1999eb}. There are two fixed points on the $x^6$ circle,
which determine the horizontal coordinates of the fixed points in the dimer diagram. As an example, we
take the orientifold of SPP in \fref{spp_elliptic_ori}. This has been discussed in detail in
\cite{Park:1999eb} (see Figure 3 there). The four fixed points in the tiling are presented in
\fref{spp_ori}. We immediately see that $a$ and $b$ are related to the O6-plane passing through
the NS-brane in \fref{spp_elliptic_ori}. Similarly, $c$ and $d$ correspond to the O6 between the
two NS$'$-branes.

Intuitively, the fork configuration corresponds to the case when we set opposite charges for
those orientifolds in the tiling which then become the same O6 in the Type IIA picture after
T-duality (i.~e. they are at the same horizontal position).

Finally, we briefly describe an alternative perspective on these theories, given shivers in the mirror. The toric diagram contains
no internal points, and thus the Riemman surface $\Sigma_0$ of the mirror is a sphere. The azimuthal angle of this sphere can be
identified with $x^6$.

\begin{figure}[!htp]
\begin{center}
  \epsfxsize = 8cm
  \centerline{\epsfbox{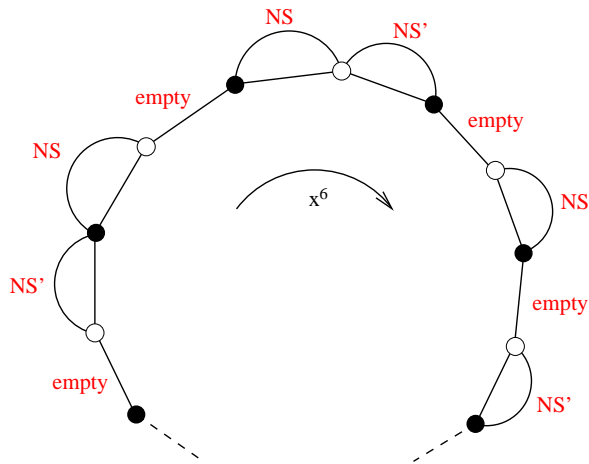}}
  \caption{Shiver for a generic $L^{aba}$. The diagram contains $a+b$ ``bubbles''.}
  \label{lababub}
\end{center}
\end{figure}

The shiver is an even sided polygon with double lines. It looks like a ring with ``bubbles''
(see \fref{lababub}). There is one bubble for each NS and NS$'$ brane. The parity of the position
of the bubble on the polygon distinguishes between the two fivebranes. The conifold
shiver (\fref{coniriem}) consists of two bubbles. On the other hand, the SPP shiver
(\fref{trivinvfig0}) has three bubbles and an empty place. Seiberg duality moves the bubbles
around while keeping the number of each type fixed. This nicely matches the Type IIA picture
where the NS5-branes are moving.

Orientifolding the shiver is relatively straightforward: the orientifold line cuts the ring into
two identical pieces. This can only be achieved if the bubbles are arranged via Seiberg duality
in a symmetric fashion. We leave the details to the interested reader.

%-------------------------------------------------------------------------------------------
%--------------------------------------------------------------------------

\end{document}